\documentclass[11pt, a4paper, logo, onecolumn, copyright]{googledeepmind}

\usepackage[authoryear, sort&compress, round]{natbib}
\usepackage{subfigure}
\usepackage{xspace}
\usepackage{microtype}
\usepackage{color,soul}
\usepackage{overpic}
\usepackage{longtable}
\usepackage{textcomp}
\usepackage{caption}
\usepackage{fontawesome5}
\newcommand{\external}[1]{#1}

\title{\textsc{\textls[50]{SynthID-Image}}: Image watermarking at internet scale}

\correspondingauthor{sgowal@google.com, pushmeet@google.com}

\keywords{provenance, watermarking, robustness}
\paperurl{}

\reportnumber{} %

\author[1,*,$\mathparagraph$]{Sven Gowal}

\author[1,*]{Rudy Bunel}
\author[1,*]{Florian Stimberg}
\author[1,*]{David Stutz}
\author[1,*]{Guillermo Ortiz-Jimenez}
\author[1,*]{Christina Kouridi}
\author[1,*]{Mel Vecerik}
\author[1,*]{Jamie Hayes}
\author[1,*]{Sylvestre-Alvise Rebuffi}

\author[1,$\mathsection$]{Paul Bernard}
\author[1,$\mathsection$]{Chris Gamble}
\author[1,$\mathsection$]{Miklós Z. Horváth}
\author[1,$\mathsection$]{Fabian Kaczmarczyck}
\author[1,$\mathsection$]{Alex Kaskasoli}
\author[1,$\mathsection$]{Aleksandar Petrov}
\author[1,$\mathsection$]{Ilia Shumailov}
\author[1,$\mathsection$]{Meghana Thotakuri}
\author[1,$\mathsection$]{Olivia Wiles}
\author[1,$\mathsection$]{Jessica Yung}

\author[1,$\dagger$]{Zahra Ahmed}
\author[1,$\dagger$]{Victor Martin}
\author[1,$\dagger$]{Simon Rosen}
\author[1,$\dagger$]{Christopher Sav\v{c}ak}
\author[1,$\dagger$]{Armin Senoner}
\author[1,$\dagger$]{Nidhi Vyas}

\author[1,$\mathparagraph$,$\ddagger$]{Pushmeet Kohli}

\affil[*]{Core contributor (randomized order)}
\affil[$\mathsection$]{Contributor}
\affil[$\dagger$]{Support}
\affil[$\ddagger$]{Sponsor}
\affil[$\mathparagraph$]{Lead}
\affil[1]{Google DeepMind}

\usepackage[acronym]{glossaries}
\newacronym{dwt}{DWT}{discrete wavelet transform}
\newacronym{dct}{DCT}{discrete cosine transform}
\newacronym{roc}{ROC}{receiver operating characteristic}
\newacronym{auc}{AUC}{area under the curve}

\newcommand{\synthid}{\textsc{\textls[50]{SynthID}}\xspace}
\newcommand{\synthidimage}{\textsc{\textls[50]{SynthID-Image}}\xspace}
\newcommand{\synthidom}{\textsc{\textls[50]{SynthID-O}}\xspace}
\newcommand{\imagen}{\textsc{\textls[50]{Imagen}}\xspace}

\newcommand{\gaussianshading}{\textsc{\textls[50]{GaussianShading}}\xspace}
\newcommand{\stablesignature}{\textsc{\textls[50]{StableSignature}}\xspace}
\newcommand{\trustmarkp}{\textsc{\textls[50]{TrustMark-P}}\xspace}
\newcommand{\trustmarkq}{\textsc{\textls[50]{TrustMark-Q}}\xspace}
\newcommand{\trustmarkb}{\textsc{\textls[50]{TrustMark-B}}\xspace}

\newcommand{\trustmark}{\textsc{\textls[50]{TrustMark}}\xspace}
\newcommand{\stegastamp}{\textsc{\textls[50]{StegaStamp}}\xspace}
\newcommand{\invismark}{\textsc{\textls[50]{InvisMark}}\xspace}
\newcommand{\wam}{\textsc{\textls[50]{WAM}}\xspace}
\newcommand{\videoseal}{\textsc{\textls[50]{VideoSeal}}\xspace}
\newcommand{\videosealzero}{\textsc{\textls[50]{VideoSeal-0.0}}\xspace}
\newcommand{\videosealone}{\textsc{\textls[50]{VideoSeal-1.0}}\xspace}
\newcommand{\imagenet}{\textsc{\textls[50]{ImageNet}}\xspace}

\usepackage{amsmath,amsfonts,bm}
\usepackage{bbm}

\def\1{\bm{1}}

\def\vc{{\bm{c}}}

\def\vx{{\bm{x}}}

\DeclareMathAlphabet{\mathsfit}{\encodingdefault}{\sfdefault}{m}{sl}
\SetMathAlphabet{\mathsfit}{bold}{\encodingdefault}{\sfdefault}{bx}{n}

\def\sA{{\mathbb{A}}}

\def\sP{{\mathbb{P}}}

\newcommand{\R}{\mathbb{R}}

\DeclareMathOperator{\sign}{sign}

\def\R{\mathbb{R}}

\begin{document}

\begin{abstract}

\external{We introduce \synthidimage, a deep learning-based system for invisibly watermarking AI-generated imagery.
This paper documents the technical desiderata, threat models, and practical challenges of deploying such a system at internet scale, addressing key requirements of effectiveness, fidelity, robustness, and security.
\synthidimage has been used to watermark over ten billion images and video frames across Google's services and its corresponding verification service is available to trusted testers.
For completeness, we present an experimental evaluation of an external model variant, \synthidom, which is available through partnerships.
We benchmark \synthidom against other post-hoc watermarking methods from the literature, demonstrating state-of-the-art performance in both visual quality and robustness to common image perturbations.
While this work centers on visual media, the conclusions on deployment, constraints, and threat modeling generalize to other modalities, including audio.
This paper provides a comprehensive documentation for the large-scale deployment of deep learning-based media provenance systems.
}
\end{abstract}

\maketitle
\section{Introduction}
\label{sec:introduction}

In the wake of powerful generative artificial intelligence (AI) systems and their accessibility in products such as Gemini, ChatGPT, Midjourney or ElevenLabs
including their corresponding open-source alternatives \citep{RombachARXIV2022,TouvronARXIV2023,LescaoARXIV2022}, leading AI companies including Google committed to advancing responsible practices in deploying AI.
One such practice consists of establishing the provenance of generated media artifacts.
In this context, the Coalition for Content Provenance and Authenticity (C2PA; \citealp{c2pa2023}),
describes provenance as serving two main purposes: allowing leading AI companies to responsibly disclose that content is AI-generated and allowing users to verify the authenticity of these artifacts.
The latter, in particular, has received considerable attention due to worries of widespread misinformation \citep{ZellersNIPS2019,WeidingerARXIV2021} or impersonation (e.g., so-called ``deep fakes''; \citealp{AgarwalCVPR2019,CarliniARXIV2020,Vaccari2020,YuICCV2021b}).
Importantly, we note that establishing provenance is materially different from detecting AI-generated content, as the public discourse tends to focus on.\footnote{\tiny\url{https://www.technologyreview.com/2023/11/06/1082996/the-inside-scoop-on-watermarking-and-content-authentication}}
Not only are such detectors\footnote{\tiny\url{https://contentatscale.ai/ai-content-detector/}, \url{https://www.compilatio.net/en/blog/best-ai-detectors}, \url{https://gptzero.me/}} generally less reliable, but even if we could reliably detect AI-generated content, they do \emph{not} necessarily establish provenance.
Fine-grained origins are typically still unknown and we believe that the increase of fidelity of future generative models will also increase the complexity of establishing provenance through detection alone.

While C2PA establishes a standard for encoding and verifying provenance through metadata that is attached to the content, this approach is vulnerable to removing said metadata \citep{ZhaoARXIV2024}.\footnote{Metadata is often stripped accidentally and can also be trivially removed.}
\external{Watermarking is a promising responsible disclosure mechanism that allows to embed provenance information within the generated content directly.}
As visible watermarks \citep{BraudawayICIP1997} have been shown to be ineffective in the age of deep learning \citep{DekelCVPR2017}, there has been a recent revival of deep learning based methods for \emph{invisible} watermarking.
Originating in the information hiding community \citep{Anderson1996,Cox2007}, invisible watermarks aim to hide information within a multimedia artifact.
Classical methods use wavelet or Fourier transforms for this purpose and recently deep learning based methods have been introduced for this task \citep{ZhuECCV2018,AdiUSENIX2018} as they promise to store more information with higher \emph{robustness} to removal attempts.
However, this exposes these methods to a wide array of attacks
that exploit vulnerabilities in the underlying deep neural networks.
Furthermore, many methods still lack robustness to common post-processing or manipulations of the content and, unsurprisingly, they are vulnerable to adversarial attempts to remove or fake the watermark, e.g. by utilizing a rich body of work on adversarial examples \citep{BiggioECMLPKDD2013,SzegedyICLR2014}.
Building (adversarially) robust watermarking systems remains an extremely challenging problem.
More importantly, there is little to no guidance on how to deploy such systems in a robust manner at internet scale.

To successfully enable provenance of AI-generated content at internet scale, a watermarking system needs to balance a number of potentially conflicting desiderata.
As the main objective during the development of generative AI models is \textbf{quality}, it is crucial for watermarking systems not to impact it.
This is commonly implied by demanding invisible watermarking.
Then, we want the produced watermarks to be \textbf{effective} in verifying provenance which also includes \textbf{robustness} to expected and unexpected changes in the underlying media artifact due to \textit{every day use}.
Deviating from every day use, we explicitly want watermarks to be \textbf{secure} against \textit{malicious use}, ranging from watermark removal or forgery attacks to attempts on model extraction.
To enable watermarking at internet scale, it also needs to be sufficiently \textbf{efficient} in terms of runtime, with minimal overhead during generation and high throughput during detection and verification.
A rich literature on watermarking, see surveys by \cite{WanNEUROCOMPUTING2022,Cox2007} and benchmarks by \cite{AnARXIV2024,PetitcolasSPM2000}, hints that many of these desiderata are conflicting and trade-offs have to be made.
This is particularly important in light of our final desiderata, \textbf{large-scale deployment}. This means that, ultimately, decisions have to be made on how to handle trade-offs and deal with imperfect robustness and security in practice.

With \synthidimage,\footnote{\tiny\url{https://deepmind.google/technologies/synthid/}} we built a watermarking system that is currently watermarking all of Google's AI-generated imagery.
Throughout development, we found that many existing surveys, benchmarks and methods do not realistically capture the above desiderata and their trade-offs, often ignoring key threat models, evaluations or deployment considerations. In this paper, we want to share a detailed account of our learnings from developing, deploying and maintaining \synthidimage since its first public announcement in August 2023.\footnote{\tiny\url{https://deepmind.google/discover/blog/identifying-ai-generated-images-with-synthid/}}
\external{Moreover, with this work, we evaluate our external variant of \synthidimage which is available via partnerships.}
Our contributions can be summarized along the structure of this paper:
\begin{description}
\item[\synthidimage (Section \ref{sec:problem})] follows a post-hoc and model-independent approach to watermarking. This means the watermark is applied on top of the AI-generated content using an encoder, not as part of the generation process, and detected using a corresponding decoder. This makes \synthid applicable to any generative model and thus maximizes utility, deployability and organizational flexibility at the cost of stringent constraints on quality and efficiency.
\item[Invisibility (Section \ref{sec:quality}):] Because quality dictates many trade-off among relevant desiderata, we meticulously benchmark the watermarks impact on quality by predominantly using human judgement on corner-case content. We frequently run internal studies and complement this with external ones. Framing these studies correctly is key. These studies are complemented using proxy metrics that enable auto-evaluation.
\item[Robustness (Section \ref{sec:robustness})] requires comprehensive and reproducible benchmarking of numerous possible transformations. This is informed by manually deciding on appropriate ranges, essentially defining when transformed content can visibly be determined as heavily tampered or when transformations change the underlying (semantic) content significantly.
\item[Payload (Section \ref{sec:payload}):] Tracking and establishing provenance goes beyond ``just'' detecting a watermark. While many approaches allow multi-bit watermarks, it is often neglected how a multi-bit payload is best used. Adapting \synthidimage's payload to its use case allows us to serve different generative models, users and deployment scenarios.
\item[Ensuring security (Section \ref{sec:attacks})] of the watermarking system involves understanding a wide range of threat models, setting them up for benchmarking and prioritizing them. The relevance of each threat model depends on how the system is deployed, i.e., what information of is publicly available or can be extracted under reasonable assumptions. Achieving perfect security is impossible; thus, we focused our efforts on making key attacks as difficult and expensive as possible.
\item[Deployment (Section \ref{sec:deployment}):] Aiming for internet-scale deployment requires solving a wide range of additional problems, including decision making (i.e., what constitutes a positive and when to abstain), defining payload use, and putting watermarking in context of other methods such as C2PA or retrieval (e.g., reverse image search).
\external{\item[New state-of-the-art (Section \ref{sec:results}):] We present experimental results for one variant of \synthidimage. This model achieves state-of-the-art performance across the most comprehensive range of transformations to date. We also provide a usage recommendation of how to adopt this external variant securely.}
Figure \ref{fig:intro} demonstrates that our models establish a new state-of-the-art in terms of quality, detection performance and robustness across a wide range or transformations.
\end{description}
We believe this to be the first comprehensive account of deploying image watermarking models at internet scale.

\external{
\begin{figure}[t]
\centering
  \begin{minipage}[c]{0.45\textwidth}
    \includegraphics[width=\textwidth]{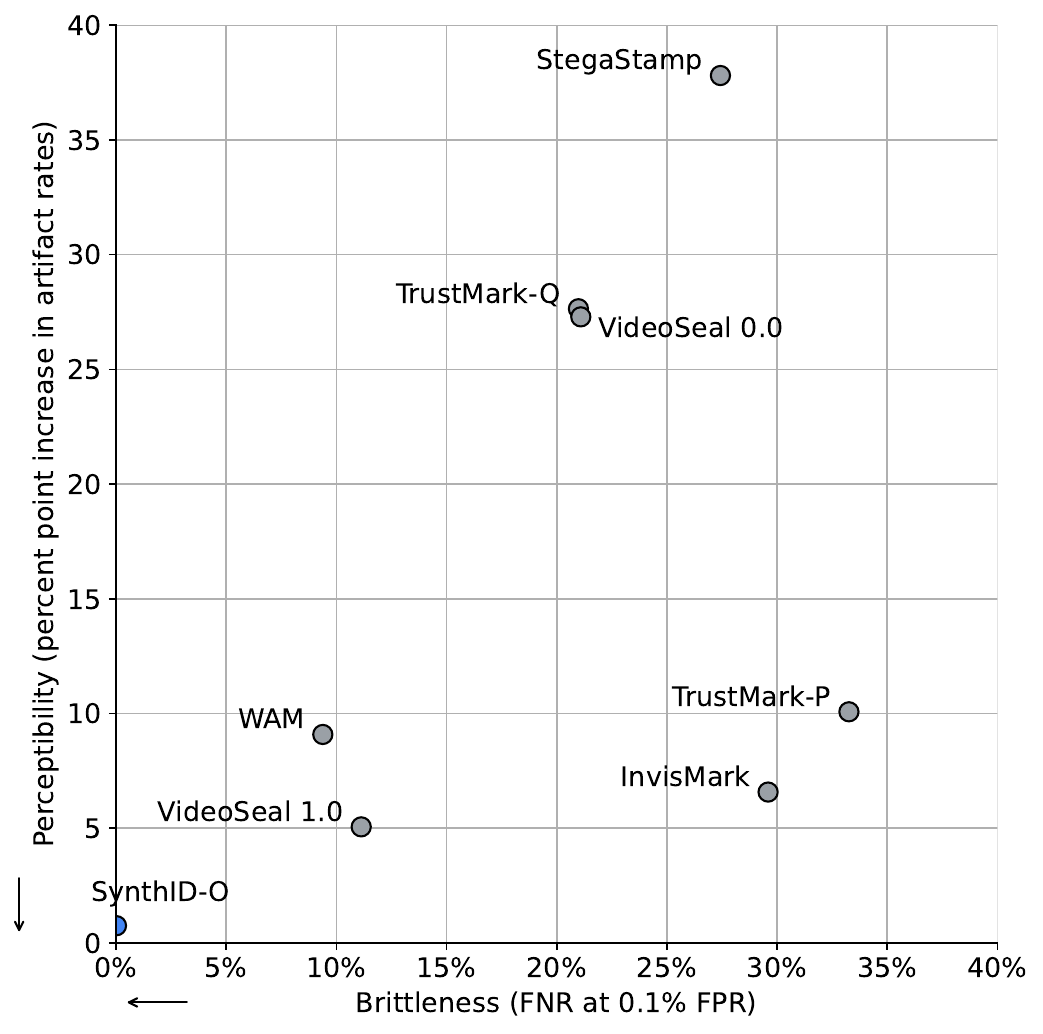}
  \end{minipage}\hfill
  \begin{minipage}[c]{\dimexpr\linewidth-0.45\textwidth-2\tabcolsep\relax}
    \caption{Quality as evaluated by human raters against FNR at 0.1\% FPR averaged across various image transformations. Each model  is calibrated to obtain 0.1\% FPR (more details to follow in Section~\ref{sec:results}). The quality is measured on the y-axis by looking at the difference in artifact rates between watermarked and non-watermarked content. \synthidom (our external variant of \synthidimage available via partnerships) achieves the highest quality (i.e., lowest perceptibility) and robustness (i.e., lowest brittleness) compared to other baselines. \label{fig:intro}}
  \end{minipage}
\end{figure}
}

\section{Our approach}
\label{sec:problem}

Digital watermarking is commonly defined as \emph{``the practice of hiding a message about an image, audio clip, video clip, or other work of media within that work itself \citep{Cox2007}.''} Generally, this implies the ability to detect and extract this hidden message, which we refer to as the watermark \emph{payload}, in a way that is robust to post-processing of the watermarked artifact.
While watermarking can be used for many modalities~(e.g., text; \citealp{dathathri2024scalable}), in this paper, we solely consider multimedia watermarking and focus on images.
Besides its original use for \emph{steganography} (i.e., the practice of \emph{``undetectably embedding a secret message''}; \citealp{Cox2007}), watermarking in the context of generative AI is primarily associated with tackling misinformation, impersonation \citep{YuICCV2021b} and copyright protection \citep{KadianWPC2021,Xuehua2010}.

The application to AI-generated content provenance is the dominant motivation of recent deep learning based watermarking methods.
It deviates significantly from earlier research \citep{WanNEUROCOMPUTING2022}.
For example, we exclusively consider \emph{blind watermarking} which assumes that we only have access to the watermarked content for detection, not pairs of original and watermarked content.
In the context of text-to-image models, blind watermarking also excludes knowledge of the used prompt.
Moreover, in contrast to steganography, we may not care about users or the general public knowing that content gets watermarked.
In fact, companies have been vocal about their plans to watermark AI-generated content as demonstrated through the ongoing dialogue with legislators and participation in C2PA.\footnote{In the context of soft-binding schemes ({\tiny\url{https://c2pa.org/specifications/specifications/1.0/specs/C2PA_Specification.html##_soft_bindings}}).}

Irrespective of the approach, there are several desiderata that we expect \synthidimage to fulfill. Most prominently, these include \textbf{quality} preservation, \textbf{robustness} to every-day changes to the watermarked content, and hiding a sufficiently large \textbf{payload} in the watermark.
The difficulty of developing any watermarking system stems from unfavorable trade-offs between these desiderata, namely the quality-robustness-payload trade-off.
For example, without requiring an invisible watermark, it becomes trivial to perform robust watermarking with large payload.
Thus, in the following, we start with an overview of these desiderata and trade-offs.

\subsection{Desiderata}
\label{subsec:problem-desiderata}

\paragraph{Quality and diversity.}

Product and research teams (especially those that build generative models) are keen to demonstrate and release their most powerful models.
This race to higher quality content means that watermarking can be an afterthought and must, as a result, maintain the quality and diversity of the generated content.
While some ad-hoc watermarking schemes are provably lossless \citep{ChristARXIV2023,YangARXIV2024}, they can reduce diversity \citep{gunn2024undetectable}.
Post-hoc schemes do not suffer from diversity degradation by construction, but may affect quality.
In practice, all methods require careful empirical assessment.
Unfortunately, quality is the most difficult requirement to properly measure and ensure.
Common \textbf{automated metrics} are unreliable and can only be used as proxies during development. Thus, in developing \synthidimage, we focused heavily on \textbf{human studies} to evaluate quality, as discussed in detail in Section \ref{sec:quality}.

\paragraph{Effectiveness and robustness.}

It is important to properly define under which assumptions a watermarking system is expected to work reliably.
For a start, the watermark should always be perfectly detectable when the watermarked content is not transformed or manipulated.
Good detection performance is commonly evaluated using the \textbf{true positive rate (TPR)} at a specific (low) \textbf{false positive rate (FPR)}. Achieving high TPR at low FPR is what we call an \emph{effective} watermarking scheme.
Beyond effectiveness, we require watermark detection to be robust to general purpose, everyday changes such as compression, quantization, basic geometric transformations or image effects.
While evaluating watermarking schemes against various such transformations is common practice, as shown in older as well as more recent benchmarks \citep{PetitcolasSPM2000,AnARXIV2024,PietARXIV2023}, there is little agreement on which exact transformations to consider (across modalities) and evaluations are often not exhaustive, leading to misleading claims about the effectiveness of watermarking \citep{ZhangARXIV2023b}.
For \synthidimage, we hand-picked and -tuned a large variety of transformations, as detailed in Section \ref{sec:robustness}.

\paragraph{Payload.}

Beyond detection, we require watermarks to include an additional multi-bit payload. Again, we expect this payload to be robust to everyday content changes. This is commonly evaluated using \textbf{bit accuracy}, i.e., accuracy of predicting individual bits, or \textbf{code accuracy}, i.e., accuracy of predicting a full multi-bit code. While most state-of-the-art methods \citep{FernandezICCV2023,ZhangARXIV2019e,YangARXIV2024} directly tackle this multi-bit watermarking problem, the exact use of the payload at test time is rarely discussed and thus evaluation usually focuses on detection only. For \synthidimage, we tackle the detection and payload problems separately and  found the payload to play a crucial role in how provenance is established, see Section \ref{sec:payload}.

\paragraph{Security.}

The security of a watermarking system entails both worst-case robustness of the watermark as well as security of the watermarking scheme itself. This includes a variety of threat models, including evasion of watermark detection, watermark forgery, model stealing or secret extraction attacks. Depending on the threat model, ensuring security might result in a number of secondary desiderata. Avoiding evasion and forgery attacks on the decoder generally requires an adversarially robust detector (see Section \ref{subsec:problem-formalization}). Thus, adversarial robustness is an important but not the only part of security. For example, to avoid model extraction attacks we may require generated watermarks to be sufficiently diverse across different payloads or hard to train surrogate models on \citep{OrekondyICLR2020}. Some aspects of security can also be addressed using infrastructure changes such as limiting the number of queries per second for the detector or providing only an encoder to trusted customers. We give a comprehensive overview of threat models and their practical relevance for \synthidimage in Section \ref{sec:attacks}.

\paragraph{Efficiency.}

Deploying watermarking at scale requires reasonably low latency for both \emph{encoding} and \emph{decoding} the watermark. For encoding, this is usually measured relative to the latency of content generation. Because this is often user-facing, we aim for a single-digit percentage of additional overhead. For decoding, it is important to realize that it may be run orders of magnitude more frequently than encoding. This fact is commonly ignored in the literature where decoding is framed as an ad-hoc action initiated by a user who wants to check for the watermark. However, we expect watermark decoders to be run at increasingly large scale to check the majority of content that is shared on the web, e.g., as part of social media, web search and scraping, training data creation, etc. Thus, decoding efficiency might often be the bottleneck. It is also important to realize that decoding may be run in addition to other checks such as verifying C2PA \citep{c2pa2023} metadata and content pre- or post-processing (e.g., video encoding/decoding). We benchmark efficiency in terms of throughput per second and hardware unit (e.g., GPU). At encoding time, batch size will be pre-determined by generation requests; for decoding, batch size can be chosen to optimize throughput.

\paragraph{Deployment.}

As \synthidimage is intended for internet-scale deployment, it clearly distinguishes itself from prior work (see Section \ref{sec:related-work} for a discussion). This introduces a number of additional requirements not sufficiently considered previously. Most importantly, different deployment settings -- including internal deployment, watermarking-as-a-service or open watermarking models -- have significant influence on the above desiderata and key design choices. This includes proper decision making, versioning of watermarks and how \synthidimage fits into a larger landscape of methods for provenance, e.g., C2PA or search-based methods \citep{BalanCVPRWORK2023,BharatiTIFS2021,Zhang2020,NguyenICCV2021}. We found that such deployment considerations had the largest impact on deciding what approach to use and which operating point in terms of the above desiderata to choose.

\subsection{Post-hoc, model-independent watermarking}

Watermarking schemes can be categorized along several, partly orthogonal dimensions.
Because we focus on blind watermarking only, watermarking has to shift the distribution of generated content in a detectable manner. Thus, blind watermarking is also referred to as distributional watermarking. Additionally, recent literature typically distinguishes between ad-hoc and post-hoc watermarking depending on whether the process of watermarking happens during the generative process (ad-hoc) \citep{FernandezICCV2023,WenARXIV2023,YangARXIV2024} or as a post-processing step (post-hoc) \citep{WenARXIV2019,HayesNIPS2017,HayesARXIV2020b,TancikCVPR2020,LuoCVPR2020,BuiARXIV2023}. Ad-hoc methods are typically integrated into the corresponding generative model and thus require at least partial access to the underlying generative model. Post-hoc watermarking, in contrast, can be performed without any knowledge or assumptions on the generative model and is thus model-independent.
We deliberately designed \synthidimage as a post-hoc, model-independent approach, a choice largely based on deployment considerations while ensuring the best trade-off between the remaining desiderata.

\paragraph{Advantages.}

First and foremost, \synthidimage must consistently watermark all of Google's current and future AI-generated content, stemming from a variety of different models.
We believe that the promise of watermarking in the context of preventing misinformation stems from this consistency, i.e., being able to mark \emph{all} content using a single (or as few as possible) watermarking scheme(s).
Because ad-hoc approaches are inherently tied to the respective generative model architectures and versions, they are unable to fulfill this promise at scale.
For example, having different watermarks for different generative models or versions necessitates running many multiple detectors or cleverly training ``meta-detectors''.
With a post-hoc approach, this can be handled using a payload.
We also found that this creates less organizational dependencies and allows more flexibility in how the watermarking system is deployed internally and externally.
For example, in contrast to most ad-hoc approaches, \synthidimage could be open-sourced or deployed as-a-service without implications for any generative model.
Moreover, it is easier to debug, update or simply turn on or off.
We appreciate that some ad-hoc methods can be used in a post-hoc way \citep{FernandezICCV2023}; but the dependence on details of the underlying generative model remains, combining the disadvantages of ad- and post-hoc approaches rather than their advantages.

\paragraph{Drawbacks.}

The main disadvantage of post-hoc approaches is that they are inherently lossy, i.e., they impact the quality of the generated content.
This is in contrast to ad-hoc approaches which can be completely lossless \citep{YangARXIV2024}. However, we found that reaching a level of quality where the watermark is essentially invisible to (expert) users is possible in practice.
We also believe that the impact of watermarking on quality is also easier to measure for post-hoc approaches.
This is because we can directly compare watermarked and non-watermarked content to judge invisibility, as done in our human studies, see Section \ref{sec:quality}.
For ad-hoc approaches, by construction, judging quality is more entangled with the quality of the underlying generative model.
Ad-hoc approaches risk reducing diversity of generations~\citep{gunn2024undetectable}, which is incredibly difficult to test for.

\noindent There are various other implications of our choice.
For example, post-hoc watermarking causes additional computational overhead for adding the watermark compared to ad-hoc approaches.
This overhead can be limited using architectural changes, but cannot be entirely eliminated.
This is rather critical for image generation where watermarking has to be performed \emph{after} generation; for audio and video, generation and watermarking can be parallelized to a large extent. On the other hand, watermark detection can be less efficient for ad-hoc methods, for example, if they require expensive inversion of the diffusion process \citep{WenARXIV2023,YangARXIV2024}.
As we expect detection to be run several orders of magnitude more often, this is an important consideration.
We also noticed that robustness of post-hoc approaches is easier to benchmark, since content to be watermarked can be fixed independent of the generative model, and more controllable.
Watermark robustness of ad-hoc approaches may be entangled with internal representations of the underlying generative models.

\subsection{Training formulation}
\label{subsec:problem-formalization}

As a post-hoc scheme, we used an encoder-decoder approach for \synthidimage following prior work \citep{WenARXIV2019,HayesNIPS2017,HayesARXIV2020b,TancikCVPR2020,LuoCVPR2020,BuiARXIV2023,BuiCVPR2023}.
To be more specific, let $\vx\in\mathcal{X}$ be an image, where $\mathcal{X}\subseteq\R^n$ represents the set of all images (in practice, the set of all images is finite and countable).
We assume there exists a metric $d:\mathcal{X}\times\mathcal{X}\to\R_+$ that encodes the perceptual similarity between two images.
We will use this metric to formalize the concept of \emph{invisibility} of a watermark.
A post-hoc watermarking scheme is a pair $\langle f, g \rangle$ consisting of an encoder function $f:\mathcal{X}\to\mathcal{X}$, which adds an identification mark, and a decoder function $g:\mathcal{X}\to\{\pm 1\}$, which tries to detect if the mark is present.
Let $\mathbb{P}_{\mathcal{X}}$ be a target distribution that we would like to watermark.
Given a perceptual threshold $\epsilon > 0$, the optimal watermarking scheme on $\mathbb{P}_{\mathcal{X}}$ solves
\begin{align}
        \min_{f, g} \quad & \mathbb{E}_{\vx\sim\mathbb{P}_{\mathcal{X}}} \left[\ell(g(f(\vx)), +1) + \ell(g(\vx), -1)\right] \label{eq:learning_loss}\\
    \operatorname{s.t.} \quad & d(\vx, f(\vx))\leq \epsilon, \nonumber
\end{align}
where $\ell$ represents a suitable classification loss function.
Assuming that $g$ makes a decision by thresholding some logit using a threshold $\kappa$, the classification loss we are ideally interested in is $\ell(g_\kappa(f(\vx)), k) = \mathbbm{1}[\text{sign}(g_\kappa(f(\vx))) = k]$.
In practice, an appropriate approximation such as cross-entropy is used to train $g$ and the threshold $\kappa$ is calibrated for a target FPR post-training.
Minimizing this loss will ensure that the watermarking scheme is \textbf{effective}, i.e., is able to watermark content and detect them.
The constraint on $d$, in contrast, will ensure \textbf{quality}, i.e., the watermark being invisible.
It is important to note that Equation~\eqref{eq:learning_loss} assumes that images to watermark and non-watermarked images come from the same distribution $\mathbb{P}_{\mathcal{X}}$.
While we make this assumption to keep our watermarking scheme future-proof and universal across various generative models, this is clearly sub-optimal when targeting specific AI-generated content (e.g., coming from a family of text-to-image models).
We account for this discrepancy during the calibration process which is applied post-training.

We are particularly interested in \textbf{robust} watermarking schemes.
That is, schemes that are robust to everyday alterations of watermarked samples that still preserve the usefulness and semantic content of the samples.
We define a distribution of valid transformations $\mathbb{P}_\mathcal{T}$ that semantically preserve the content and require robustness \emph{on average}, i.e.,
\begin{align}
    \min_{f, g} \quad & \mathbb{E}_{\vx\sim\mathbb{P}_{\mathcal{X}},\tau\sim\mathbb{P}_\mathcal{T}}\left[\ell(g(\tau(f(\vx))), +1) + \ell(g(
    \tau(\vx)), -1)\right] \label{eq:watermarking} \\
    \operatorname{s.t.} \quad & d(\vx, f(\vx))\leq \epsilon. \nonumber
\end{align}
The difficulty of this problem is clearly determined by the perceptual threshold $\epsilon$ and the distribution over transformations $\mathbb{P}_\mathcal{T}$, resulting in a robustness-quality trade-off.
Clearly, allowing more visible watermarks will render the system more robust to stronger transformations.
Note again that Equation~\eqref{eq:watermarking} assumes that the same transformations are applied to watermarked and non-watermarked images.
This is not necessarily the case in practice but it simplifies analysis and evaluation of the resulting scheme.

For a watermarking scheme to be \textbf{secure}, we might also need to be robust to specific malicious modifications of the content.
Attackers may want to manipulate the images $\vx$ or $f(\vx)$ to provoke false positives and false negatives, respectively.
This leads to a \emph{worst-case} formulation of the above optimization problem that requires adversarial training \citep{MadryICLR2018} which remains largely unsolved.
For \synthidimage, we found that it is sufficient to enforce adversarial robustness only in settings where adversaries have limited information (e.g., black-box access) and constrained compute.
As training a perfectly robust and secure watermarking scheme may be infeasible, this shifts the focus to making malicious attacks as difficult as possible.

An important aspect of a useful provenance system is the ability to discern content from different actors. As such, having the ability to embed a hidden message, called \textbf{payload}, in the watermark is fundamental.
In this setting, training a  watermarking scheme becomes a multi-bit classification problem rather than a detection problem.
More formally, let $\vc\in\{\pm1\}^C$ be a $C$-bit binary payload.
The encoder function $f:\mathcal{X}\times\{\pm1\}^C\to\mathcal{X}$ now crafts a watermark based on both $\vx$ and $\vc$ and the decoder function $g:\mathcal{X}\to\{\pm1\}^C$ is now tasked with recovering the payload $\vc$ rather than detecting the presence of the watermark alone.
The robust watermarking problem then becomes:
\begin{align}
    \min_{f, g} \quad & \mathbb{E}_{\vx\sim\mathbb{P}_{\mathcal{X}},\vc\sim\mathbb{U}_{\{\pm1\}^C},\tau\sim\mathbb{P}_\mathcal{T}}\left[\ell(g(\tau(f(\vx, \vc))), \vc)\right] \label{eq:payload} \\
    \operatorname{s.t.} \quad & d(\vx, f(\vx))\leq \epsilon, \nonumber
\end{align}
where $\mathbb{U}_\sA$ represents the uniform distribution over the set $\sA$.
Clearly, this problem becomes more difficult for longer payloads, i.e., larger $C$.
This means we are now dealing with a \textbf{robustness-quality-payload} trade-off.
For \synthidimage, we solve both the detection and the payload problem, i.e., Equations \eqref{eq:watermarking} and \eqref{eq:payload}, separately, without making any assumptions on the payload $\vc$ such that it can serve a variety of purposes.

\subsection{Benchmarking to control trade-offs}
\label{subsec:problem-discussion}

\paragraph{Primary trade-off.}
The most important trade-off between our desiderata is the robustness-quality-payload trade-off.
More visible watermarks are easier and more robust to detect and can store more bits.
In a nutshell, we want the watermark to be as robust and contain as many bits as possible while staying invisible. As we will see in Section \ref{sec:quality}, the main bottleneck of this approach is objectively ensuring invisibility through appropriate metrics and studies.
In practice, we control these trade-offs using a combination of different losses, data augmentation methods and payload encoding strategies.
Moreover, the data distribution has a large impact on this trade-off.
For example, requiring invisible watermarks can be harder for some images (e.g., with less structure).
Requiring high quality on these examples leads to less robustness overall.
Thus, deciding to not watermark specific content can improve the trade-off significantly.

\paragraph{Secondary trade-offs.}
Besides the robustness-quality-payload trade-off, there are various secondary trade-offs. For example, requiring adversarial robustness or preventing watermark extraction or exchange attacks may worsen quality.
Another example is using a cascade of decoders, with more efficient decoders offering less robustness first, which allows to trade-off robustness and efficiency.

\noindent For \synthidimage, we found that the most effective way of controlling these trade-offs is proper end-to-end benchmarking of all desiderata of interest. While we will discuss appropriate metrics in detail in the following sections, we want to emphasize that end-to-end benchmarking in itself is a non-trivial endeavor. Critically, benchmarking needs to include all key metrics and threat models to highlight relevant trade-offs. This requires a high degree of automation and tooling around model development.
Training and evaluation will also inherently include randomness.
It is important to be able to control this randomness appropriately for debugging and reproducibility.
For transformations and attacks this also includes appropriately tuning hyper-parameters. Moreover, the choice of datasets for evaluation is important: key considerations are whether to benchmark on natural or generated images and how to include corner-case content, especially for quality evaluation.

\section{Invisibility}
\label{sec:quality}

Quality is the most difficult requirement to properly evaluate in practice because ``invisibiliy'' or any quality change in generated content is tricky to objectively quantify.
So far, in Section~\ref{subsec:problem-formalization}, we worked with a perceptual metric $d$ (many prior publications also work with an unknown oracle instead; \citealp{ZhangARXIV2023b}). Of course, there exist automated metrics and proxies, including PSNR, SSIM \citep{WangTIP2004} or pre-trained perceptual models such as L-PIPS \citep{ZhangARXIV2018}, FID \citep{HeuselNIPS2017}, CLIP \citep{RadfordICML2021} or CMMD \citep{Jayasumana2023RethinkingFT}.
However, none of these metrics can ultimately replace human evaluation. This is particularly relevant when dealing with a wide distribution of generated content that might not match benign, natural content.
Learned metrics will always be biased to the content distribution that these metrics have been developed for.
In the case of images, these are typically ImageNet-style natural images.
For \synthidimage, we decided to go with rather basic metrics such as PSNR or SSIM, using them with extreme caution and mostly to perform relative comparisons while relying heavily on human evaluation, both internally and externally, to decide invisibility in absolute terms. The latter also includes hand-tuning relevant transformations considered during robustness evaluation, as detailed in Section \ref{sec:robustness}. While we also rely heavily on standard datasets, we took particular care to run human studies on actual AI-generated content from state-of-the-art generative models, using a diverse set of prompts.

\paragraph{Setups.}
Independent of the metric, there are two modes of quality evaluation. First, we can compare pairs of original and watermarked content. This \emph{side-by-side} comparison allows to directly quantify a perceptual distance and is the default for most post-hoc approaches where we have a one-to-one correspondence between original and watermarked content.
In Equation \eqref{eq:watermarking}, this means directly evaluating the distance $d$. Second, as users do not typically have access to the original content alongside the watermarked content, we can also compare the \emph{distributions} of a metrics. This is how ad-hoc approaches are typically evaluated, e.g., by computing FID or CLIP scores for original and watermarked content and performing hypotheses tests to see if both distributions are distinguishable \citep{YangARXIV2024}.
However, using CLIP also assumes access to the corresponding prompts and is thus limited to ad-hoc approaches.
This approach is typically more difficult as we cannot directly judge, for a fixed prompt or image, to what extent quality changes.
For \synthidimage, we generally found the first approach to be more appropriate for internal evaluation as it allows for more scrutiny. For studies with external users, in contrast, we follow the second approach because it represents the more realistic setting.

\paragraph{Internal rater studies.}

More specifically, during fast-paced development, we used a reasonably small set of generated images to run side-by-side comparison studies within the team for most model candidates.
This involves showing the original and watermarked content side-by-side, randomizing which is the watermarked and asking ourselves to identify the watermarked one. Across the team, we usually target a performance close to random chance $50\% \pm 10\%$.
This is a particularly high bar since watermarks are easier to spot side-by-side, with the ability to zoom in and out of the image.
Moreover, throughout development, we learned to spot watermarks well as we usually know how watermarks of specific model versions look.
Over time, as watermarks got less visible, we also added an option to abstain.
We included hand-picked examples of generated content, often those that we found difficult to watermark or informed by direct feedback from colleagues working on the corresponding generative models. This also contributed to these studies being extremely conservative. 
In between these studies, we typically compare models in relative terms using PSNR, SSIM or similar automatic metrics, especially to hill-climb on quality or other desiderata.

\begin{figure*}
    \begin{overpic}[width=0.19\textwidth]{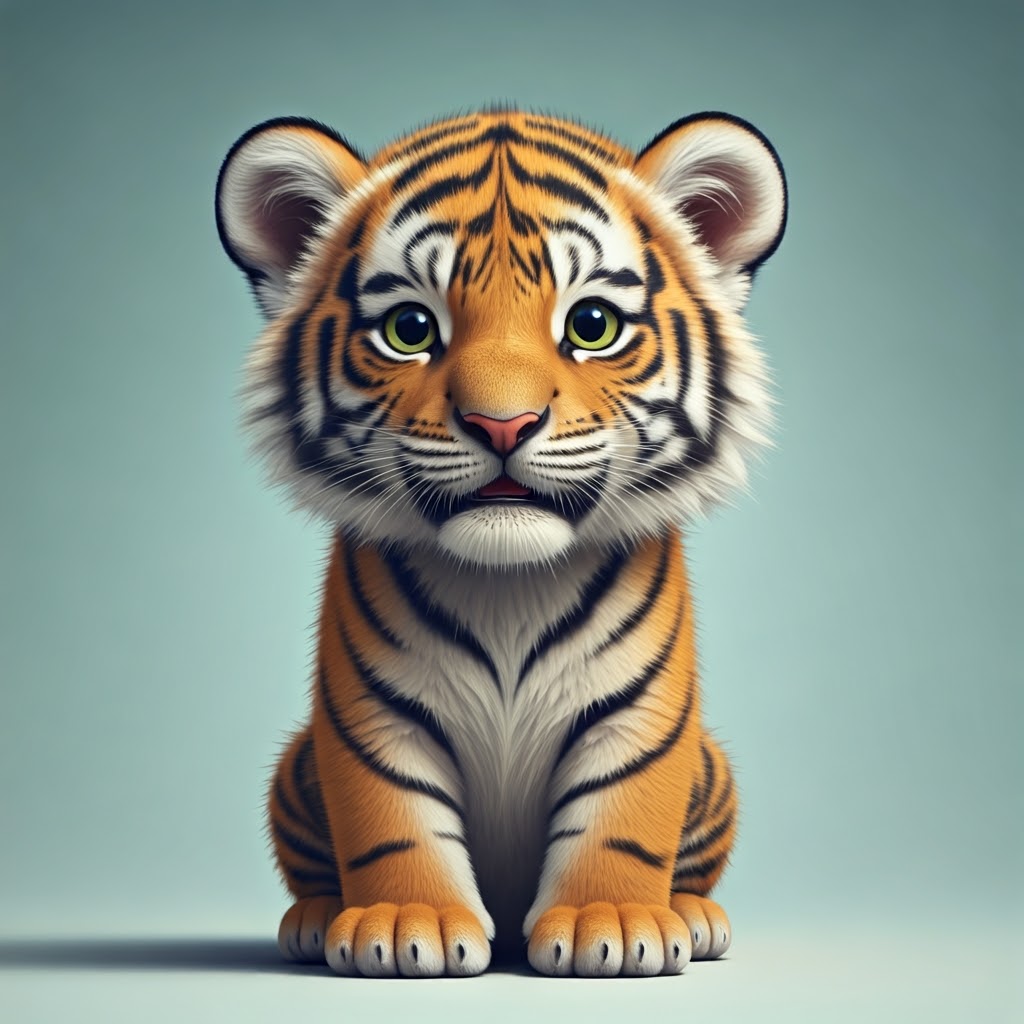}
    \put(0,102){\tiny \bf 3D Rendering}
    \end{overpic}
    \begin{overpic}[width=0.19\textwidth]{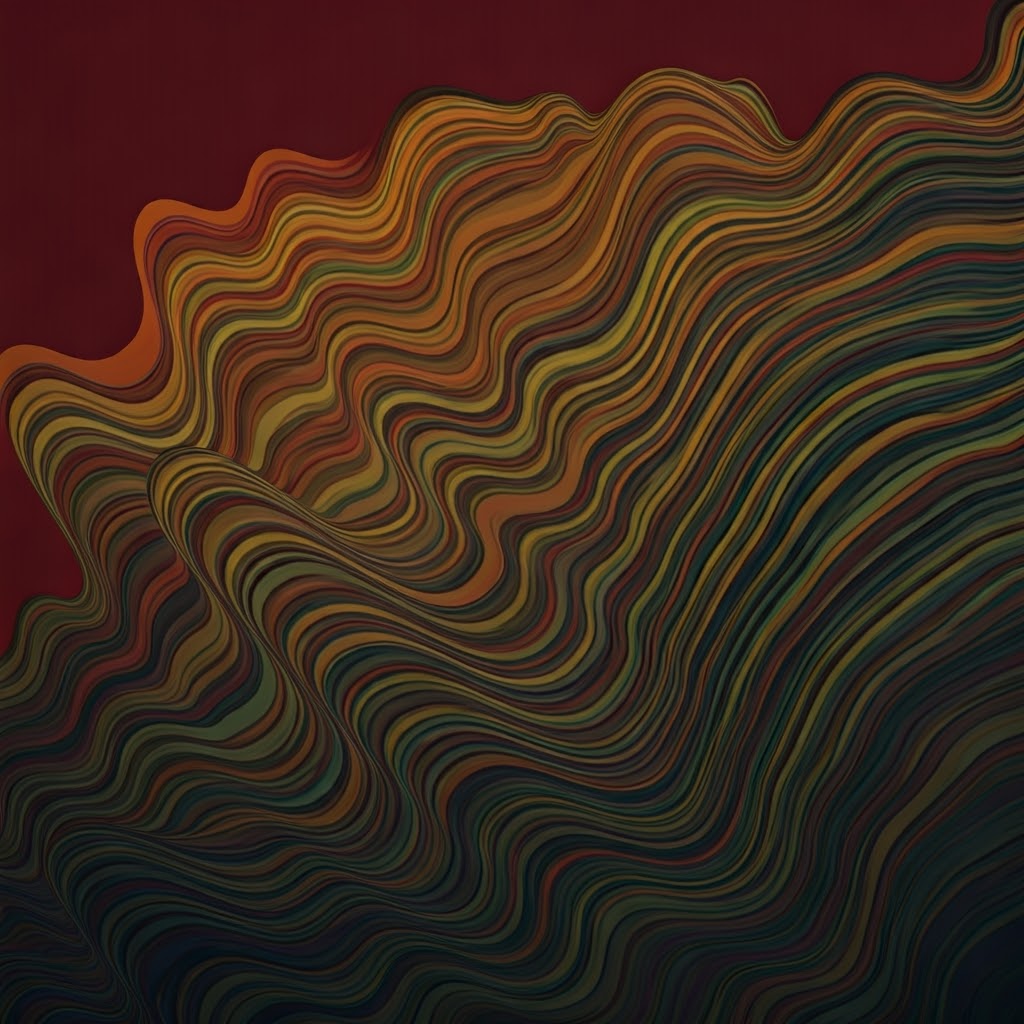}
    \put(0,102){\tiny \bf Abstract art}
    \end{overpic}
    \begin{overpic}[width=0.19\textwidth]{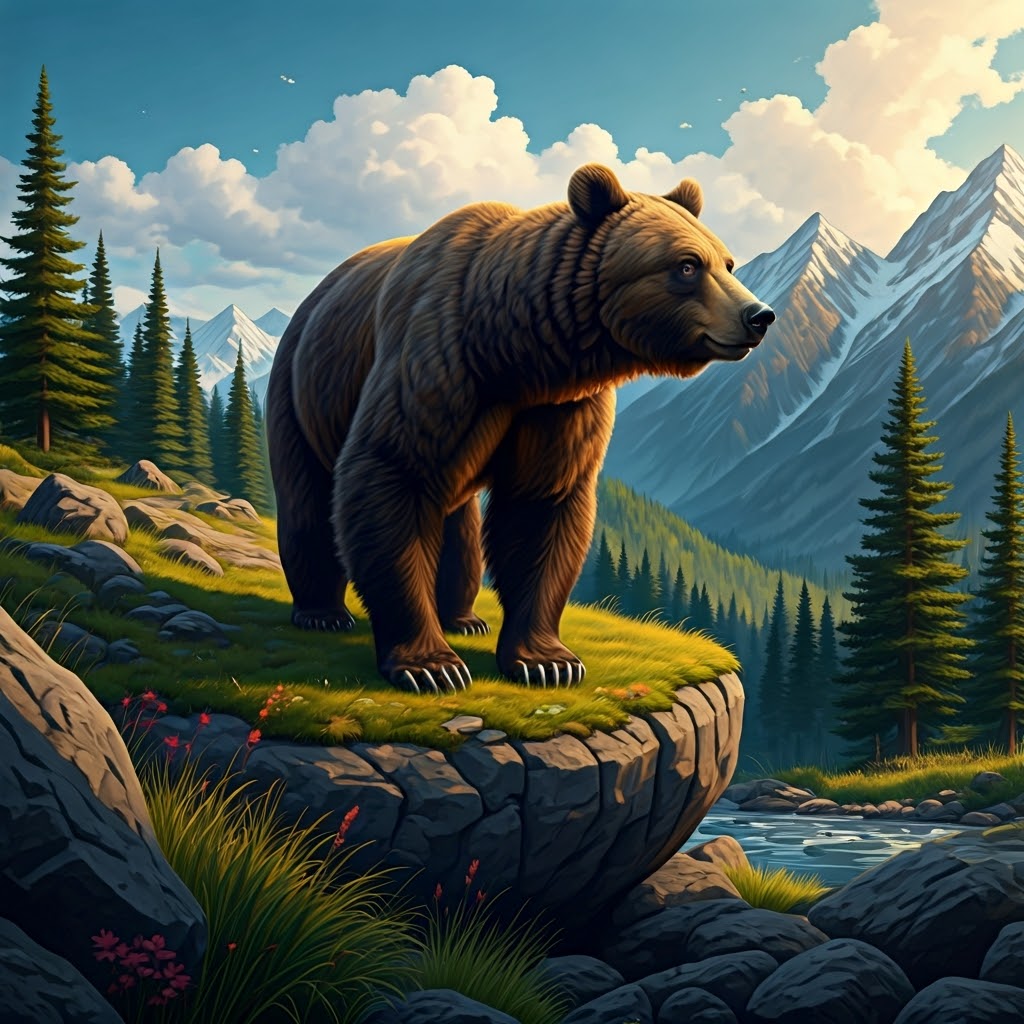}
    \put(0,102){\tiny \bf Animal Painting}
    \end{overpic}
    \begin{overpic}[width=0.19\textwidth]{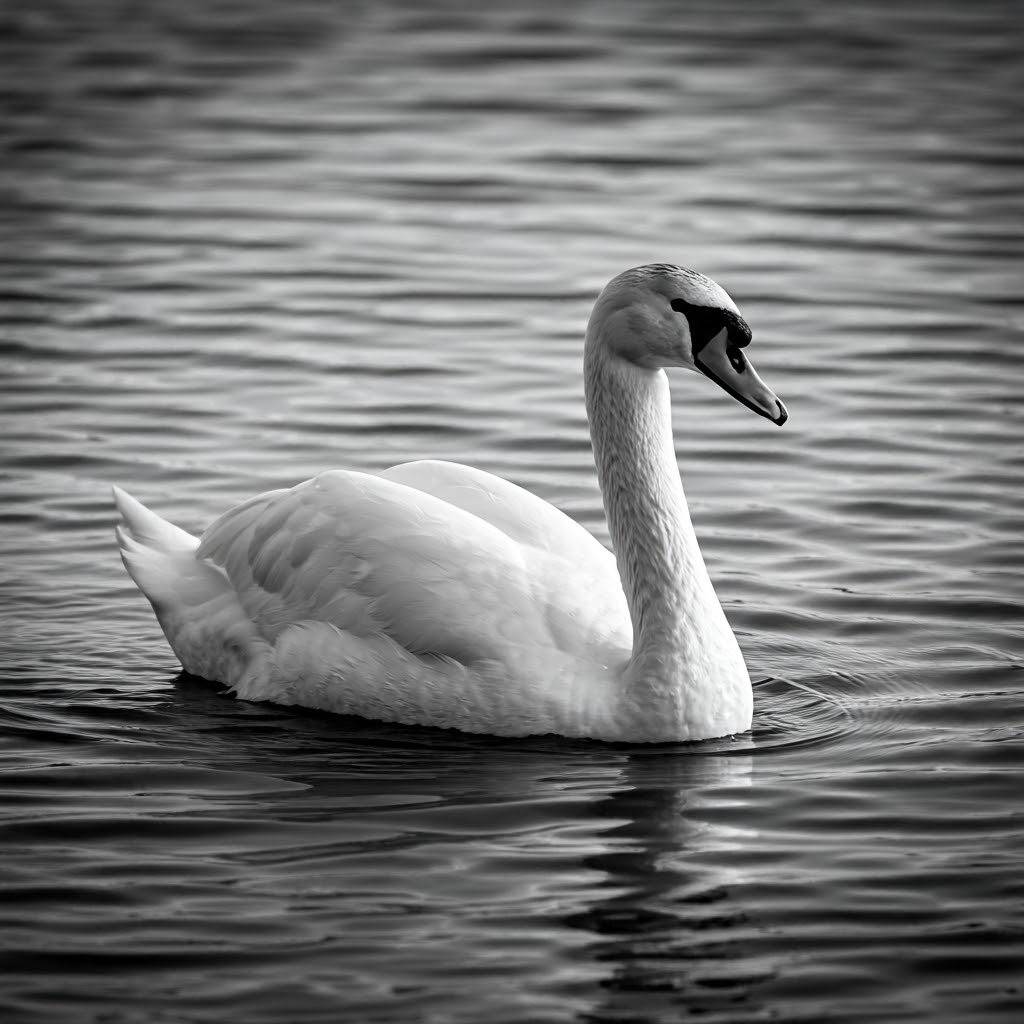}
    \put(0,102){\tiny \bf Animal Photo}
    \end{overpic} \vspace{0.6em}%
    \begin{overpic}[width=0.19\textwidth]{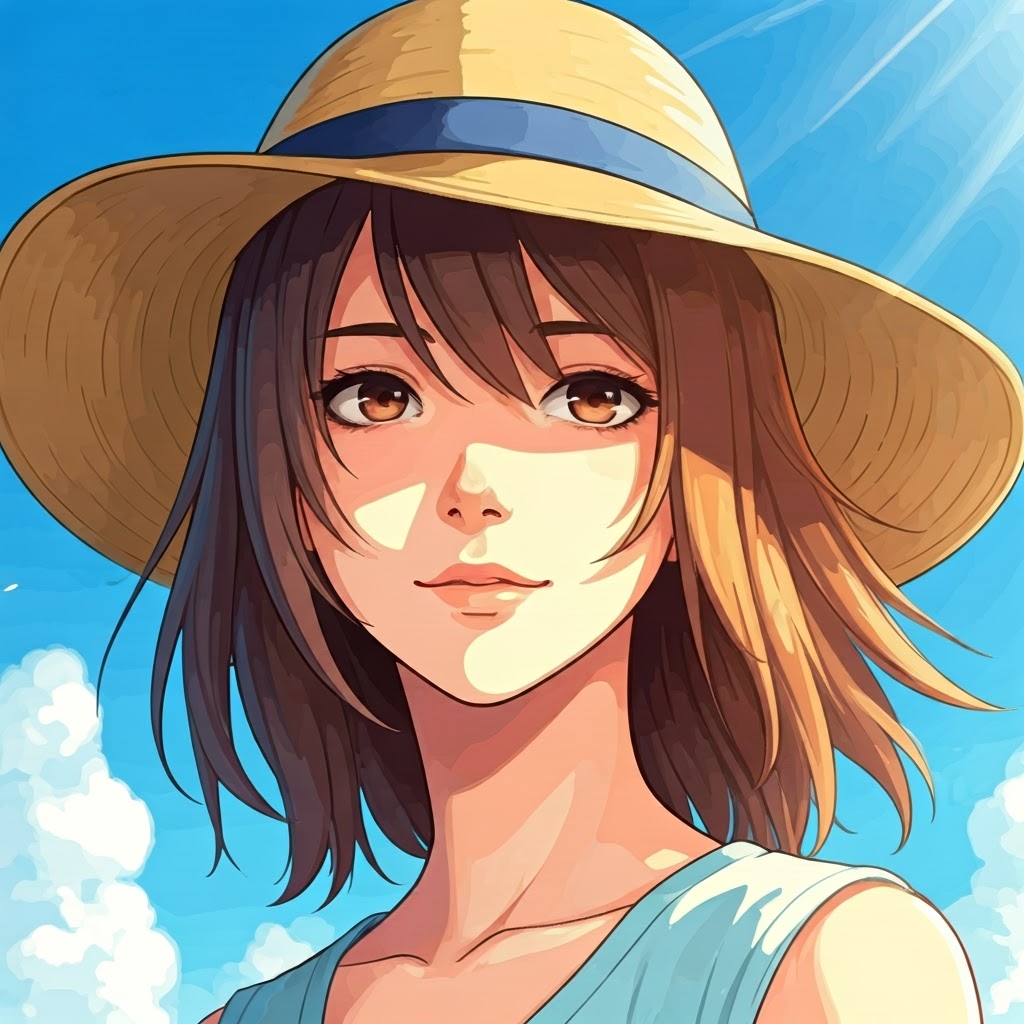}
    \put(0,102){\tiny \bf Cartoon}
    \end{overpic} \\\vspace{0.6em}%
    \begin{overpic}[width=0.19\textwidth]{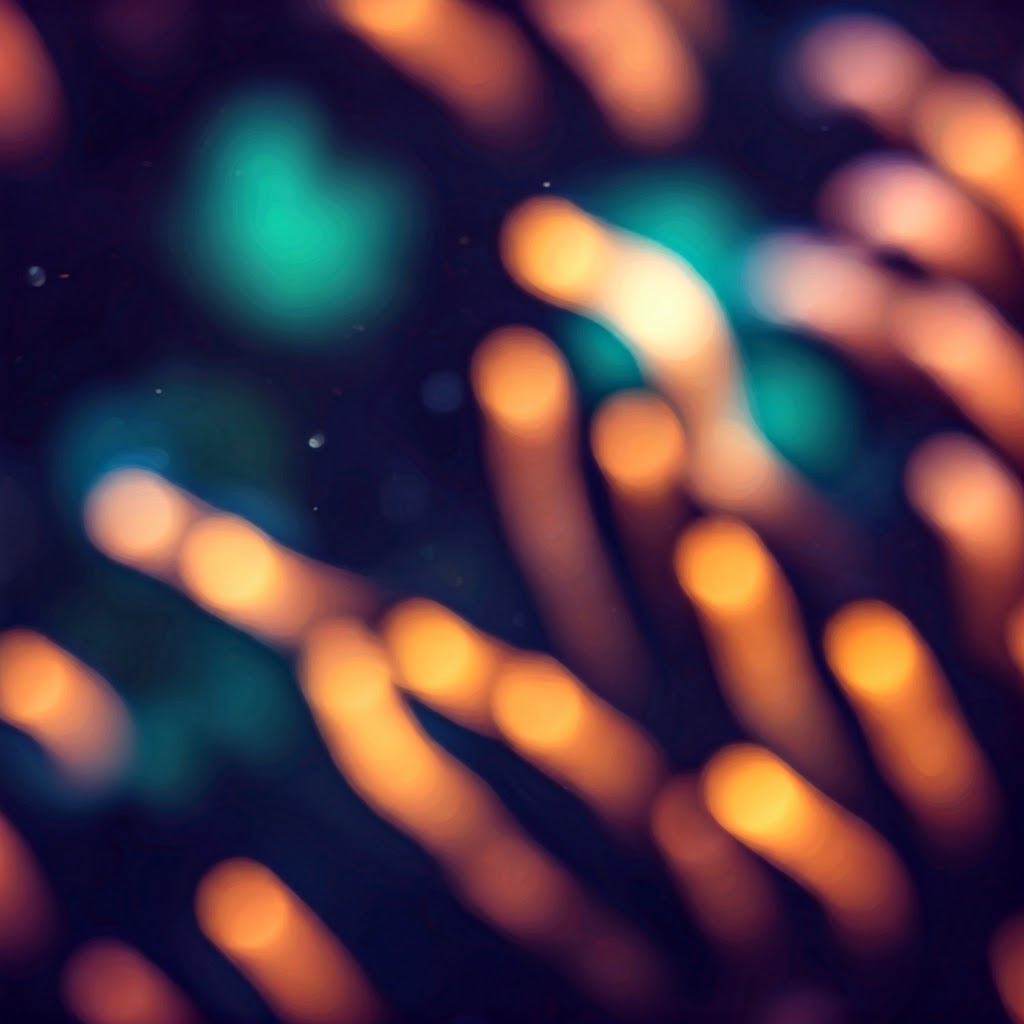}
    \put(0,102){\tiny \bf Defocused Blurry}
    \end{overpic}
    \begin{overpic}[width=0.19\textwidth]{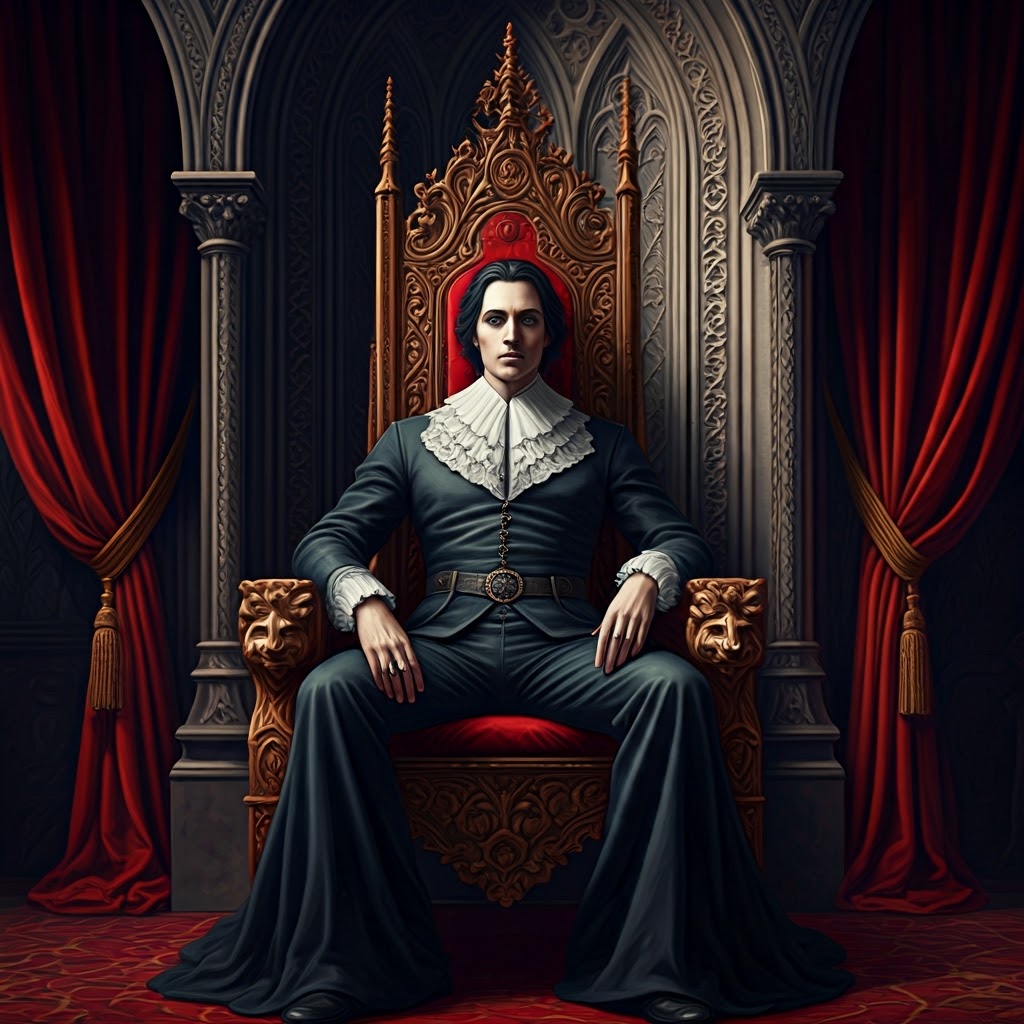}
    \put(0,102){\tiny \bf Fantasy Painting}
    \end{overpic}
    \begin{overpic}[width=0.19\textwidth]{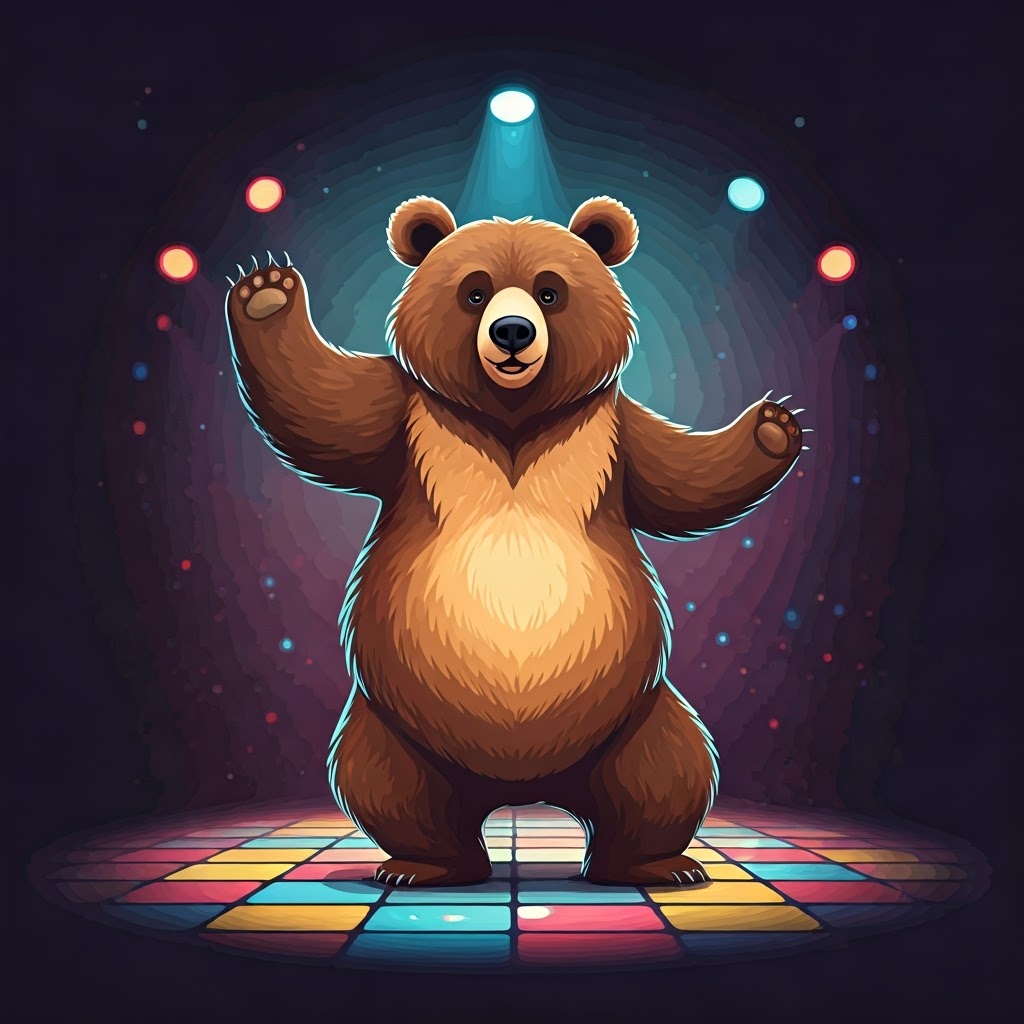}
    \put(0,102){\tiny \bf Flat art}
    \end{overpic}
    \begin{overpic}[width=0.19\textwidth]{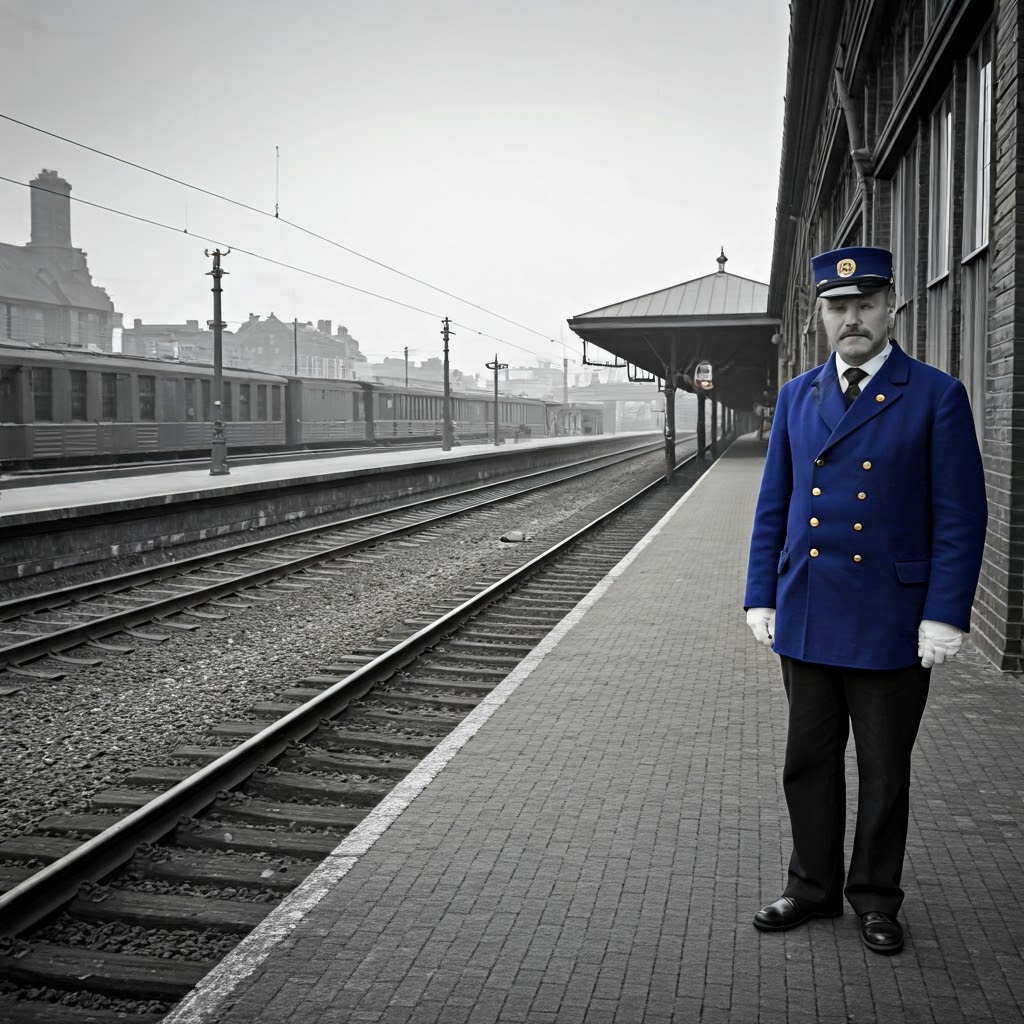}
    \put(0,102){\tiny \bf Historic Photo}
    \end{overpic}
    \begin{overpic}[width=0.19\textwidth]{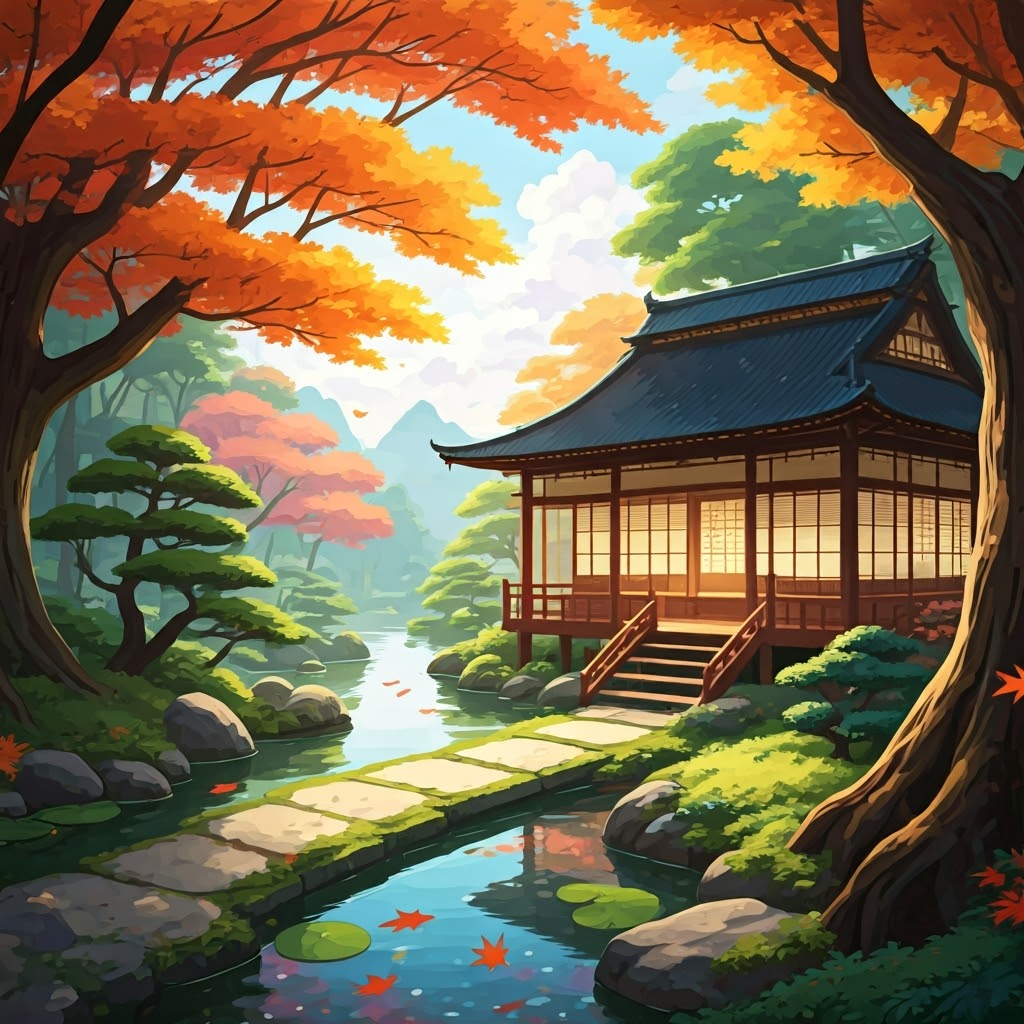}
    \put(0,102){\tiny \bf Landscape Painting}
    \end{overpic} \\\vspace{0.6em}%
    \begin{overpic}[width=0.19\textwidth]{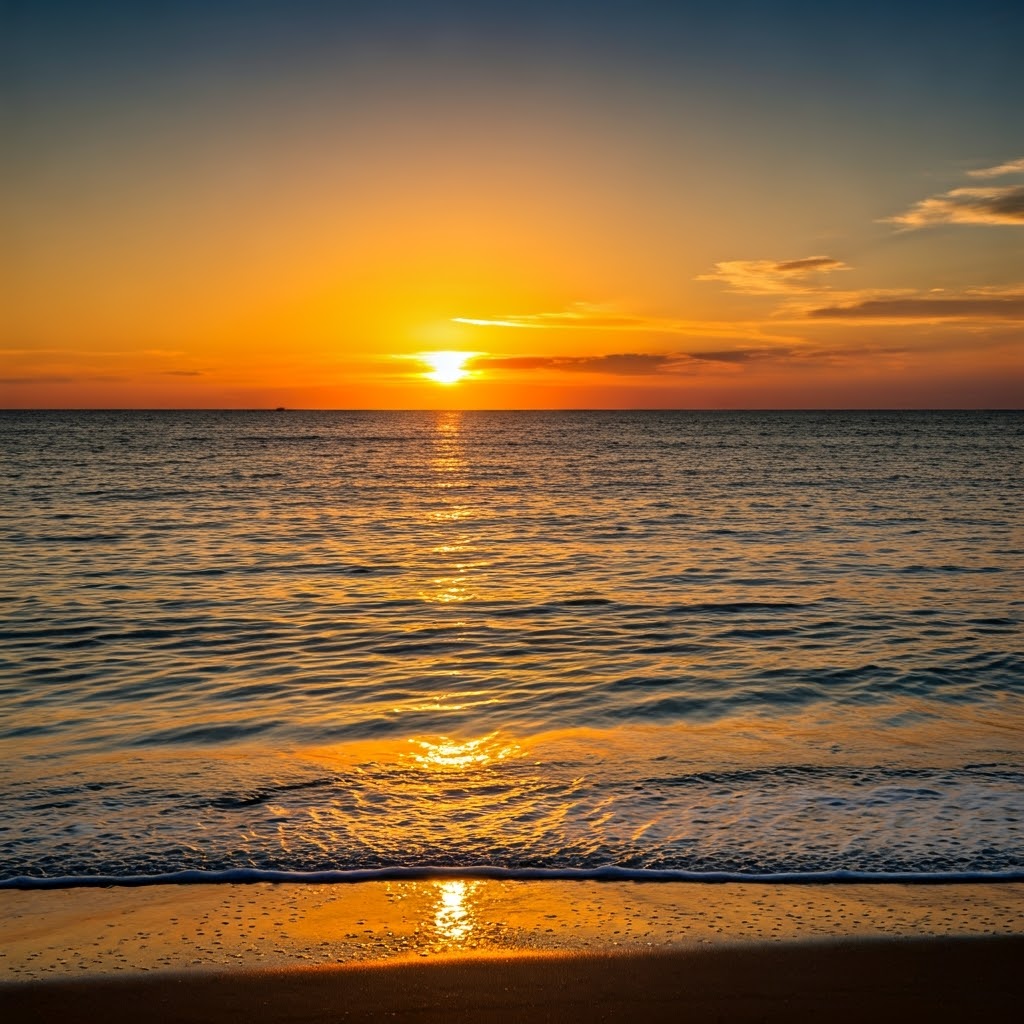}
    \put(0,102){\tiny \bf Landscape Photo}
    \end{overpic}
    \begin{overpic}[width=0.19\textwidth]{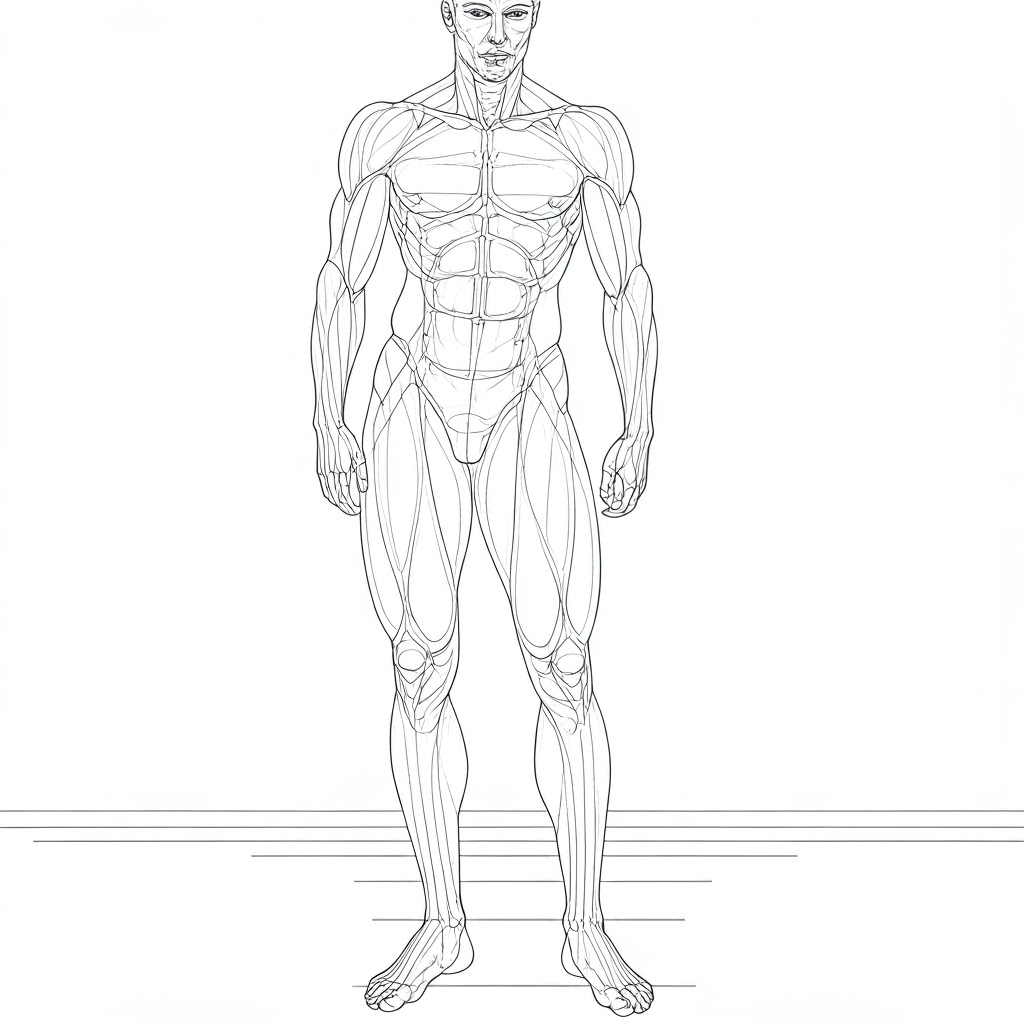}
    \put(0,102){\tiny \bf Line Drawing}
    \end{overpic}
    \begin{overpic}[width=0.19\textwidth]{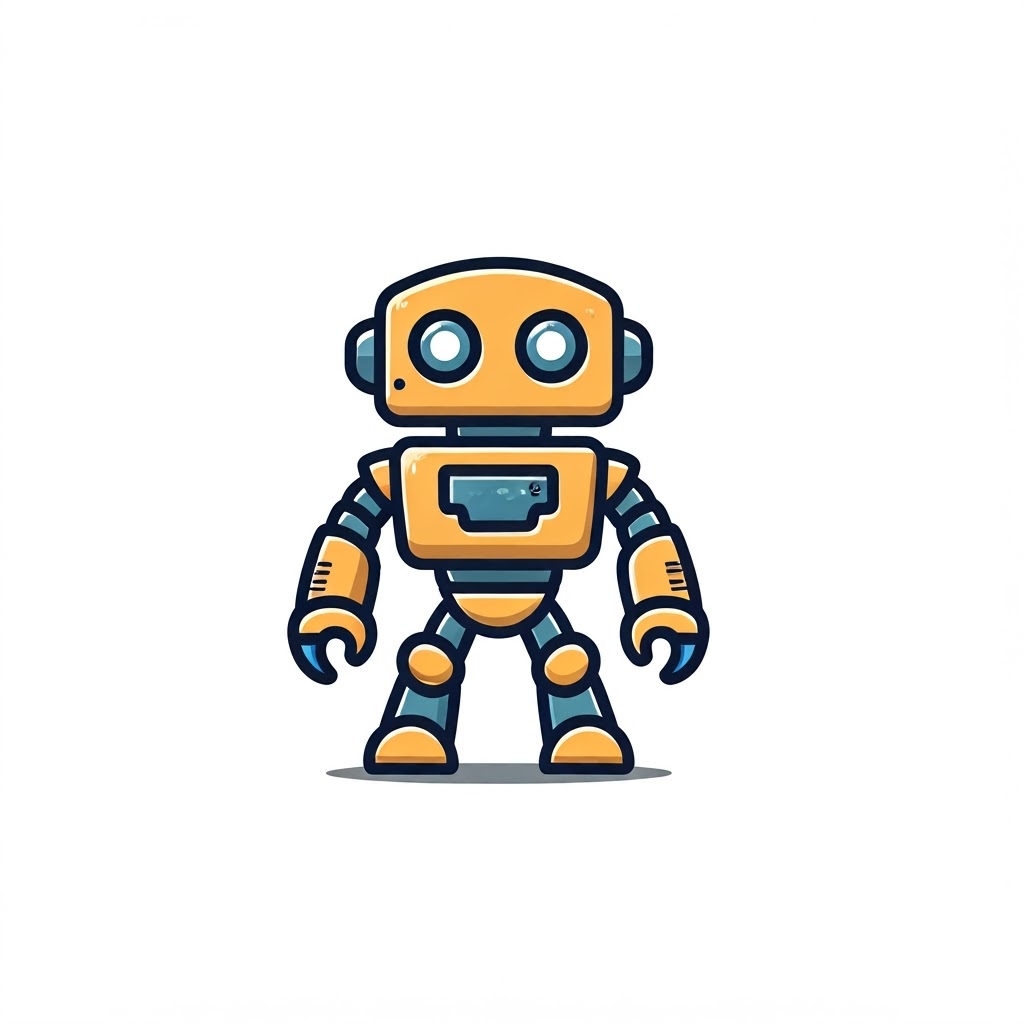}
    \put(0,102){\tiny \bf Logo}
    \end{overpic}
    \begin{overpic}[width=0.19\textwidth]{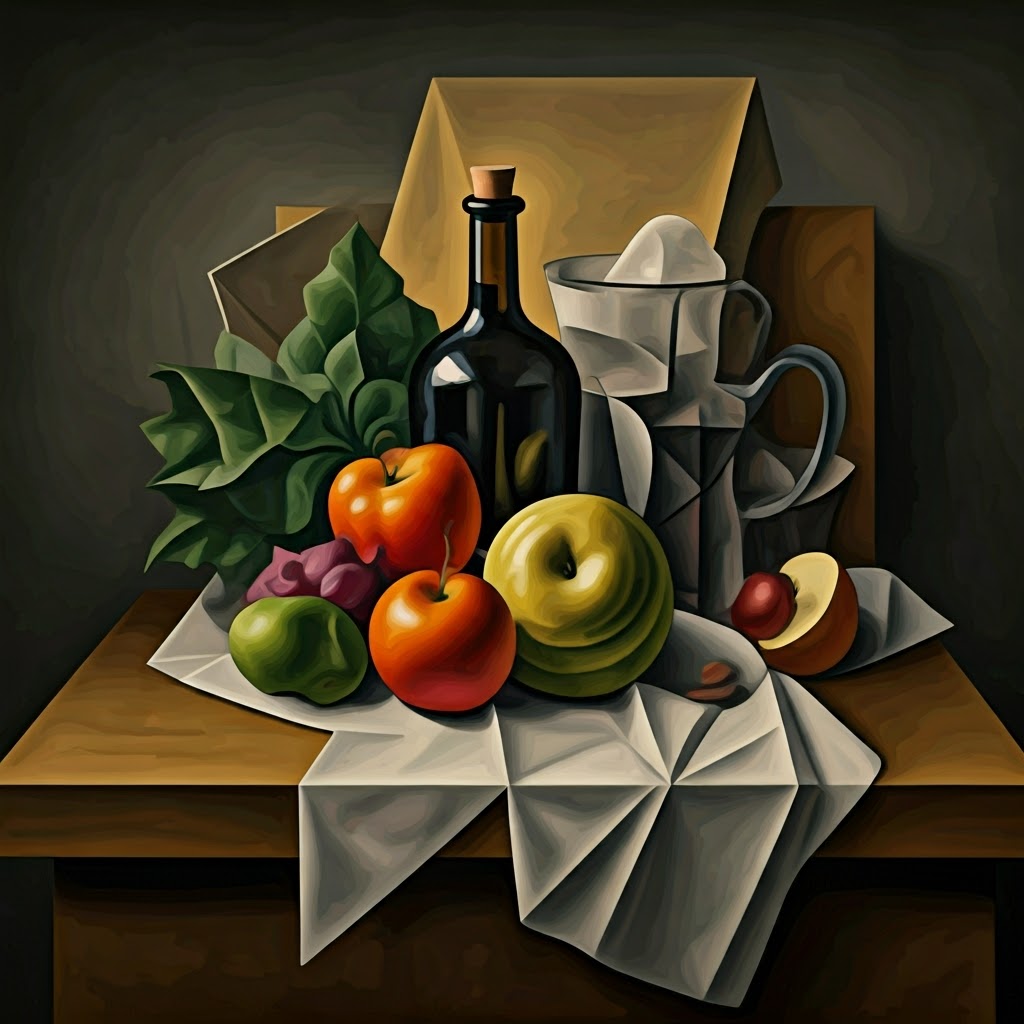}
    \put(0,102){\tiny \bf Object Painting}
    \end{overpic}
    \begin{overpic}[width=0.19\textwidth]{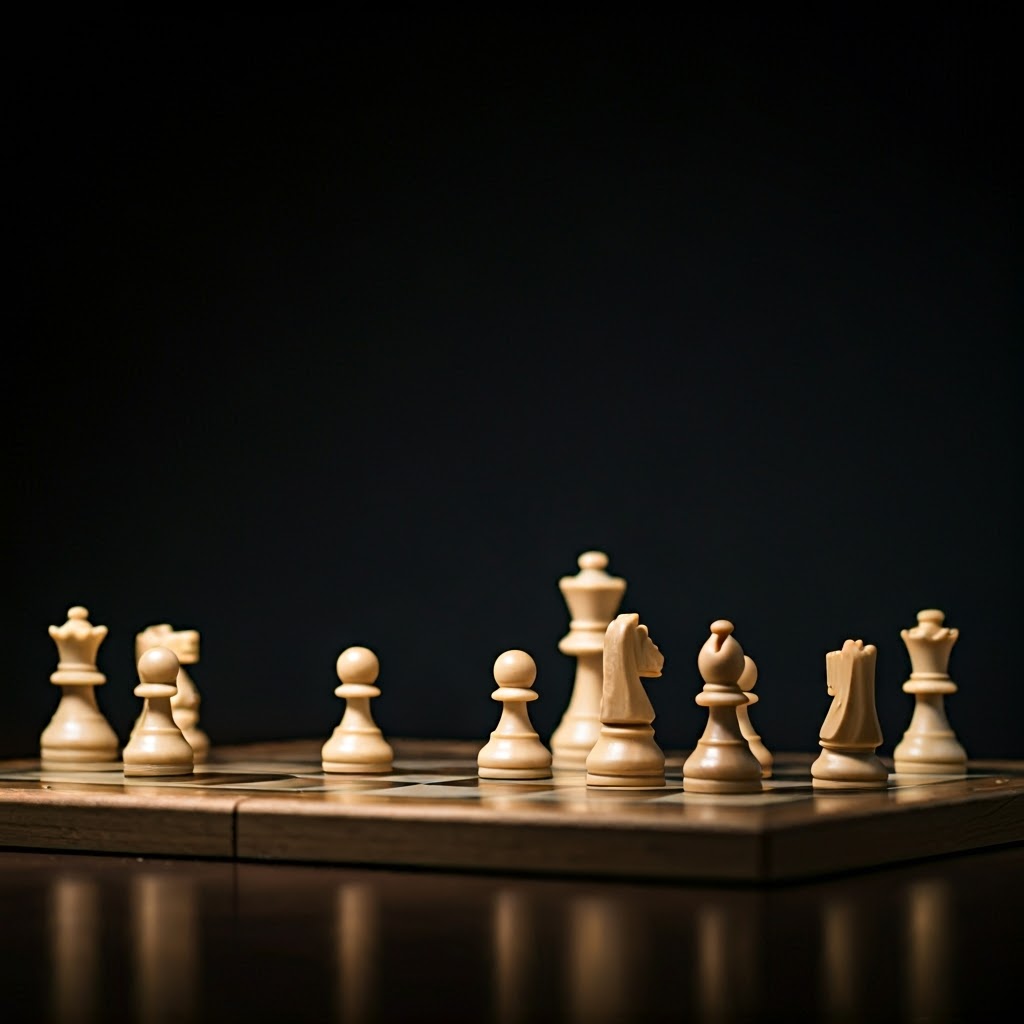}
    \put(0,102){\tiny \bf Object Photo}
    \end{overpic} \\\vspace{0.6em}%
    \begin{overpic}[width=0.19\textwidth]{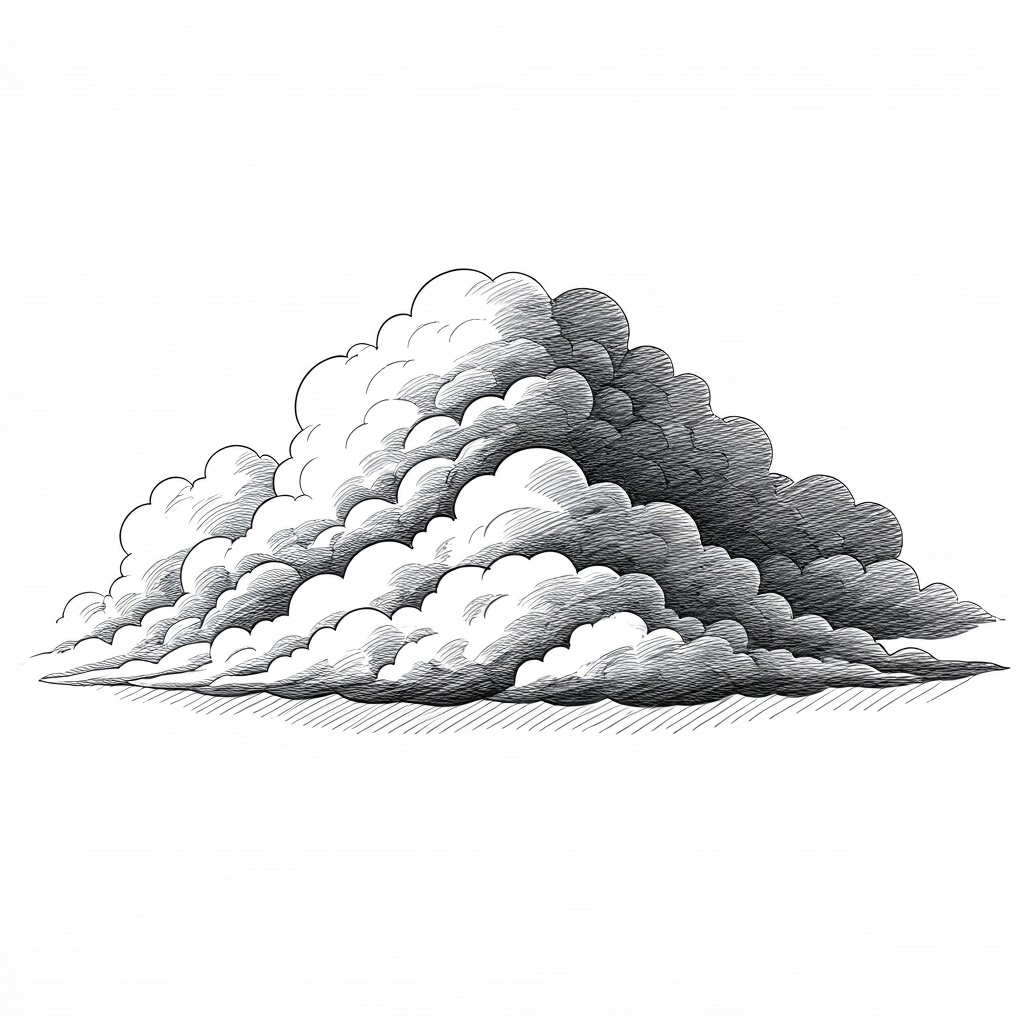}
    \put(0,102){\tiny \bf Pencil Drawing}
    \end{overpic}
    \begin{overpic}[width=0.19\textwidth]{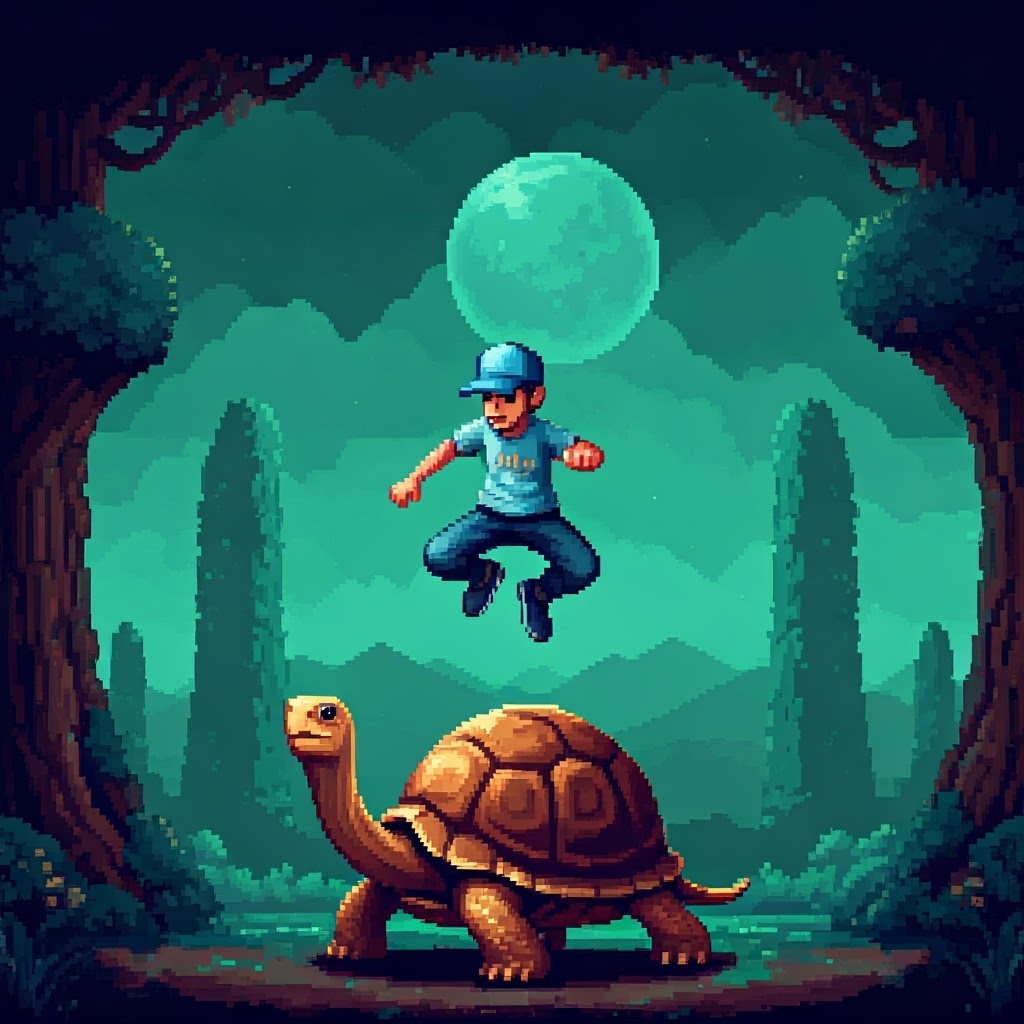}
    \put(0,102){\tiny \bf Pixel art}
    \end{overpic}
    \begin{overpic}[width=0.19\textwidth]{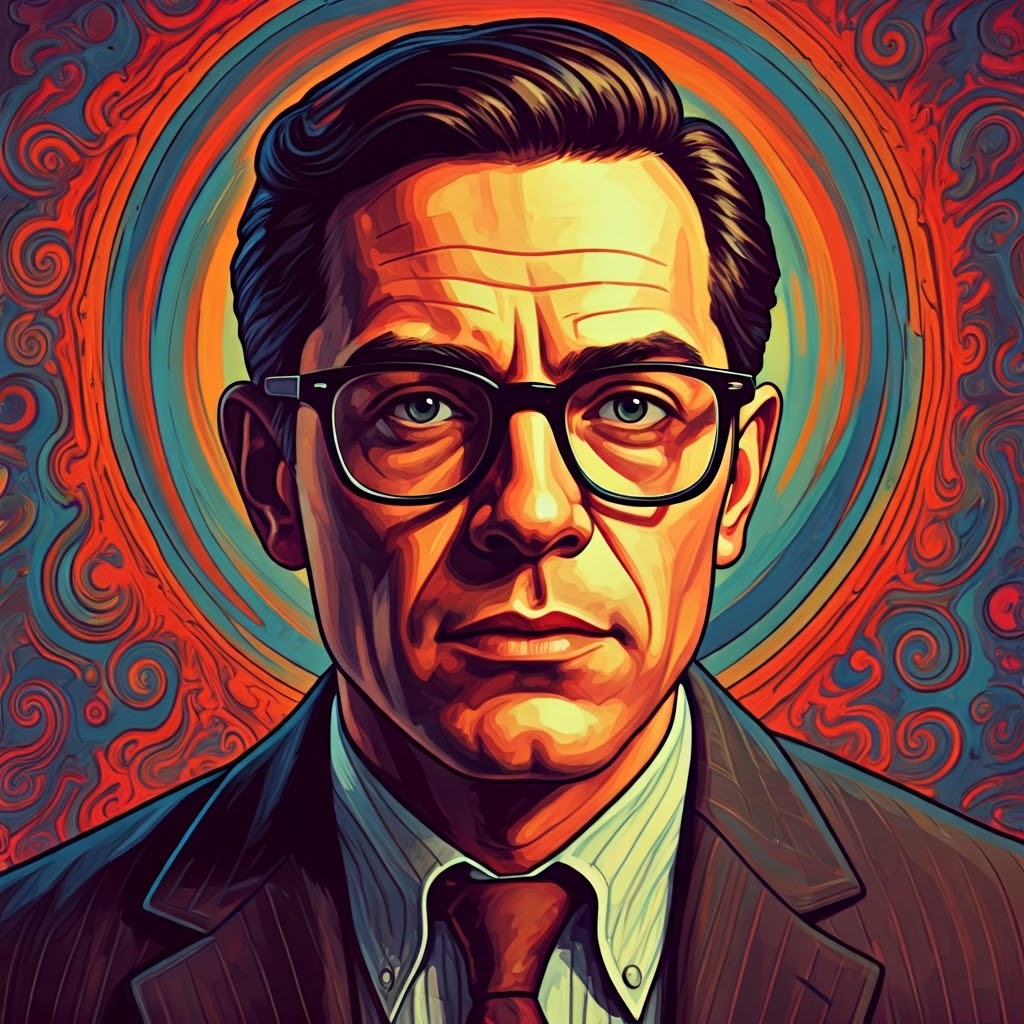}
    \put(0,102){\tiny \bf Portrait Painting}
    \end{overpic}
    \begin{overpic}[width=0.19\textwidth]{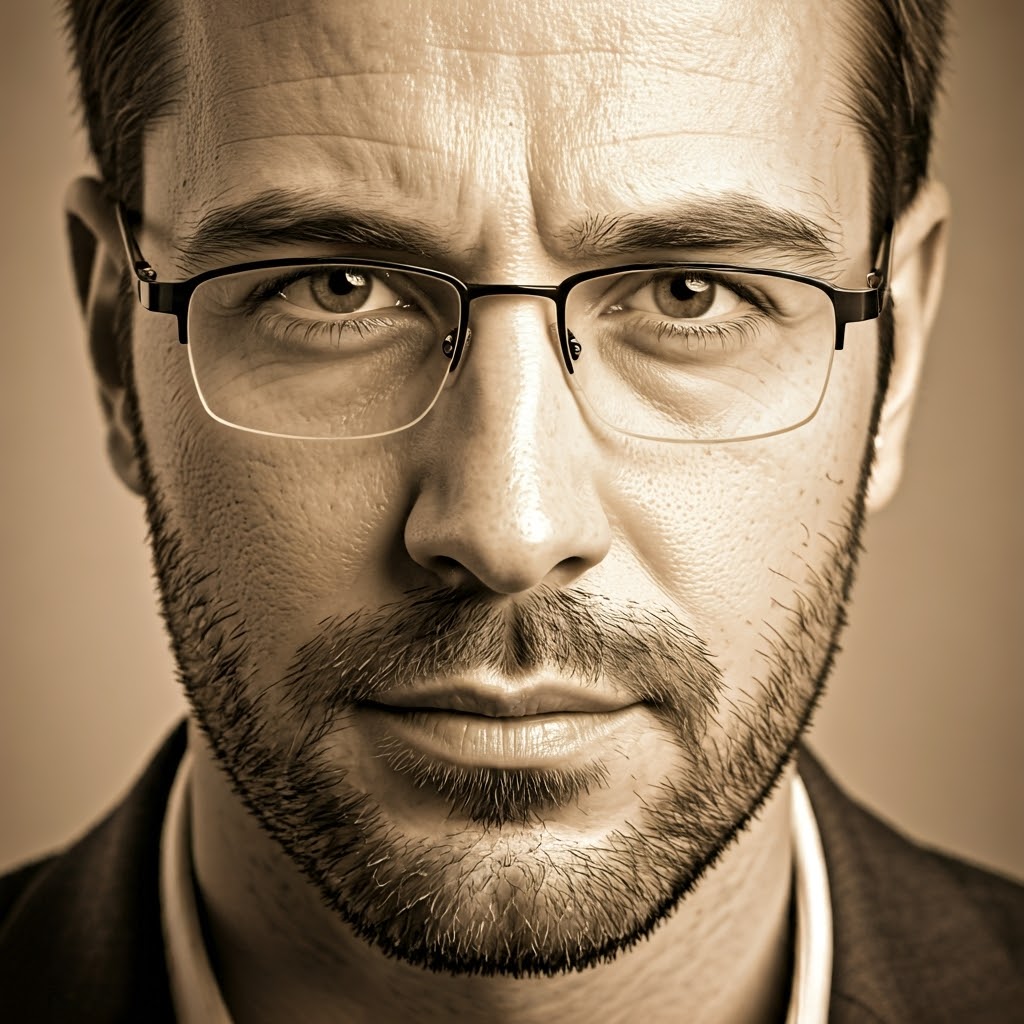}
    \put(0,102){\tiny \bf Portrait Photo}
    \end{overpic}
    \begin{overpic}[width=0.19\textwidth]{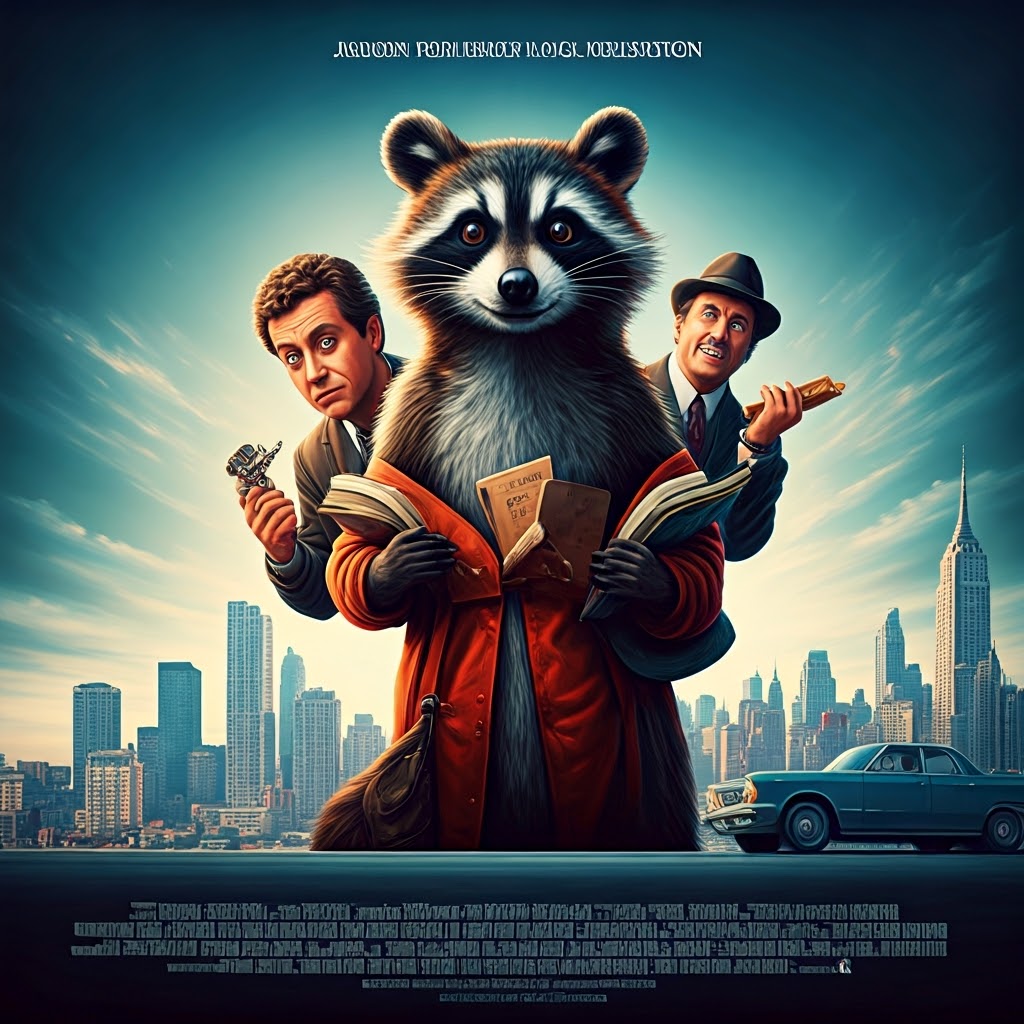}
    \put(0,102){\tiny \bf Poster Art}
    \end{overpic} \\\vspace{0.6em}%
    \begin{overpic}[width=0.19\textwidth]{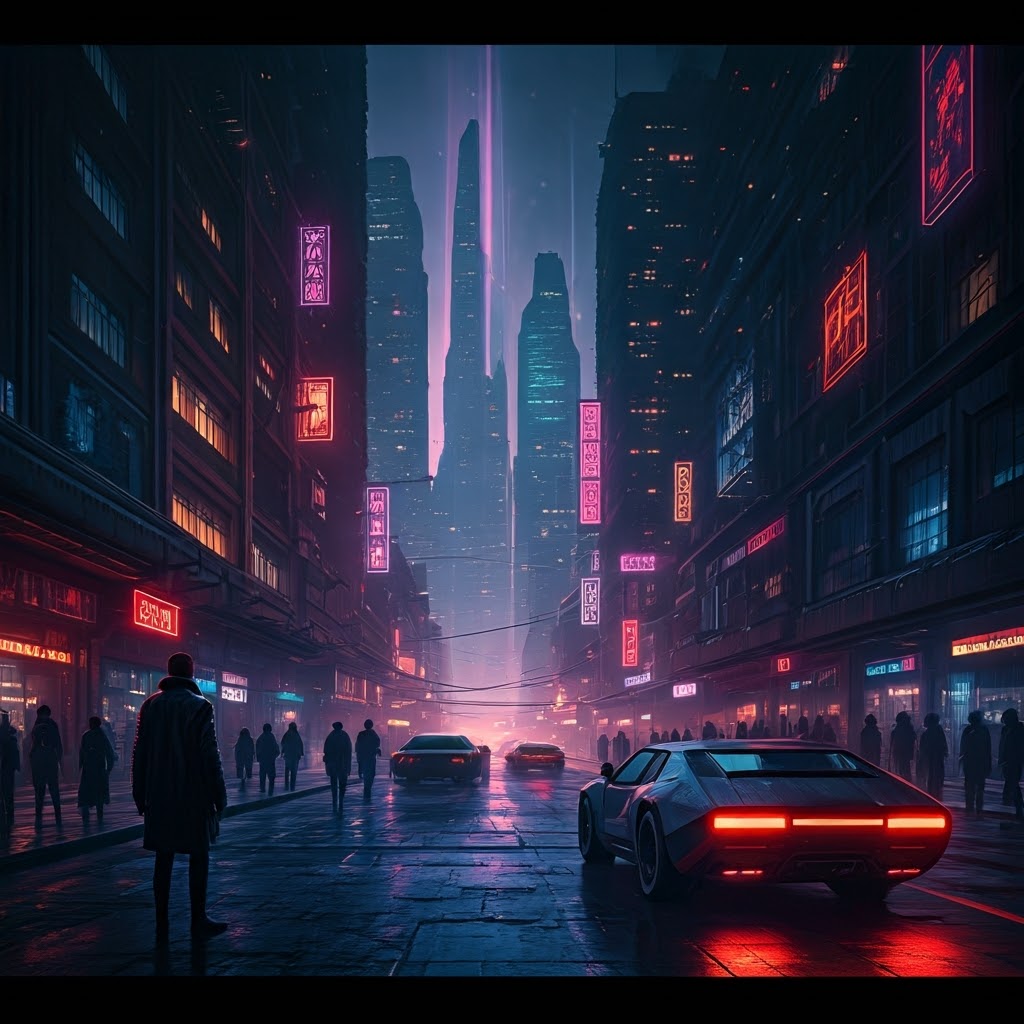}
    \put(0,102){\tiny \bf SciFi Photo}
    \end{overpic}
    \begin{overpic}[width=0.19\textwidth]{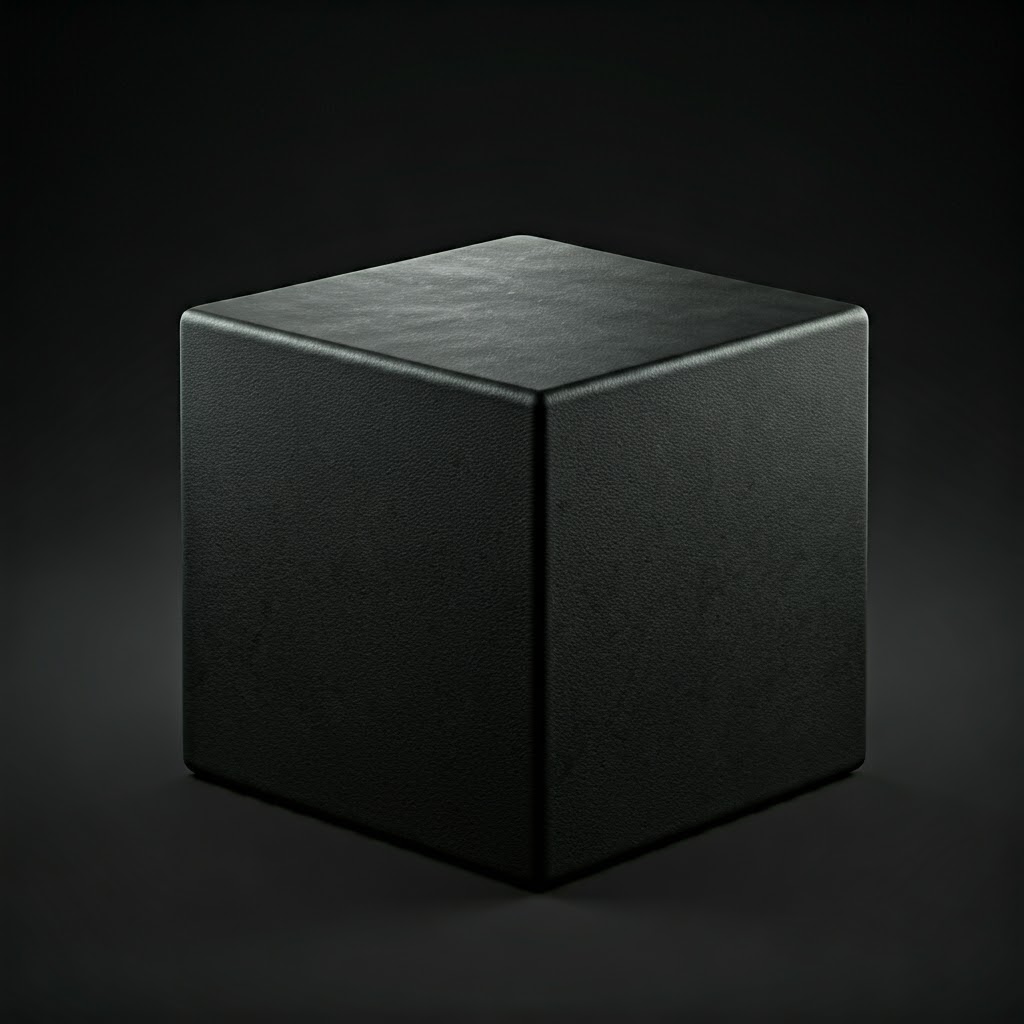}
    \put(0,102){\tiny \bf Shapes}
    \end{overpic}
    \begin{overpic}[width=0.19\textwidth]{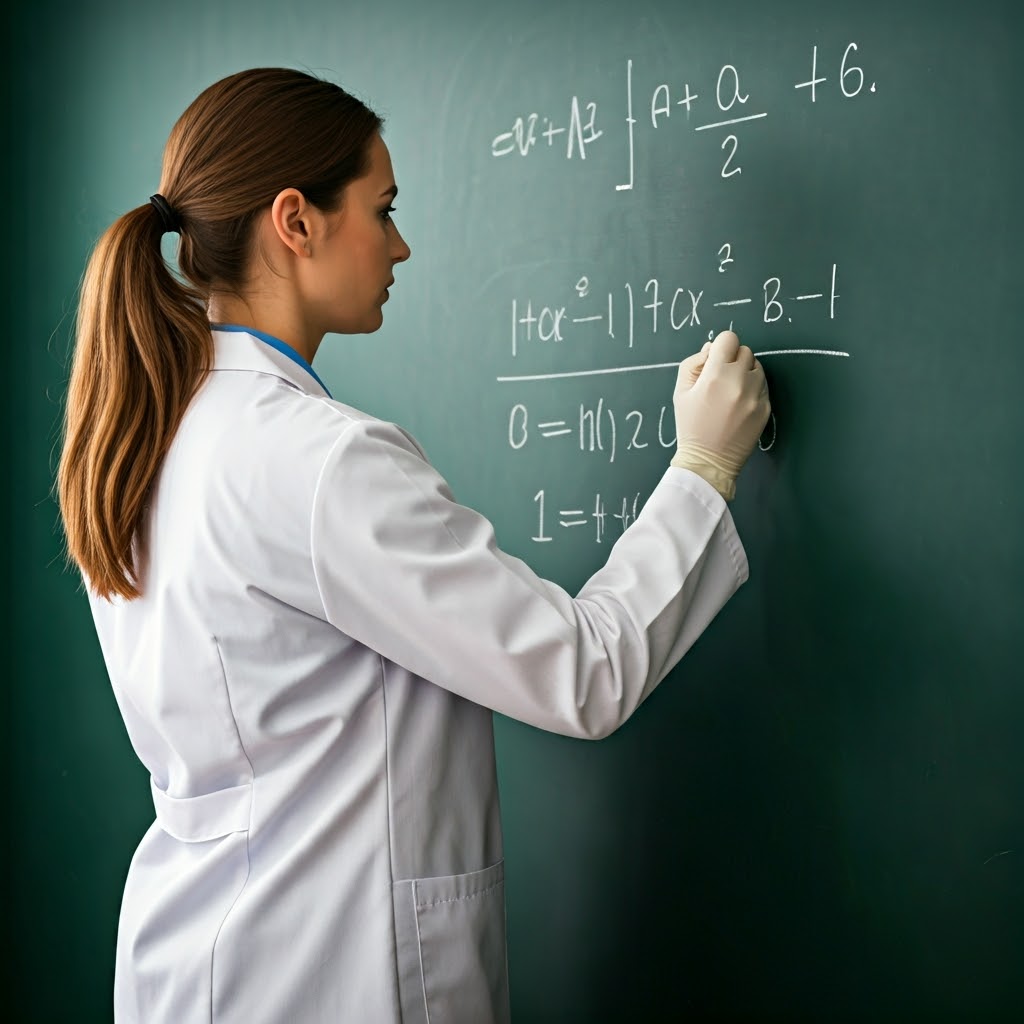}
    \put(0,102){\tiny \bf Stock Photo}
    \end{overpic}
    \begin{overpic}[width=0.19\textwidth]{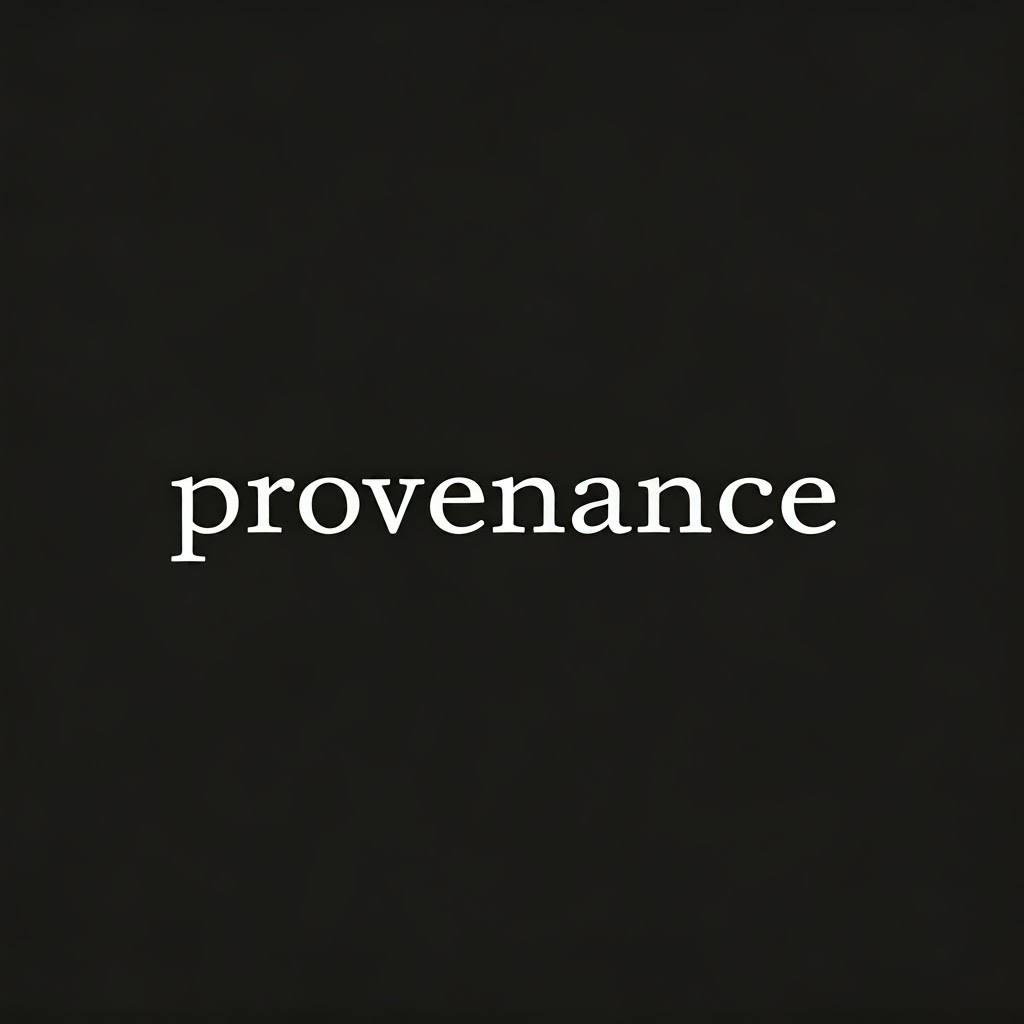}
    \put(0,102){\tiny \bf Text}
    \end{overpic}
    \begin{overpic}[width=0.19\textwidth]{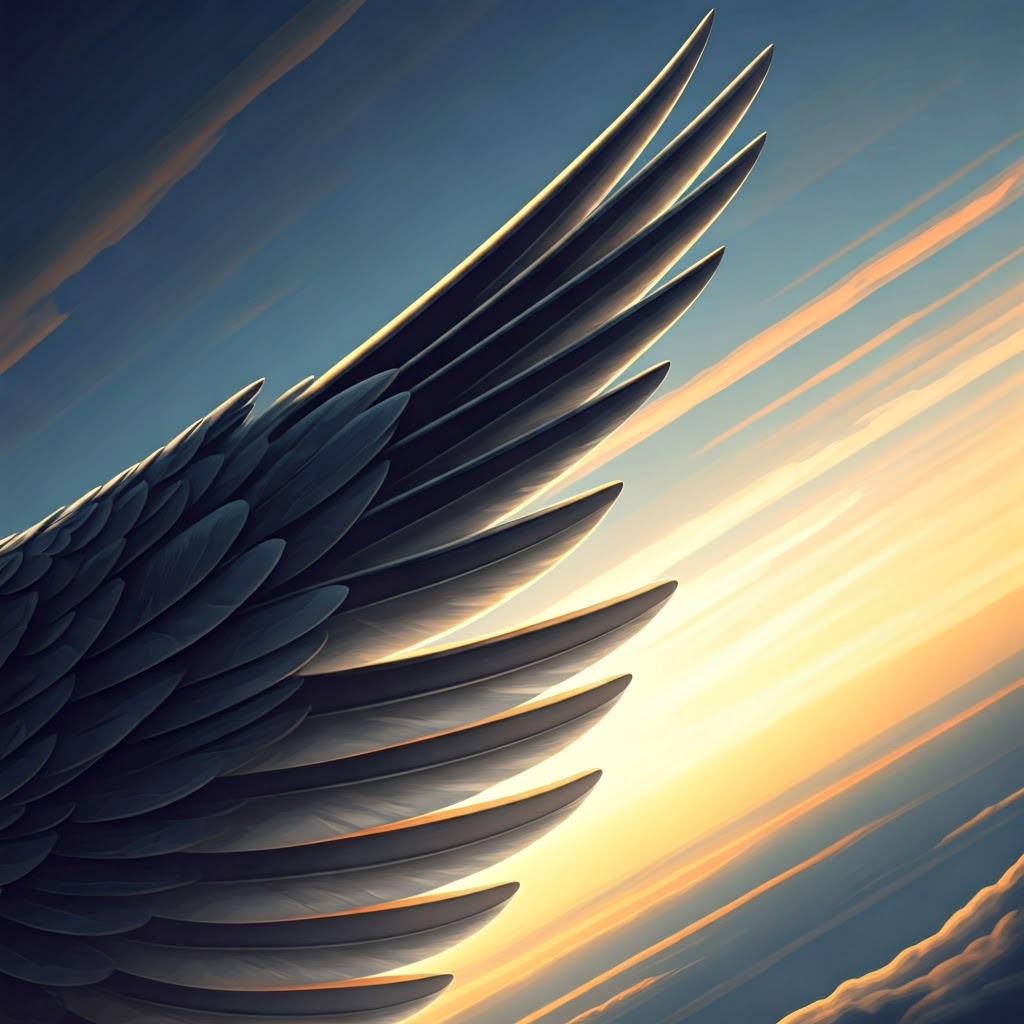}
    \put(0,102){\tiny \bf Wallpaper}
    \end{overpic}
    \caption{Illustrative images generated using \imagen for different categories of prompt. These images (either watermarked or not) are shown to external human raters and include corner-case images such as grayscale photographs, sketches or close-to-uniform color images.}
    \label{fig:categories}
\end{figure*}

\paragraph{External rater studies.}

For evaluation using external raters, this side-by-side approach was replaced by showing either the original or the watermarked content and asking raters to find ``artifacts''. This is the more realistic setup, simulating that users usually do not have access to pairs of original and watermarked content. In contrast to the very specific, hand-picked content examples in our internal evaluation, we constructed a set of hundreds of prompts across various categories. For our image watermark evaluation, example categories are shown in Figure~\ref{fig:categories} with images generated (and watermarked) using \imagen and \synthidimage.
These categories also illustrate that the distribution of generated images tends to be significantly different from natural images; this, however, reflects many use-cases of such generative models. In the study, we also included training examples for the raters to highlight common types of artifacts. Here, it was important to include both watermark-related artifacts as well as generation-related artifacts. We also had to use dummy examples to double-check that raters pay attention. If, after the study, we cannot distinguish the rate of artifacts between the watermarked and non-watermarked with statistical significance, we concluded the watermark to be invisible.

\paragraph{Additional considerations.}

There are several considerations that affect quality.
For \emph{images}, this mainly concerns resolution and color space.
While resolution could be set by the corresponding generative model whose output must be watermarked; we intend \synthidimage to be independent of such dependencies as generative models will likely change going forward.
While higher resolution allows less visible watermarks at similar robustness and payload size, we found that efficiency constraints ended up determining resolution.
This choice required taking care of not introducing any artifacts due to re-sizing during encoding.
The color space largely impacts which pixel-level changes are actually perceptible.
We found this matters particularly for black-and-white images or images with few colors or low contrast.

\paragraph{Corner-cases.}

An important realization in terms of quality, but also relevant for robustness and security, is the frequency of corner-case imagery when working at internet scale.
We already mentioned the tricky case of black-and-white images -- typically ignored in the literature.
Other examples include logos with very few but uniform coloring, images with slow gradients, abstract paintings, pixel art, sparse drawings, etc.
See Figure~\ref{fig:categories} for some examples. These types of images are rarely included in standard vision datasets that perceptual metrics are trained on.
As \emph{corner-cases}, we generally define any category of content that poses a challenge to watermarking in terms of quality and is not well represented in many standard datasets. This might mean that such content is difficult to watermark in an invisible way, or that the watermark becomes fragile in terms of robustness.
For \synthidimage, we found it important to identify and track such content throughout development. Large parts of the content used for internal studies included such corner cases.

\section{Transformation robustness}
\label{sec:robustness}

To consistently detect watermarks, it is important that they are robust to common transformations and content post-processing.
This generally includes a pre-determined set of transformations that we expect to be readily available on personal computers or smartphones, including resizing or cropping, quantization and compression, or common image processing filters (e.g. Instagram's photo filters).
This is similar to the notion of corruption robustness in the computer vision literature \citep{HendrycksARXIV2018,HendrycksARXIV2019b} and has to be contrasted with adversarial transformations \citep{EngstromARXIV2017,AlaifariARXIV2018} that assume malicious intent. Concretely, this means that we expect robustness to a random transformation with reasonable strength.

The main metrics to evaluate detection performance are \textbf{true} and \textbf{false positive rates} (TPR and FPR) with and without transformations.
Formally, given a distribution $\sP_\mathcal{T}$ over content transformations, we can measure the watermarking scheme's nominal rates as
\begin{align*}
\mathbb{E}_{\substack{\vx \sim \mathbb{P}_{\mathcal{X}}, \tau \sim \sP_\mathcal{T}}} \left [ g_\kappa(\tau(\vx)) \right ] \quad \textrm{(TPR)}\quad\quad\quad
\mathbb{E}_{\substack{\bar{\vx} \sim \mathbb{P}_{\mathcal{\bar{X}}}}} \left [g_\kappa(\bar{\vx}) \right ] \quad \textrm{(FPR)}
\end{align*}
with $\sP_{\bar{\mathcal{X}}}$ representing the distribution over all possible non-watermarked content and $g_\kappa$ indicating the decoder at the operating point (commonly a threshold) $\kappa$.
Note that we can also measure the worst-case across transformations for analysis purposes (replacing $\mathbb{E}_{\tau \sim \sP_\mathcal{T}}$ with $\max_{\tau \sim \sP_\mathcal{T}}$).
Some papers also report the area under the receiver operating characteristic (AUROC), considering TPR and FPR across various detection thresholds $\kappa$.
However, we found that AUROC is not sensitive enough in practice. This is because we are particularly interested in operating points with very low FPR (i.e., $\ll 1\%$), disregarding the majority of the ROC curve.
Note that prior work \citep{FernandezICCV2023,YangARXIV2024} often simulates results at such low FPRs based on assumptions of the detection/payload logits distribution.
But this has been found to be misleading \citep{FernandezICCV2023} which is why we rely purely on sufficiently large datasets to avoid simulation.
For evaluating TPR and FPR consistently across transformations, it is important to choose the operating point $\kappa$ consistently across transformations.
This is because, for deployment, we have to choose a single $\kappa$; evaluation on different subsets of transformations with different operating points can thus be misleading.
To the best of our knowledge, this is neglected in prior work, resulting in overly optimistic robustness evaluations.

\begin{figure*}
\centering
\begin{overpic}[width=0.19\textwidth]{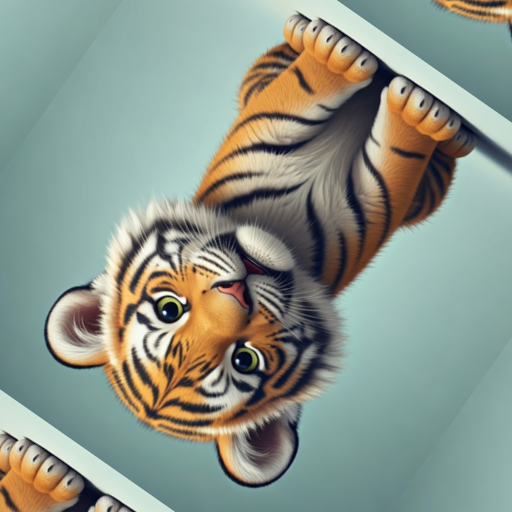}\put(0,102){\tiny \bf All rotations}\end{overpic}
\begin{overpic}[width=0.19\textwidth]{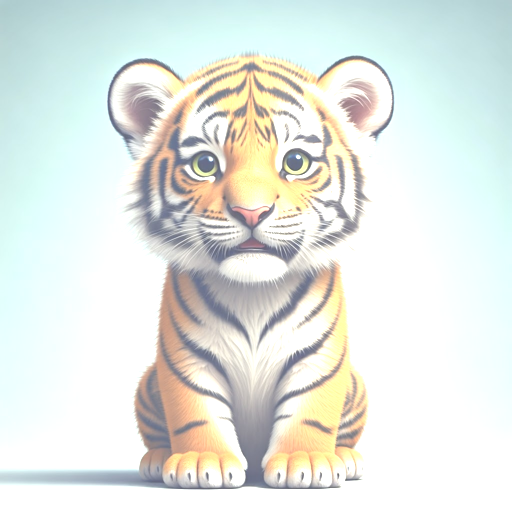}\put(0,102){\tiny \bf Brightness}\end{overpic}
\begin{overpic}[width=0.19\textwidth]{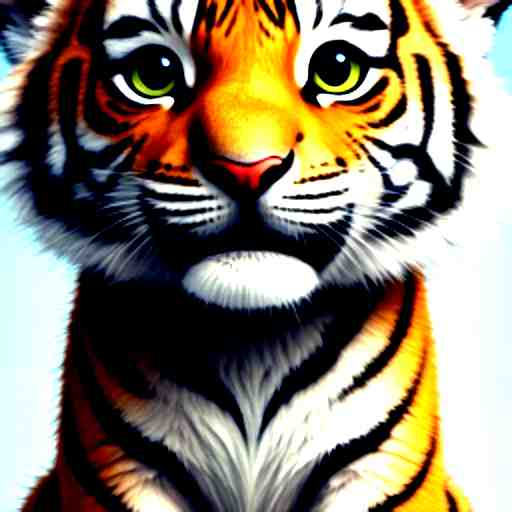}\put(0,102){\tiny \bf Combined}\end{overpic}
\begin{overpic}[width=0.19\textwidth]{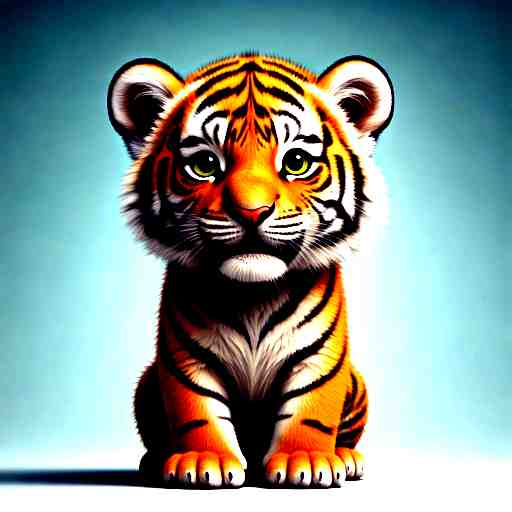}\put(0,102){\tiny \bf Combined NoCrop}\end{overpic}
\begin{overpic}[width=0.19\textwidth]{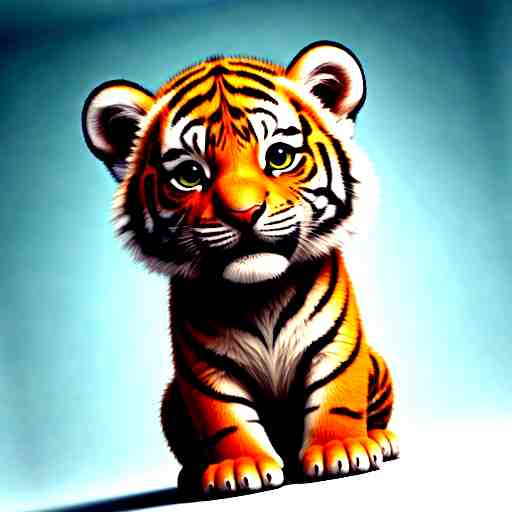}\put(0,102){\tiny \bf Combined NoCrop Rotate}\end{overpic}
\\\vspace{0.6em}
\begin{overpic}[width=0.19\textwidth]{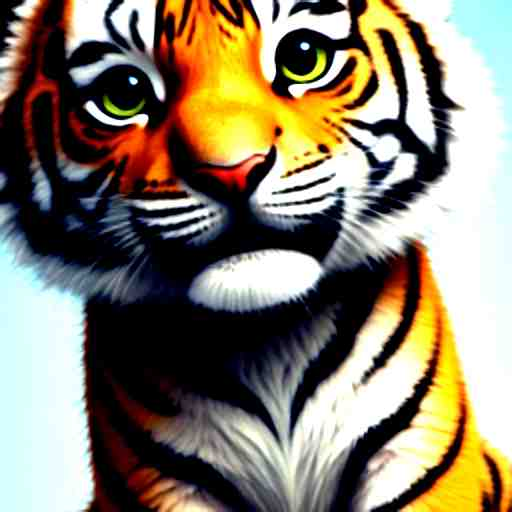}\put(0,102){\tiny \bf Combined Rotate}\end{overpic}
\begin{overpic}[width=0.19\textwidth]{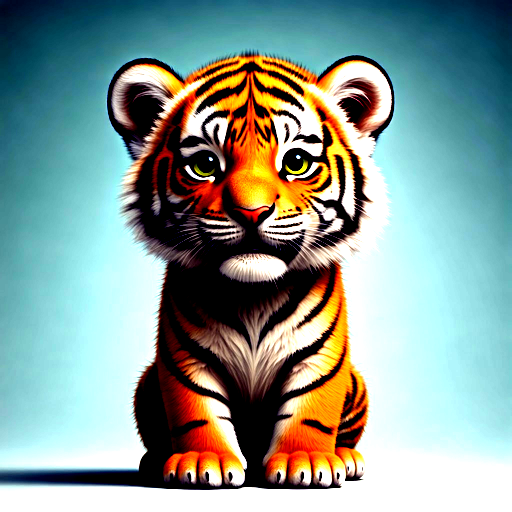}\put(0,102){\tiny \bf Contrast}\end{overpic}
\begin{overpic}[width=0.19\textwidth]{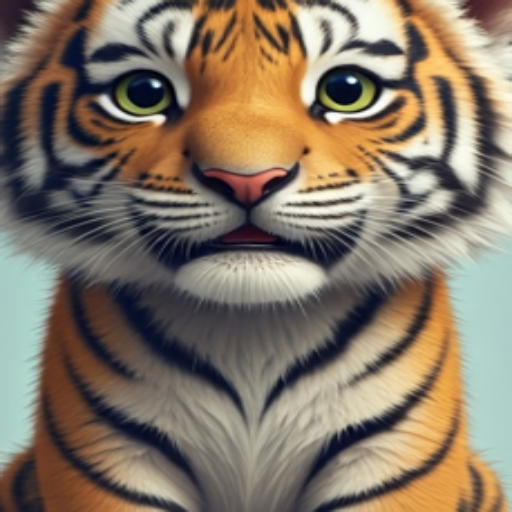}\put(0,102){\tiny \bf Crop resize}\end{overpic} 
\begin{overpic}[width=0.19\textwidth]{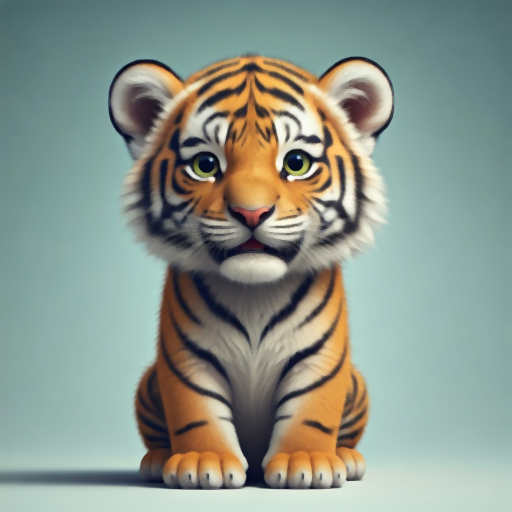}\put(0,102){\tiny \bf Denoise}\end{overpic}
\begin{overpic}[width=0.19\textwidth]{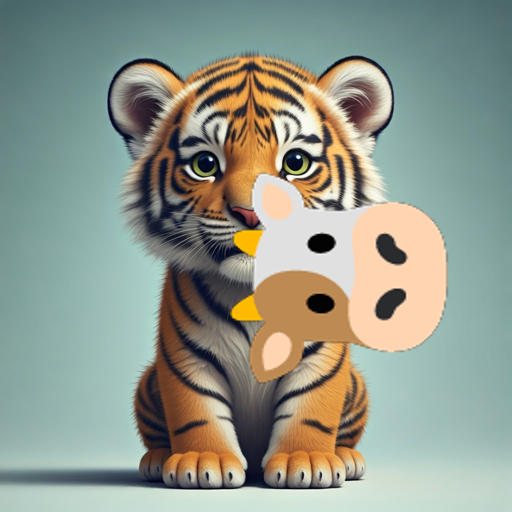}\put(0,102){\tiny \bf Emoji overlay}\end{overpic} \\\vspace{0.6em}
\begin{overpic}[width=0.19\textwidth]{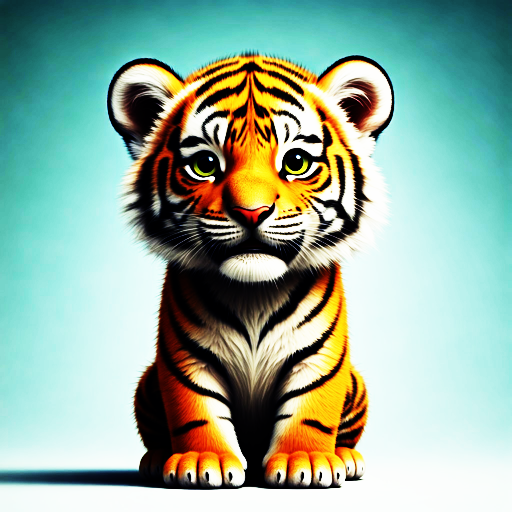}\put(0,102){\tiny \bf Exposure}\end{overpic}
\begin{overpic}[width=0.19\textwidth]{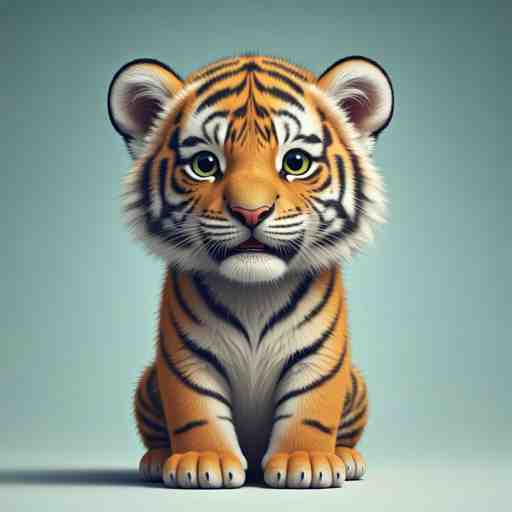}\put(0,102){\tiny \bf File format}\end{overpic}
\begin{overpic}[width=0.19\textwidth]{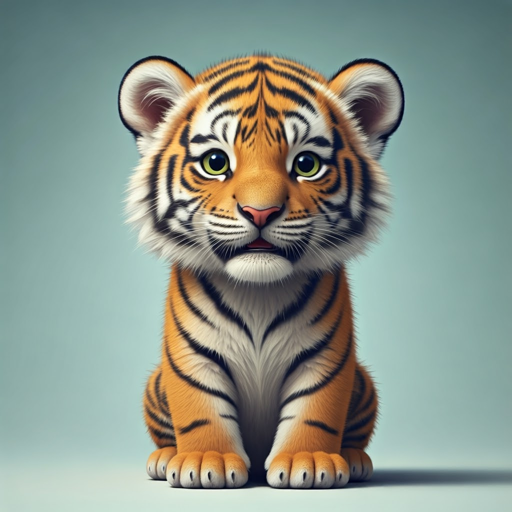}\put(0,102){\tiny \bf Flip left-right}\end{overpic} 
\begin{overpic}[width=0.19\textwidth]{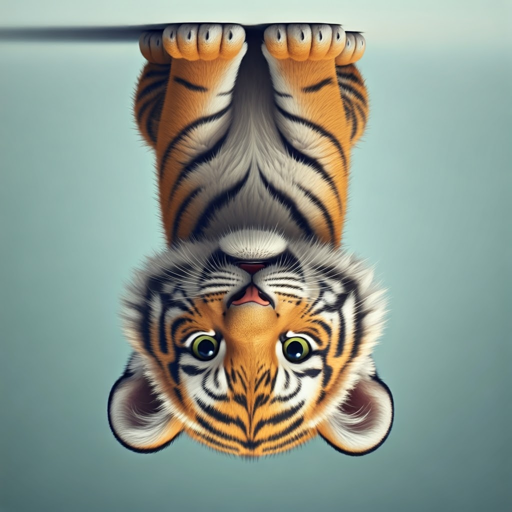}\put(0,102){\tiny \bf flip up-down}\end{overpic}
\begin{overpic}[width=0.19\textwidth]{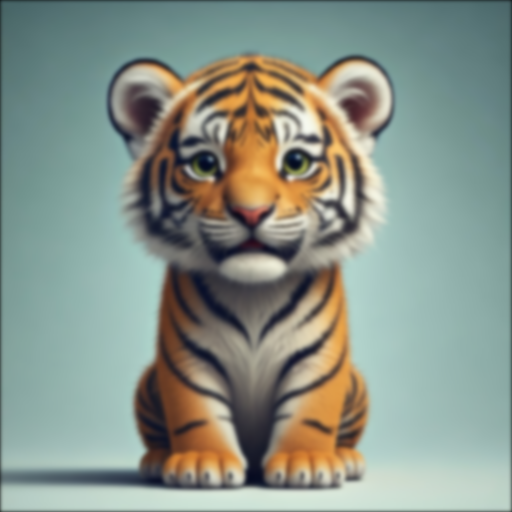}\put(0,102){\tiny \bf Gaussian blur}\end{overpic} \\\vspace{0.6em}
\begin{overpic}[width=0.19\textwidth]{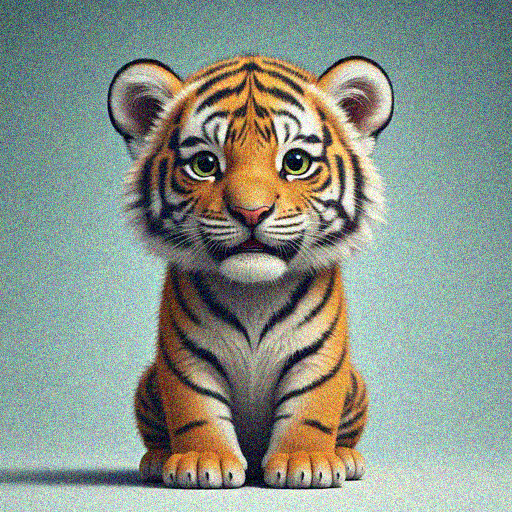}\put(0,102){\tiny \bf Gaussian noise}\end{overpic}
\begin{overpic}[width=0.19\textwidth]{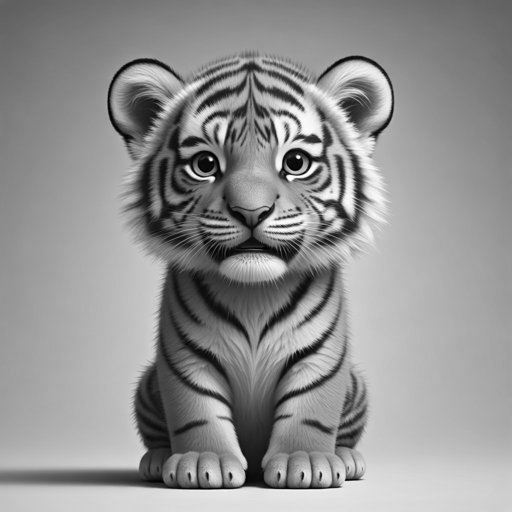}\put(0,102){\tiny \bf Grayscale}\end{overpic}
\begin{overpic}[width=0.19\textwidth]{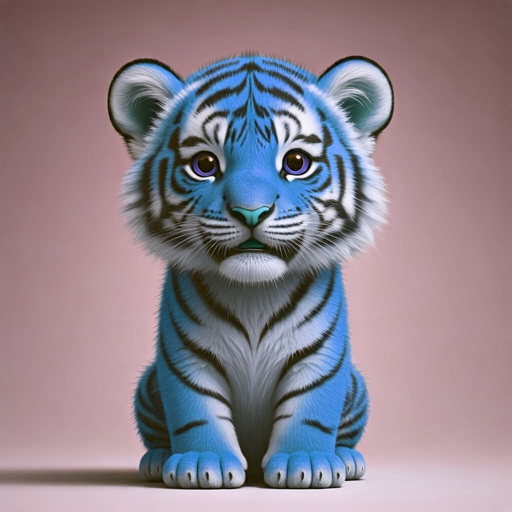}\put(0,102){\tiny \bf Hue}\end{overpic} 
\begin{overpic}[width=0.19\textwidth]{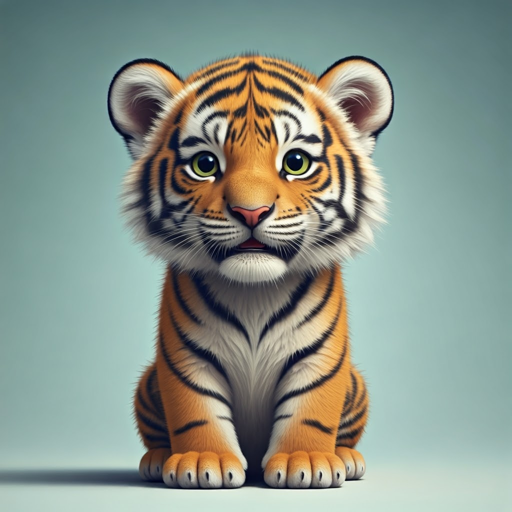}\put(0,102){\tiny \bf Identity}\end{overpic}
\begin{overpic}[width=0.19\textwidth]{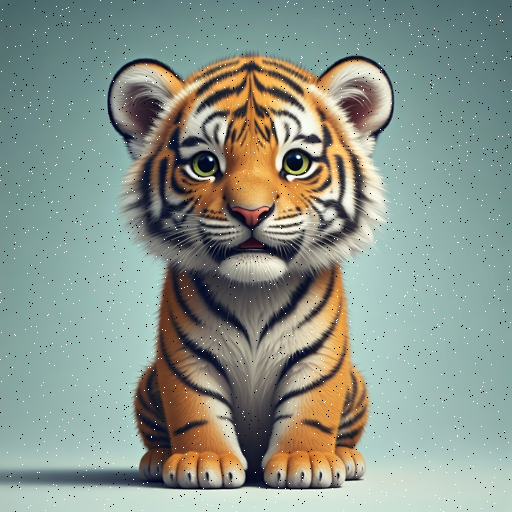}\put(0,102){\tiny \bf Impulse noise}\end{overpic} \\\vspace{0.6em}
\begin{overpic}[width=0.19\textwidth]{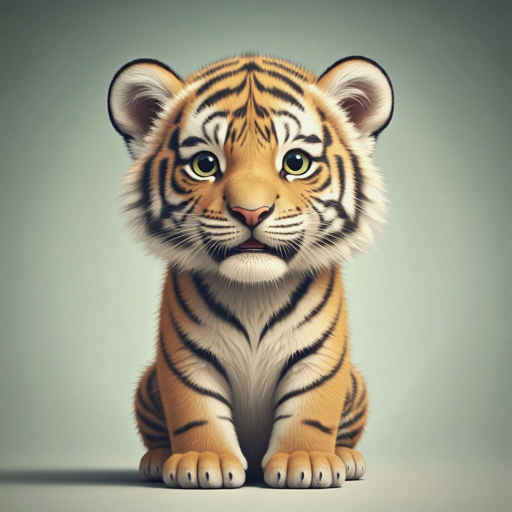}\put(0,102){\tiny \bf Instagram}\end{overpic}
\begin{overpic}[width=0.19\textwidth]{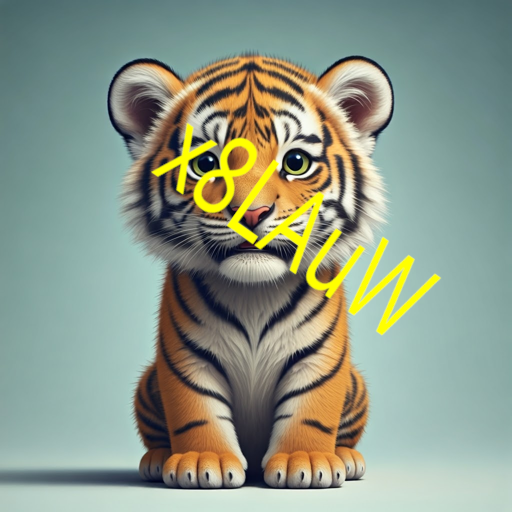}\put(0,102){\tiny \bf Light text overlay}\end{overpic}
\begin{overpic}[width=0.19\textwidth]{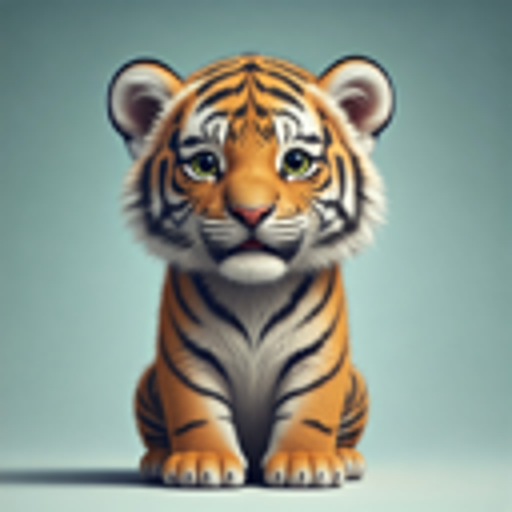}\put(0,102){\tiny \bf Resize}\end{overpic} 
\begin{overpic}[width=0.19\textwidth]{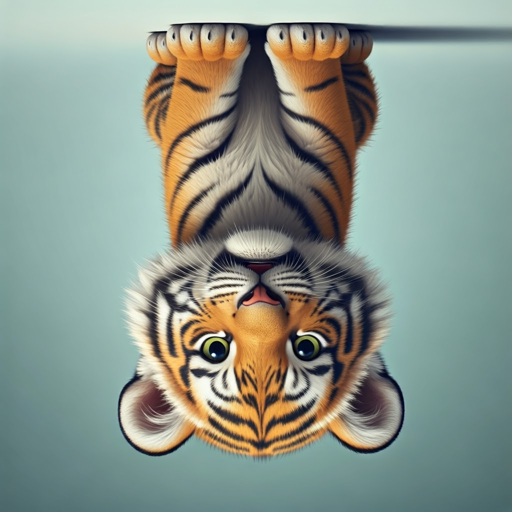}\put(0,102){\tiny \bf Rotation}\end{overpic}
\begin{overpic}[width=0.19\textwidth]{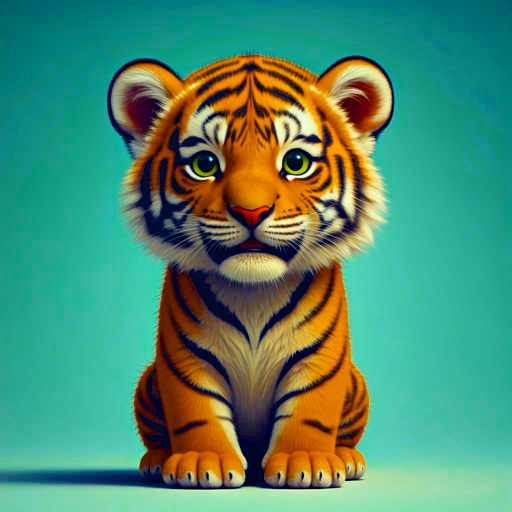}\put(0,102){\tiny \bf Saturation}\end{overpic} \\\vspace{0.6em}
\begin{overpic}[width=0.19\textwidth]{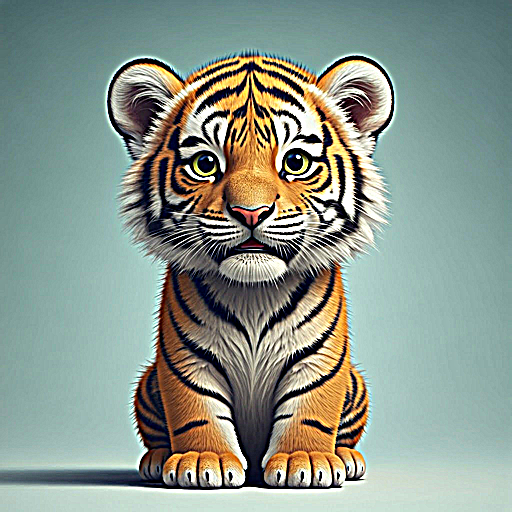}\put(0,102){\tiny \bf Sharpness}\end{overpic}
\begin{overpic}[width=0.19\textwidth]{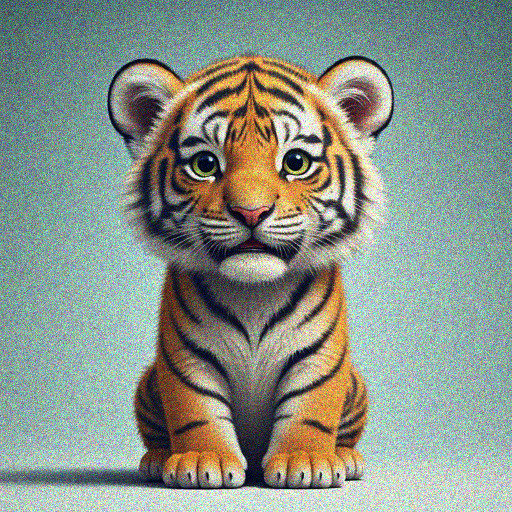}\put(0,102){\tiny \bf Shot noise}\end{overpic}
\begin{overpic}[width=0.19\textwidth]{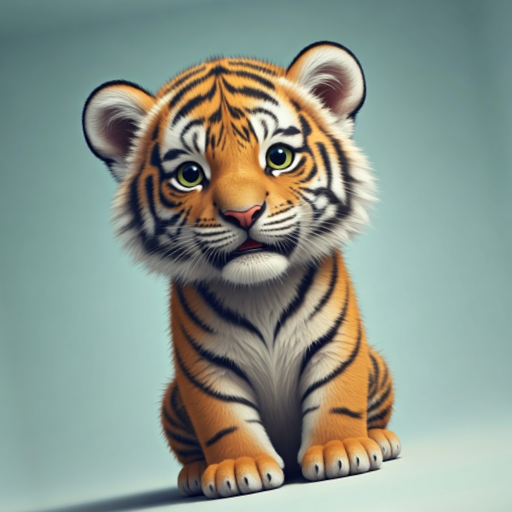}\put(0,102){\tiny \bf Small rotation}\end{overpic} 
\begin{overpic}[width=0.19\textwidth]{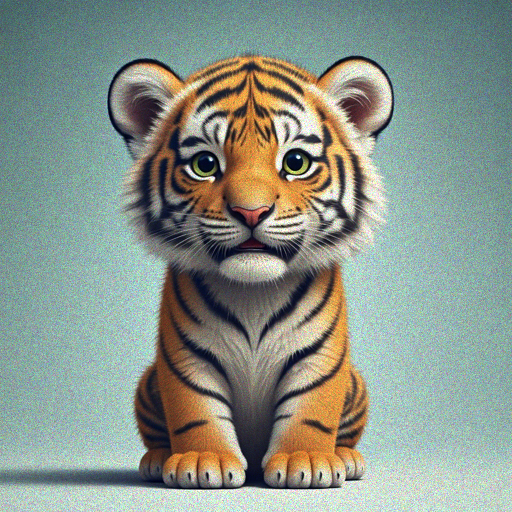}\put(0,102){\tiny \bf Speckle noise}\end{overpic}
\begin{overpic}[width=0.19\textwidth]{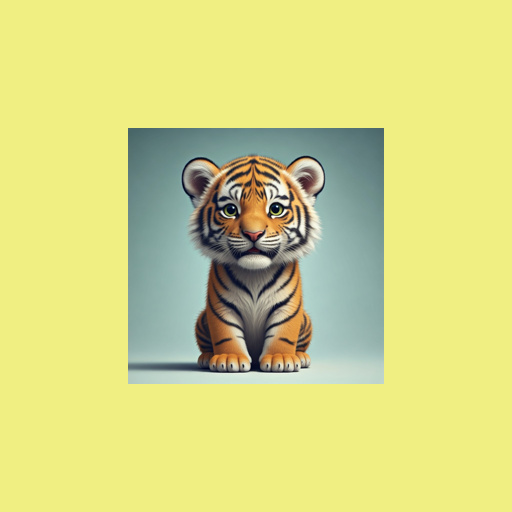}\put(0,102){\tiny \bf Zoom out}\end{overpic} \\\vspace{0.6em}
\caption{Illustrative examples of the 30 ``basic'' transformations used in our evaluation (alphabetically listed).} 
\label{fig:transformations_worst}
\end{figure*}

Prior work on watermarking quickly realized that properly specifying and training for relevant transformations is key for robustness. Thus, many previous works include concrete transformations and parameters to be tested for robustness \citep{TancikCVPR2020,WenARXIV2023,ZhangARXIV2024}.
However, we found that none of these works are exhaustive, as also seen in the surveyed methods and transforms by \cite{WanNEUROCOMPUTING2022}.
Commonly overlooked transformations can be simple ones, such as small rotations, left/right and up/down flips, or different types of noise (Gaussian, salt\&pepper, speckle).
For example, \gaussianshading \citep{YangARXIV2024} misses rotations and flips; \stablesignature \citep{FernandezICCV2023} and WAVES \citep{AnARXIV2024} ignore various noise types.
This is despite common knowledge that these type of transformations are extremely common and work on corruption robustness indicates that deep neural networks are not inherently robust to them.
Moreover, many methods are benchmarked against ``fixed-strength'' transformations, e.g., fixed rotations or crops, or a fixed JPEG quality \citep{FernandezICCV2023,SaberiARXIV2024,YangARXIV2024}.
While these usually correspond to rather strong transformations, it is not guaranteed that the methods are also the least robust against them; for example, neural networks may be more vulnerable to small rotations than 90, 180 or 270 degree rotations \citep{EngstromARXIV2017,DumontARXIV2018}.
It has also been shown that combinations of transforms can be effective ways to fool networks.
For example, \stablesignature has combinations based on cropping, brightness changes and JPEG compression.
As generalization of neural networks between transformations is still poorly understood, a key component of \synthidimage is exhaustively listing, implementing and tuning relevant transformations.
We developed a set of ``basic'' transformations that we believe any large-scale watermarking system should be robust to.
Our two guiding principles for inclusion in this basic set of transformations are \emph{accessibility} and \emph{detectability}. Rotations, flips, brightness changes, etc. are all very accessible on every smartphone camera or social media app. This also includes Instagram-like filters\footnote{\tiny\url{https://github.com/akiomik/pilgram}} or overlaying texts and logos.
To tune strengths, we consider all transformations reasonable if they have reasonable quality and preserve the underlying semantics of the content (as judged by human raters).
It is also important to consider any possible transformations that any software might automatically include as pre- or post-processing, such as compression and quantization.
All of our basic transformations are shown in Figure~\ref{fig:transformations_worst}.

\section{Payload}
\label{sec:payload}

For \synthidimage, where we aim to serve several generative models across different partners, the payload is a crucial component of turning watermarking into a tool for provenance. While some recent work \citep{YangARXIV2024} simulates multiple users, details are scarce and suggest that a simple integer user identifier is encoded in the payload, without any redundancy or additional logic.

During development, we evaluate the payload robustness in terms of bit and code accuracy, across all relevant transformations. Letting $\vc = (c_1, \ldots, c_C)$ be the originally encoded payload code and $g_i(\cdot)$ the logit corresponding to $c_i$ (with $g_0(\cdot)$ corresponding to the detection logit), we define bit accuracy as the accuracy of the predicted payload bits in expectation over bit positions:
\begin{align}
    \mathbb{E}_{\vx \sim \mathbb{P}_{\mathcal{X}},\tau\sim\mathbb{P}_{\mathcal{T}}}\left[\frac{1}{C}\sum_{i = 1}^C \mathbbm{1}[\sign(g_i(\tau(f(\vx)))) = c_i]\right]. \label{eq:bit-accuracy}
\end{align}
Bit accuracy is comparable across models with different payload sizes.
In some cases, it is also interesting to monitor bit accuracy individually as their accuracy can vary widely by location depending on the exact training procedure.
This might inform how to robustly encode information in the payload in practice.
Bit accuracy does not explicitly measure what we might be interested in, which is how often we get the entire payload right, i.e., code accuracy:
\begin{align}
    \mathbb{E}_{\vx \sim \mathbb{P}_{\mathcal{X}},\tau\sim\mathbb{P}_{\mathcal{T}}}\left[\mathbbm{1}[\forall_{i=1}^{C} \sign(g_i(\tau(f(\vx)))) = c_i]\right]. \label{eq:code-accuracy}
\end{align}
Code accuracy generally degrades with larger payload size and degrades more quickly with challenging transformations than bit accuracy. As before, it is important to monitor these metrics across transformations.
This is because bit and code accuracy are highly relevant for many deployment scenarios where pure detection performance in terms of TPR/FPR is not sufficient.
Note that both metrics are defined conditional on watermarked examples.
That is, we generally neglect what happens to the payload on non-watermarked content.
This is somewhat unique to \synthidimage, as we explicitly separate detection and payload recovery (see Section \ref{subsec:problem-formalization}).
For approaches that directly use the payload for detection, TPR and FPR needs to be defined following Equation \eqref{eq:code-accuracy}.
For example, \cite{YangARXIV2024,FernandezICCV2023} select a random but fixed payload for each watermarked example and make a detection decision by thresholding on the number of matching bits between this payload and the predicted one.
Of course, if the reference payload is not available, this test is not applicable.

In the setting from \cite{YangARXIV2024} where each user has an integer identifier, code accuracy becomes the relevant metric for provenance (in this case, watermark detection and payload recovery coincide). Essentially, this turns watermark detection into a multi-class problem, where the number of classes grows with the number of users.
For \synthidimage, where we deliberately disentangled watermark detection and payload decoding for flexibility, we do not always need to invoke the logic for decoding the payload.

\section{Ensuring security}
\label{sec:attacks}

\external{
There is a wide range of attacks on watermarking systems that \synthidimage has to defend against \citep{ZhaoARXIV2024}. In the following, we discuss the relevant threat models we considered. A threat model describes an attacker's objective, capabilities, and knowledge. Even though attackers might have different objectives, they often use a common set of tools that can inform which defenses to prioritize.

\subsection{Threat Models: Objectives, Capabilities, and Tools}

An attacker's primary objectives are either \textbf{watermark removal} (creating a false negative) to obscure content and possibly claim ownership  or \textbf{watermark forgery} (creating a false positive) to wrongly attribute ownership  of a potentially incriminating piece of content. Further objectives that aim at facilitating these goals include \textbf{model extraction} \citep{TramerUSENIX2016,OhARXIV2017b,OrekondyCVPR2019} to steal the models themselves (or information about them, e.g., hyper-parameters,architecture details), \textbf{secret extraction} \citep{JovanovicARXIV2024} to find hidden keys (e.g., used for payload encoding), or \textbf{payload attacks} to manipulate the embedded information.

The success of these attacks depends on the attacker's capabilities. This includes their level of \textbf{access}, ranging from black-box (API-only) to white-box (full model weights and architecture). An attacker may also have privileged access to secrets (e.g., private keys) or possess \textbf{paired information} (both original and watermarked content), which significantly simplifies attacks, especially for post-hoc systems like \synthidimage. Finally, their available \textbf{resources}, such as query limits and computational power, constrain the feasibility of an attack.
To execute an attack, adversaries can employ various tools, largely from the adversarial machine learning community \citep{PapernotEUROSP2018}.
\begin{description}
    \item[(Universal) adversarial examples] These are inputs with small, often imperceptible, perturbations designed to cause misclassification \citep{SzegedyICLR2014,CarliniSP2017}. For watermarking, they can be used to make the decoder fail (removal) or fire incorrectly (forgery). While defenders aim for invisibility, attackers may tolerate visible artifacts, making attacks easier \citep{BrownARXIV2017}.
    \item[Surrogate models] Since adversarial examples often transfer between different models, an attacker can train a local substitute model to craft attacks in a black-box setting \citep{PapernotASIACCS2017,TramerUSENIX2016}.
    \item[Re-generation attacks] These attacks use other powerful generative models (like diffusion models) to reconstruct a watermarked image, potentially washing out the watermark in the process \citep{AnARXIV2024,ZhaoARXIV2023a}.
\end{description}

\subsection{Defenses and discussion}

The main difficulty of defending against many of the above threat models is anticipating future attacks within relevant threat models \citep{PapernotEUROSP2018}.
These generally depend on how the watermarking scheme is deployed in practice (see Section \ref{sec:deployment}).
Our strategy for \synthidimage has been to defend against attacks that are stronger than what we expect in a real-world deployment.
For example, ensuring robustness against white-box attacks provides a strong defense against more realistic black-box threats \citep{CroceICML2020}.
We identified relevant attacks through red-teaming and integrated them into our automated evaluation benchmarks to monitor trade-offs.

\paragraph{Robust training and fine-tuning.}

The foundation of our defense is robust training.
The watermark encoder and decoder are trained on a wide range of data augmentations and transformations to help them withstand common image manipulations~\citep{ZhuECCV2018,HayesARXIV2020}.
Robust training may include adversarial training of the decoder \citep{MadryICLR2018,KurakinICLR2017b}, which aims to harden it against constrained adversarial manipulations.
Additionally, for \synthidimage, we specifically tested and ensured robustness against off-the-shelf weak re-generation attack models (e.g., using variational autoencoders).
Ultimately, robustness to a large range of different threat models has been shown to be particularly tricky \citep{TramerARXIV2019,MainiARXIV2019,StutzICML2020}, as this will either reduce performance or require a more visible watermark.
However, since \synthidimage is deployed in a proprietary setting, our main goal is to make black-box attacks computationally infeasible at scale, a more achievable goal than defending against a determined white-box adversary.

\paragraph{Randomness.}

Appropriate randomness in the watermarking process is crucial for countering certain extraction or exchange attacks.
For a given generative model, using the same seed and prompt should ideally produce the exact same watermarked output every time.
This deterministic behavior prevents an attacker from re-generating the image to reveal differences in watermark patterns, a technique known as a collusion attack.
On the other hand, for content that is highly predictable (e.g., due to severe memorization) or simple, an attacker might anticipate the output and attempt to forge or remove a watermark.
To mitigate this, the watermark generation process should incorporate stochastic elements, ensuring sufficient diversity in the watermarks applied to semantically similar but non-identical images.

\paragraph{Specificity.}

To prevent watermark exchange attacks, the watermark pattern should be inextricably linked to the content it marks.
We enforce this property by making the watermark generation process content-dependent.
During training, we introduce specific loss functions designed to minimize the chance that a watermark extracted from one piece of content can be successfully applied to another. This ensures the watermark is not merely a generic stamp but a specific signature of the content itself.

\paragraph{Filtering.}

While detecting and filtering various types of adversarial attacks have been shown ineffective \citep{CarliniARXIV2017,BryniarskiARXIV2021,StutzICML2020}, we can decide not to watermark specific content. In text watermarking \citep{KirchenbauerICML2023}, for example, zero-entropy generations (e.g., due to memorization) are automatically not watermarked. Similarly, we can decide not to watermark specific corner-case content such as nearly uniform images.
Such content can easily be filtered by looking at basic statistics of the content.
}

\section{Internet-scale deployment}
\label{sec:deployment}

Our final desiderata, requiring us to actually deploy \synthidimage in various settings at scale, implies various practical considerations that we found rarely discussed in prior work. Some of these depend on the actual deployment scenario, while others address practicalities like decision making, versioning and combinations with other tools for provenance.

\paragraph{Deployment settings.}

For \synthidimage, we considered the following three deployment scenarios:
\begin{description}
\item[As-an-internal-service:] Here, either encoder, decoder or both are purely used for internal purposes. A common construction is to allow watermarking of content generated by internal models (even if the content is then publicly served) and have detection run only internally, too. A variation of this is to allow internal watermarking but expose detection externally to trusted partners or even the end-user. Note that paired information is only available internally and access to queries and secrets can be tightly controlled.
\item[As-a-service:] Here, both encoder and decoder are offered as-a-service to third party companies. In this setting, third parties have access to paired information and potential secrets. Moreover, payloads need to be used to distinguish between different customers all using the same watermarking system. Again, third parties can decide to open the detection part to the end user.
\item[Open models:] In this setting, either encoder, decoder or both are publicly available. This makes paired information more widely available and payloads derived from private keys become more important.
\end{description}
In all these scenarios it is important to realize that \synthidimage has two different components, the encoder and the decoder that are deployed differently. Also, models from the same model family, i.e., multiple \synthid-based models, might be deployed in different scenarios.

The deployment specifics also affect the relevance of certain threat models in practice. This is because the attacker's knowledge and level of access changes. Internal use has the smallest attack surface as access to models, paired information, and secret keys can be controlled tightly.
Giving customers the possibility to produce paired information or opening encoder and/or decoder models increases the attack surface.
Along this, threat models like watermark removal/false negatives become less relevant as they are particularly easy to provoke using (universal) adversarial examples against open models.
False positives, in contrast, gain in importance and can be reduced through clever payload selection.
In the setting where multiple similar models are released through different avenues, open models may also increase the risk to internal models by simplifying surrogate and transfer-based attacks.

\paragraph{Decision-making.}

Previous work on watermarking used statistical tests for decision making on top of the decoder's predictions. These approaches are often based on assuming Bernoulli distributions on the individual bits \citep{FernandezICCV2023,YangARXIV2024,LukasUSENIX2023,YuICCV2021b,RomanARXIV2024} or exploit that diffusion models start from a Gaussian distribution \citep{ZhangARXIV2024,WenARXIV2023}. These are essentially parametric approaches; it has already been shown that individual bits can usually not be modeled using fair coin flips \citep[Appendix B.5]{FernandezICCV2023}. Nevertheless, we believe that proper decision-making is important for deploying watermarking systems at large scale in a responsible and reliable manner. To avoid any parametric assumptions, we found conformal $p$-values particularly scalable as they are non-parametric and recent work showed how to incorporate robustness considerations \citep{StutzTMLR2023}. Moreover, as this approach is based on a held-out set of calibration examples, it also offers an additional level of configuration before deployment. This can be important when having to react to distributional changes or minor issues for which updating the model is not an option (see \textbf{Versioning} below).

We define the following hypotheses: $H_0$: ``input is \emph{not} watermarked'' and $H_1$: ``input is watermarked''. We then compute valid $p$-values following \cite{BatesICMLWORK2021,BatesARXIV2022} for both hypotheses using a held-out set of calibration examples $\{\vx^{(0)}_i\}_{i=1}^n$ and $\{\vx^{(1)}\}_{i=1}^n$ of non-watermarked and watermarked samples, respectively:
\begin{align}
    \rho_k(\vx) = \frac{1 + \sum_{i = 1}^n \delta[(2k - 1) f_0(\vx) \geq (2k - 1) f_0(\vx^{(k)}_i)]}{n + 1}
\end{align}
where $f_0(\cdot)$ correspond to the detection logit. As these are valid $p$-values, i.e., $\rho_k$ is uniformly distributed over $[0, 1]$ iff $H_k$ is true, thresholding $\rho_0$ ($\rho_1$) at $\alpha$ guarantees a FPR (TPR) of at most $\alpha$ (at least $1 - \alpha$). We can control FPR \emph{and} TPR simultaneously by allowing a variable abstention rate. The calibration set can be updated online if necessary and include different transformations of the same image for robustness.\footnote{Following \cite{StutzTMLR2023}, this reduces the guaranteed FPR (TPR) to $2\alpha$ ($1 - 2\alpha$); however, this drop is never observed empirically.}

Assuming an individual logit $f_j(\cdot)$ for each bit $c_j \in \{\pm1\}$, $j \in \{1, \ldots, C\}$, we can also compute $p$-values for each bit $c_j$ being positive or negative. Given an encoded payload $\vc = (c_1,\ldots, c_C)$, we can then test whether the payload is likely the original one using any standard multiple hypothesis testing procedure \citep{SimesB1986,HochbergB1988,Holm1979}. These methods usually come with a confidence level $\alpha$ which controls the rate of correctly accepting the true payload. However, with larger $C$, these methods become more conservative. Especially in light of transformations this leads to possibly multiple payloads being accepted. This can be countered using redundant coding techniques. Again, an abstention mechanism can be created easily: if all used payloads $\vc$ can be enumerated, we can abstain if multiple payloads are accepted and the number of accepted payloads is often a good mirror of whether the artifact has been tampered with or not.

\paragraph{Versioning.}

Versioning is an important aspect of deploying at scale and for the long term. In the case of \synthidimage, the encoder always determines the main version. This is because once watermarked content is out in the wild, we have to be able to detect it. The decoder, in contrast, can be updated much more easily. In a sense this means that we need strict versioning for the encoder while the decoder can be updated continuously. This also implies that threat models that can be addressed in the encoder are more important. For example, once an encoder has been deployed, it is difficult to increase the diversity of watermarks or prevent watermark extraction. The decoder, in contrast, can be updated on the fly to address new attacks. Having the encoder determine the version means that eventually there will be multiple versions in production, assuming that watermarking quality, robustness and payload size will be improved over the coming years. This entails having to detect multiple versions of watermarks over time. This adds additional resource requirements, complicates decision making, and vulnerability might be ``inherited'' between versions.

\paragraph{Combination with search and C2PA.}

For tackling provenance of AI-generated content, watermarking is a promising but not the only tool at our disposal~\citep{collomosse2024authenticity}.
Generally, we expect watermarking to be deployed alongside a metadata-based standard like C2PA which major companies vowed to implement.
Moreover, there are also technologies based on search and retrieval.
For \synthidimage, we explored various combinations, and focused on fingerprinting, which refers to the practice of computing low-dimensional embeddings for multimedia artifacts and storing them in an internal database.
Learning such embeddings in a robust way \citep{GowalICLR2021} allows verifying the origin of AI-generated content using similarity search.
Various works have applied variations of this idea to attribution and content provenance in the widest sense \citep{SaberiARXIV2024,BalanCVPRWORK2023,BharatiTIFS2021,Zhang2020,NguyenICCV2021}.
In comparison to watermarking, similarity search is more prone to false positives rather than false negatives.
On its own, fingerprinting suffers from increased cost for storing embeddings (on top of the embedding cost) and poor scaling as false positives increase with database size.
This means that there needs to be retention rules that e.g. store embeddings of generated content only within a specific time-frame.
However, as fingerprinting has complementary failure modes, it is natural to combine both approaches.
For example, to mitigate forgery attacks, an image must not only contain the correct watermark payload but also match one of the stored embeddings.

\section{Experimental results}
\label{sec:results}

\external{
We evaluate the robustness and quality of our external \synthidimage watermarking model variant, denoted \synthidom .
\synthidom can encode 136-bit payloads within 512$\times$512 pixel images.
We compare this variant against a wide range of baselines.
Our experiments show that \synthidom establishes new state-of-the-art watermarking performance. Specifically, we find that it outperforms others in terms of quality, considering an extensive human rating study, and shows superior detection performance across a comprehensive list of image transformations.
}
Results are summarized in Figure~\ref{fig:intro}.
In this section, we provide further details.

\paragraph{Baselines.}

To measure detection performance of a given model, we leverage the model's detection score if available or use the number of correct payload bits as a score proxy (as proposed previously, e.g., \citealp{FernandezICCV2023}).
To measure payload recovery performance, we directly measure the bit accuracy for models that provide it.\footnote{Note that bit accuracy performance depends heavily on the payload length $C$ and this might not be a useful point of comparison.}
We evaluate all models without any error code correction.
We compare against the following:
\begin{description}[noitemsep,topsep=0pt,labelsep=-0.5em]
\item[\stegastamp]~\citep{TancikCVPR2020}: We use the model checkpoint released in the official repository.\footnote{\tiny\url{https://github.com/tancik/StegaStamp}} It can encode 100-bit payloads within 400$\times$400 pixel images.
\item[\trustmark]~\citep{BuiARXIV2023}: Our primary results leverage the \trustmarkq and \trustmarkp variants. \footnote{\tiny\url{https://github.com/adobe/trustmark/tree/main/python}} We chose \trustmarkq over \trustmarkb because, even though it prioritizes image quality over robustness compared to \trustmarkb, our evaluation found it to be more robust overall. All \trustmark variants can encode 100-bit payloads within 256$\times$256 pixel images.
\item[\invismark]~\citep{xu2024invismark}: We use the model checkpoint released in the official repository.\footnote{\tiny\url{https://github.com/microsoft/InvisMark}}  It can encode 100-bit payloads within 256$\times$256 pixel images.
\item[\wam]~\citep{sander2024watermark}: We use the model checkpoint released in the official repository.\footnote{\tiny\url{https://github.com/facebookresearch/watermark-anything}} It can encode 32-bit payloads within 256$\times$256 pixel images. Note that \wam can be used to watermark individual areas within an image, we use it here to watermark full images. As detection score, we use the area of the detection mask returned by the model. For payload recovery, we use the ``semi-hard'' strategy.
\item[\videoseal]~\citep{fernandez2024video}: We use the model checkpoints released in the official repository.\footnote{\tiny\url{https://github.com/facebookresearch/videoseal}} \videosealzero and \videosealone can encode 96-bit payloads and 256-bit payloads within 256$\times$256 pixel images, respectively. Note that \videoseal was primarily developed to add watermarks to videos. However, it can be used for images. As detection score, we use the number of correct payload bits as using the available detection logit performs badly for the available checkpoints.
\end{description}
For all methods, we use the default parameters provided by the authors in their individual code repositories.
We use the watermark rescaling method presented by~\cite{BuiCVPR2023} with bicubic resizing (which we found to outperform bilinear resizing consistently across all methods).

\subsection{Quality evaluations}

\paragraph{Setup.}
\external{As explained in Section~\ref{sec:quality}, we perform human evaluations to assess the quality of \synthidom and the models listed above.}
For each model, we collect 10,000 ratings (5,000 for watermarked and 5,000 for non-watermarked samples) using 1,000 images generated by \imagen.\footnote{\tiny\url{https://deepmind.google/technologies/imagen-2/}}
Unlike \imagenet which is used in the robustness evaluation, all images have a resolution of 1536$\times$1536, which reflects real-world product conditions with the focus placed on quality (as desired by creators).
Images are generated across 25 high-level creative categories shown in Figure~\ref{fig:categories}.
Each high-level category is further sub-divided into different styles such as ``black \& white'' and ``colorful'' for ``abstract art''.
All prompts are listed in the appendix (see Section~\ref{sec:prompts}).
We introduce 10 additional dummy images that exhibit clear artifacts.
Ratings from raters with less than 70\% success rate at correctly identifying dummy images are excluded from the study.

\paragraph{Results.}
Figure \ref{fig:intro} and \ref{fig:quality_metrics} (top-left) report the average difference in human identified artifact rate between watermarked and non-watermarked samples.
A 5 percent point increase indicates that the corresponding watermarking method creates newly visible artifacts in at least 5\% of the images (it is a lower bound).
\external{\synthidom performs} best, followed by \videosealone, \invismark and \wam. \stegastamp has the largest reported artifacts rate, validating our observation that its high robustness relative to existing watermarking models comes at the cost of a large degradation in quality.
We also report computational quality metrics on the same clean and watermarked image pairs in Figure~\ref{fig:quality_metrics}.
While computational metrics tend to correlate with human judgment, they fail to capture obvious artifacts (especially when those are small in size).
This is clearly noticeable for \videosealzero and \trustmarkq which exhibit strong performance on computational metrics, but do poorly on the human evaluation.\footnote{It is worth noting that in side-by-side evaluations where trained raters are asked to tell which one between two almost identical images is watermarked, all methods (except \synthid) do poorly.}

\external{
\begin{figure}[t]
    \centering
    \includegraphics[width=\textwidth]{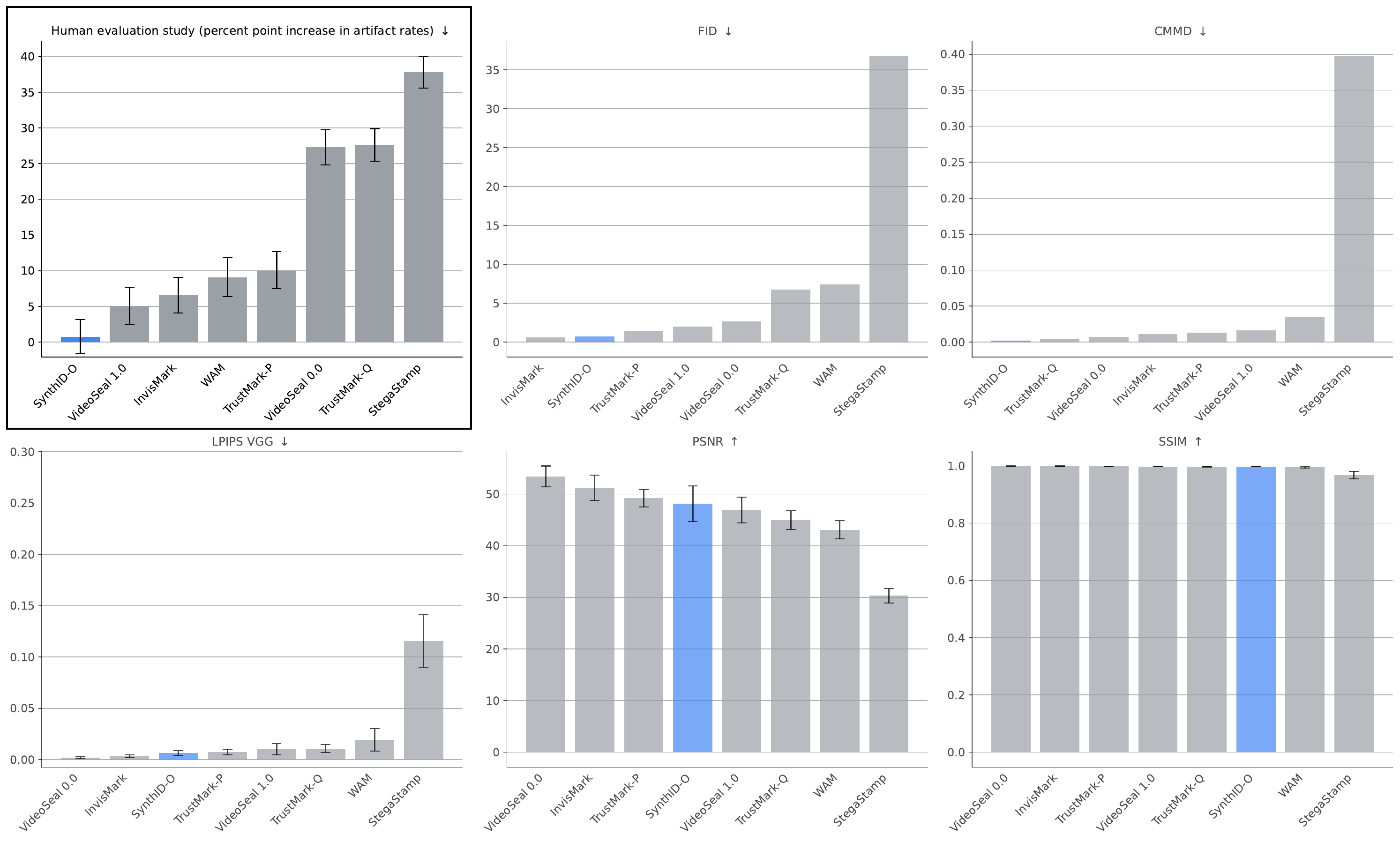}
    \caption{Quality metrics for all methods computed on 1,000 watermarked and unwatermarked images. The top-left panel shows the average  difference in identified artifact rate between watermarked and non-watermarked samples (lower is better). A 5 percent point increase indicates that the watermarking method creates visible artifacts in 5\% of the images. For FID, CMMD and LPIPS the lower is better and for PSNR and SSIM the higher is better. Overall, we see little correlation between computational metrics and human ratings.}
    \label{fig:quality_metrics}
\end{figure}
}

\subsection{Robustness evaluation}

\paragraph{Setup.}
We evaluate all models on a consistent set of 10,000 images from \imagenet's validation set.
We focus this evaluation on photo-realistic content which is the most likely source of misinformation.
The robustness of detection and payload recovery is assessed under 30 common image transformations (shown on Figure~\ref{fig:transformations_worst}).
In this section, we evaluate images at a resolution of 512$\times$512.
Evaluations at native image resolutions are shown in Appendix~\ref{app:native_image}.
Evaluations where images are transformed after being resized at each method's preferred resolution are shown in Appendix~\ref{app:native_model}.

\noindent We group image transformations into six categories: \textit{Color}, \textit{Combination}, \textit{Noise}, \textit{Overlay}, \textit{Quality} and \textit{Spatial}. \textit{Aggregated} includes all transformations across categories, and \textit{Identity} applies no transformation. Additionally, we use two evaluation settings:
\begin{description}[noitemsep,topsep=0pt]
\item[Random:] The strength of each transformation is sampled uniformly from within a specified range.
\item[Worst:] The strength of each transformation is set to its most severe value, typically an extreme within the range described above.
\end{description}
While the \textit{random} setting simulates real-world variability, \textit{worst} setting allows us to discern the limits of model performance under extreme image manipulation.
Unless stated otherwise, the models are calibrated to obtain a 0.1\% false positive rate in average on \textit{worst}-case transformations.

\paragraph{Detection results.}

In this section, we focus exclusively on the detection performance under various image transformations.
For models that offer a detection score independent of payload recovery (i.e., \synthidom and \wam), we use this score.
For models that solely perform payload recovery, we compute the detection score by counting the number of correctly decoded bits (as done by~\citealp{FernandezICCV2023}).\footnote{This implies that, for all methods except \synthidom and \wam, we assume that there is a single, unique \textbf{known} payload we can match.}
We do not observe any significant difference when using a weighted sum over payload logits.
We also include the performance of \synthidom and \wam using payload matching in a separate table.

\noindent Table~\ref{tab:results_tpr_0.1_globalfpr_aggregated_512x512} summarizes the results.
It reports the true positive rate (TPR) at a fixed false positive rate of 0.1\% in average on \textit{worst}-case transformations.
We observe that over aggregated \textit{random} and aggregated \textit{worst} transformations, %
\external{\synthidom outperforms} all other methods by significant margins (i.e., %
\external{$+9.36$ and $+16.35$} percentage points respectively).
In the \textit{worst} setting,
\external{\synthidom is the only models surpassing 99\% TPR in aggregate and it reaches more than 98\% TPR on combinations of transformations (which is the most challenging setting).}
Overall, it is worth noting that not a single model outperforms all others across all categories.
Comparisons are made more difficult when we start considering quality: \stegastamp is the worst offender in terms of quality and, as a result, it is regularly topping individual categories in terms of robustness.

\external{
\begin{table}[t]
\resizebox{\textwidth}{!}{%
\begin{tabular}{lcccccccc}
\toprule
\rowcolor[HTML]{EFEFEF}  & \rotatebox[origin=c]{45}{~\synthidom~} & \rotatebox[origin=c]{45}{~\invismark~} & \rotatebox[origin=c]{45}{~\stegastamp~} & \rotatebox[origin=c]{45}{~\trustmarkp~} & \rotatebox[origin=c]{45}{~\trustmarkq~} & \rotatebox[origin=c]{45}{~\videosealzero~} & \rotatebox[origin=c]{45}{~\videosealone~} & \rotatebox[origin=c]{45}{~\wam~} \\
\midrule
Identity (excl. resizing) & \textbf{100.00\%} & \textbf{100.00\%} & \textbf{100.00\%} & [99.95\%, 99.96\%] & \textbf{100.00\%} & \textbf{100.00\%} & 99.98\% & \textbf{100.00\%} \\
\rowcolor[HTML]{EFEFEF} Aggregated & \textbf{99.98\%} & [70.06\%, 70.41\%] & [72.41\%, 72.57\%] & [66.25\%, 66.74\%] & [78.61\%, 79.02\%] & [78.35\%, 78.91\%] & [88.75\%, 88.87\%] & \underline{90.62\%} \\
Color & \textbf{100.00\%} & [75.34\%, 75.49\%] & \underline{99.95\%} & [79.22\%, 79.82\%] & [99.65\%, 99.69\%] & [75.97\%, 76.22\%] & [99.86\%, 99.87\%] & 81.29\% \\
Combination & \textbf{99.96\%} & [22.28\%, 23.19\%] & [49.62\%, 49.75\%] & [23.06\%, 24.46\%] & [64.00\%, 66.25\%] & [77.67\%, 79.65\%] & [84.59\%, 84.91\%] & \underline{96.08\%} \\
Noise & 99.98\% & [92.10\%, 92.73\%] & \textbf{100.00\%} & [95.10\%, 95.49\%] & \textbf{100.00\%} & [96.93\%, 97.37\%] & [99.11\%, 99.18\%] & \textbf{100.00\%} \\
Overlay & \textbf{100.00\%} & \textbf{100.00\%} & \textbf{100.00\%} & [99.71\%, 99.73\%] & 99.99\% & [99.93\%, 99.95\%] & 99.99\% & \textbf{100.00\%} \\
Quality & 99.99\% & [89.51\%, 89.78\%] & \textbf{100.00\%} & [97.56\%, 97.72\%] & \textbf{100.00\%} & [98.60\%, 98.76\%] & 99.99\% & 99.97\% \\
Spatial & \textbf{99.98\%} & [61.09\%, 61.32\%] & [25.23\%, 25.73\%] & [38.05\%, 38.32\%] & [40.75\%, 41.04\%] & [55.95\%, 56.46\%] & [67.47\%, 67.72\%] & \underline{84.34\%} \\
\rowcolor[HTML]{EFEFEF} Aggregated Worst & \textbf{99.72\%} & [60.11\%, 60.53\%] & [66.51\%, 66.79\%] & [54.15\%, 54.68\%] & [68.73\%, 68.87\%] & [61.22\%, 61.91\%] & [76.74\%, 76.89\%] & \underline{83.37\%} \\
Color Worst & \textbf{100.00\%} & [69.58\%, 69.67\%] & \underline{[99.79\%, 99.82\%]} & [66.32\%, 67.16\%] & [98.93\%, 99.05\%] & [66.19\%, 66.49\%] & [99.69\%, 99.71\%] & 84.00\% \\
Combination Worst & \textbf{98.06\%} & [6.83\%, 7.89\%] & [27.07\%, 27.83\%] & [4.16\%, 4.76\%] & [21.22\%, 21.72\%] & [27.71\%, 29.05\%] & [31.00\%, 31.38\%] & \underline{55.96\%} \\
Noise Worst & 99.96\% & [80.86\%, 82.05\%] & \textbf{100.00\%} & [88.13\%, 89.02\%] & 99.99\% & [90.77\%, 91.93\%] & [97.73\%, 97.89\%] & \textbf{100.00\%} \\
Overlay Worst & \textbf{100.00\%} & \textbf{100.00\%} & \textbf{100.00\%} & [98.76\%, 98.85\%] & [99.87\%, 99.89\%] & [99.79\%, 99.83\%] & 99.97\% & \textbf{100.00\%} \\
Quality Worst & \underline{99.99\%} & [76.24\%, 76.73\%] & \textbf{100.00\%} & [84.65\%, 85.36\%] & \underline{99.99\%} & [81.84\%, 82.97\%] & 99.98\% & 99.38\% \\
Spatial Worst & \textbf{99.97\%} & [50.06\%, 50.13\%] & [15.25\%, 15.87\%] & [25.09\%, 25.17\%] & [25.76\%, 25.92\%] & [38.89\%, 39.32\%] & [51.62\%, 51.86\%] & \underline{76.04\%} \\
\bottomrule
\end{tabular}
}
\caption{TPR at 0.1\% FPR aggregated across each transformation category when images are resized to 512$\times$512. For each model, the detection threshold is calibrated to reach 0.1\% FPR across all \textit{worst} transformations in average. When brackets are used, we display the range of TPR values that can accommodate the target FPR (this number is not unique when there are ties).\label{tab:results_tpr_0.1_globalfpr_aggregated_512x512}}
\end{table}
}

\noindent For completeness, we also report the detection performance of \synthidom and \wam if they were to use their payload for detection in Table~\ref{tab:results_tpr0.1globalfpr_aggregated_payload_512x512}.
\external{
\begin{table}[t]
\begin{minipage}[c]{0.4\textwidth}
\resizebox{\textwidth}{!}{%
\begin{tabular}{lcc}
\toprule
\rowcolor[HTML]{EFEFEF}  & \rotatebox[origin=c]{45}{~\synthidom~} & \rotatebox[origin=c]{45}{~\wam~} \\
\midrule
Identity (excl. resizing) & 100.00\% & [99.96\%, 99.98\%] \\
\rowcolor[HTML]{EFEFEF} Aggregated & [99.96\%, 99.97\%] & [81.26\%, 81.85\%] \\
Color & [99.99\%, 100.00\%] & [77.16\%, 77.31\%] \\
Combination & [99.92\%, 99.93\%] & [89.68\%, 92.56\%] \\
Noise & [99.89\%, 99.90\%] & [99.95\%, 99.98\%] \\
Overlay & 100.00\% & [99.96\%, 99.98\%] \\
Quality & [99.96\%, 99.97\%] & [99.70\%, 99.79\%] \\
Spatial & 99.98\% & [57.39\%, 57.91\%] \\
\rowcolor[HTML]{EFEFEF} Aggregated Worst & [98.84\%, 98.90\%] & [69.11\%, 69.91\%] \\
Color Worst & 99.99\% & [69.78\%, 69.82\%] \\
Combination Worst & [91.96\%, 92.36\%] & [36.25\%, 39.52\%] \\
Noise Worst & [99.74\%, 99.77\%] & [99.95\%, 99.98\%] \\
Overlay Worst & 100.00\% & [99.97\%, 99.98\%] \\
Quality Worst & 99.97\% & [95.21\%, 96.29\%] \\
Spatial Worst & 99.97\% & [48.76\%, 49.45\%] \\
\bottomrule
\end{tabular}
}
\end{minipage}\hfill
\begin{minipage}[c]{\dimexpr\linewidth-0.4\textwidth-2\tabcolsep\relax}
\caption{TPR at 0.1\% FPR aggregated across each transform category when images are resized to 512$\times$512. The detection is performed by computing the number of correct bits in the payload. For each model, the detection threshold is calibrated to reach 0.1\% FPR across all \textit{worst} transformations. Models that are not listed here are already listed in Table~\ref{tab:results_tpr_0.1_globalfpr_aggregated_512x512}. When brackets are used, we display the range of TPR values that can accommodate the target FPR (this number is not unique when there are ties). \label{tab:results_tpr0.1globalfpr_aggregated_payload_512x512}}
\end{minipage}
\end{table}
}

\paragraph{Payload recovery results.}

We report the bit accuracy measured on watermarked images (as detailed in Section~\ref{sec:payload}).
Table~\ref{tab:results_bit_accuracy_aggregated_512x512} compares  \synthidom to other models.
We observe that \synthidom often surpasses other models despite having a larger payload (except \videosealone) and better quality.
\wam which tops a few categories only has a 32-bit payload.

\external{
\begin{table}[t]
\resizebox{\textwidth}{!}{%
\begin{tabular}{lcccccccc}
\toprule
\rowcolor[HTML]{EFEFEF}  & \rotatebox[origin=c]{45}{\shortstack[c]{~\synthidom~ \\  (136 bits)}} & \rotatebox[origin=c]{45}{\shortstack[c]{~\invismark~ \\  (100 bits)}} & \rotatebox[origin=c]{45}{\shortstack[c]{~\stegastamp~ \\  (100 bits)}} & \rotatebox[origin=c]{45}{\shortstack[c]{~\trustmarkp~ \\  (100 bits)}} & \rotatebox[origin=c]{45}{\shortstack[c]{~\trustmarkq~ \\  (100 bits)}} & \rotatebox[origin=c]{45}{\shortstack[c]{~\videosealzero~ \\  (96 bits)}} & \rotatebox[origin=c]{45}{\shortstack[c]{~\videosealone~ \\  (256 bits)}} & \rotatebox[origin=c]{45}{\shortstack[c]{~\wam~ \\  (32 bits)}} \\
\midrule
Identity (excl. resizing) & \textbf{100.00\%} & 95.04\% & 99.84\% & 99.23\% & 99.91\% & 96.94\% & 99.61\% & \underline{99.96\%} \\
\rowcolor[HTML]{EFEFEF} Aggregated & \textbf{99.59\%} & 79.28\% & 85.03\% & 79.46\% & 87.27\% & 82.67\% & 89.60\% & \underline{90.62\%} \\
Color & \textbf{99.97\%} & 82.47\% & \underline{99.14\%} & 84.53\% & 98.16\% & 83.24\% & 97.60\% & 88.42\% \\
Combination & \textbf{99.12\%} & 56.57\% & 70.45\% & 58.64\% & 74.10\% & 74.75\% & 79.31\% & \underline{93.04\%} \\
Noise & 98.57\% & 85.21\% & \underline{99.61\%} & 89.26\% & 98.85\% & 87.98\% & 90.13\% & \textbf{99.94\%} \\
Overlay & \textbf{100.00\%} & 94.62\% & 99.66\% & 97.55\% & 98.93\% & 96.24\% & 99.27\% & \underline{99.96\%} \\
Quality & \underline{99.73\%} & 88.56\% & \textbf{99.75\%} & 95.30\% & 99.62\% & 93.13\% & 98.75\% & 99.65\% \\
Spatial & \textbf{99.82\%} & 76.40\% & 61.68\% & 68.09\% & 69.43\% & 74.85\% & \underline{80.47\%} & 79.83\% \\
\rowcolor[HTML]{EFEFEF} Aggregated Worst & \textbf{98.07\%} & 75.26\% & 82.58\% & 73.29\% & 82.47\% & 76.03\% & 83.53\% & \underline{85.03\%} \\
Color Worst & \textbf{99.93\%} & 78.31\% & \underline{98.19\%} & 76.86\% & 96.03\% & 78.16\% & 95.10\% & 84.29\% \\
Combination Worst & \textbf{89.46\%} & 52.60\% & 62.62\% & 51.69\% & 58.16\% & 59.06\% & 60.49\% & \underline{71.86\%} \\
Noise Worst & 97.63\% & 80.61\% & \underline{99.44\%} & 84.64\% & 98.07\% & 84.34\% & 86.10\% & \textbf{99.91\%} \\
Overlay Worst & \textbf{99.99\%} & 93.91\% & 99.04\% & 93.69\% & 94.80\% & 95.55\% & 98.52\% & \underline{99.96\%} \\
Quality Worst & \underline{99.49\%} & 84.17\% & \textbf{99.57\%} & 89.02\% & 99.05\% & 87.00\% & 97.34\% & 97.30\% \\
Spatial Worst & \textbf{99.77\%} & 72.13\% & 57.87\% & 62.33\% & 63.04\% & 68.14\% & 73.00\% & \underline{74.97\%} \\
\bottomrule
\end{tabular}
}
\caption{Payload bit accuracy aggregated across each transformation category when images are resized to 512$\times$512. Note that numbers are not directly comparable as the payload size is different for different methods.\label{tab:results_bit_accuracy_aggregated_512x512}}
\end{table}
}

\section{Related work}
\label{sec:related-work}

While this paper is not meant to be a comprehensive survey on watermarking -- we refer to \cite{WanNEUROCOMPUTING2022} for relatively a recent overview -- we want to provide a short overview of watermarking in general and recent developments in deploying watermarking alongside generative AI.
Besides presenting results for \synthidimage, this paper also touches on robustness, threat models and practical concerns. In this spirit, our paper is similar to this recent State-of-Knowledge on watermarking \citep{ZhaoARXIV2024}.

To the best of our knowledge, \cite{VanSchyndelICIP1994,BraudawayICIP1997} presented the first invisible watermarks. As visible watermarks were not yet automatically removable as easily as today \citep{DekelCVPR2017}, early work in invisible watermarking was closely entangled with steganography \citep{Cox2007}. Irrespective of the use, early methods either used the spatial or frequency domain to hide information. For example, \cite{WangPR2001} hide information in less important bits across pixels. This is not very robust to common image transformations \citep{FazliICMV2009}. Instead, a long line of work uses DCT \citep{AhmedTC1974} or DWT \citep{ShensaTSP1992} transforms. Well-received examples include \cite{BarniSP1998,ChenMMSP1998,ChenTIT2001,ParahDSP2016} and \cite{BarniTIP2001,Kashyap2012}, respectively. Approaches working in the spatial and frequency domains have also been combined, e.g., using DCT-SVD approaches \citep{SinghMTA2017,SuSC2018}. Recently, DCT and DWT based approaches have been re-popularized due to their usage by Stability AI\footnote{\tiny\url{https://github.com/Stability-AI/invisible-watermark-gpu}} despite their flaws compared to recent deep learning based methods.

While there are early experiments using neural networks for watermark detection in \citep{YuSP2001}, proper deep learning based approaches gained traction around 2017 \citep{MunARXIV2017,ZhuECCV2018}. Usually, encoder-decoder architectures (i.e., content-based, post-hoc methods) tried to improve on quality and robustness \citep{WenARXIV2019,HayesNIPS2017,HayesARXIV2020b,TancikCVPR2020,LuoCVPR2020,BuiARXIV2023,BuiCVPR2023,PanNIPSWORK2023,PanARXIV2024,xu2024invismark}. Recently, these approaches have been extended to allow localization of (multiple) watermarks \citep{sander2024watermark}. Only more recently, with the rise of diffusion models \citep{HoNIPS2020,SohlDicksteinICML2015} such as stable diffusion \citep{RombachARXIV2022}, watermarking has been integrated into the generation process; see \cite{FernandezICCV2023,WenARXIV2023,YangARXIV2024}. The pros and cons have been discussed at length in Section \ref{sec:problem}. These works usually exclusively focus on image watermarking; literature on video \citep{ZhangARXIV2019,fernandez2024video} or audio watermarking \citep{HuaSP2016,RomanARXIV2024} tends to be sparse, possibly because methods are usually general enough across these modalities.

The importance of appropriate benchmarks for watermarking has been highlighted as early as 2000 \citep{PetitcolasSPM2000} with a thorough account of possible attacks some years later by \cite{Cox2007}. More recently, \cite{AnARXIV2024} introduced an up-to-date and relatively comprehensive benchmark of deep learning based watermarking techniques. Proper benchmarking also includes considering a variety of (adaptive) attacks \citep{SaberiARXIV2023,LukasARXIV2023,ZhaoARXIV2023a}, many of which have been discussed at length in Section \ref{sec:attacks}. Meta also maintains a benchmark for image watermarking, called Omni Seal Bench.\footnote{\tiny\url{https://github.com/facebookresearch/omnisealbench}}

In terms of deployed systems at the time of writing, the solution used by Stability AI\footnote{\tiny\url{https://github.com/ShieldMnt/invisible-watermark}} has received considerable attention, but usage is difficult to track. Imatag released several models on Hugging Face, including a \stablesignature \citep{FernandezICCV2023} based one.\footnote{\tiny\url{https://huggingface.co/imatag/stable-signature-bzh-sdxl-vae-weak}} However, the corresponding proprietary service seems to focus on content tracking and information leak identification.\footnote{\tiny\url{https://www.imatag.com/}} For DALL-E 2\footnote{\tiny\url{https://openai.com/index/dall-e-2/}}, OpenAI experimented with a visible watermark referred to as ``coloured squares'' by users. While there was no official communication on this, and OpenAI's FAQs state that users are allowed to remove it,\footnote{\tiny\url{https://web.archive.org/web/*/https://help.openai.com/en/articles/6468065-dall-e-2-faq}} this can be understood as an early approach to watermarking. More recently, for DALL-E 3,\footnote{\tiny\url{https://openai.com/index/dall-e-3/}} OpenAI announced both C2PA metadata as well as proper invisible watermarking (for image and audio). To date, however, this has not been released. Instead, OpenAI showcased early results for AI-generated image detection,\footnote{\tiny\url{https://openai.com/index/understanding-the-source-of-what-we-see-and-hear-online/}} which seems rather fragile under certain transformations. Meta recently announced to follow C2PA specifications and reveal this information to users on Facebook, Instragram and Threads.\footnote{\tiny\url{https://about.fb.com/news/2024/02/labeling-ai-generated-images-on-facebook-instagram-and-threads/}} Meta seems to pursue invisible watermarking across modalities, as well, with some open-source models \citep{sander2024watermark,RomanARXIV2024,fernandez2024video} but no concrete public-facing verification system yet. Similarly, Microsoft published and open-sourced an image watermarking model \citep{xu2024invismark}.\footnote{\tiny\url{https://github.com/microsoft/InvisMark}} Overall, this makes \synthidimage the first system deployed at large-scale that consistently watermarks large amounts of generated content and discloses this information to end users.

\external{
\section{Limitations and future work}
}

Despite its success, \synthidimage alone will not solve many of the problems we set out to alleviate, including misinformation, impersonation or copyright tracking. This is because watermarking in itself does not solve the provenance problem. Instead, \synthidimage needs to become part of an ecosystem of tools, including C2PA, that are adopted by all major players across industry and governments to watermark AI-generated content and surface this information to the public. This requires both technological advancements as well as work on policy and cooperation across companies and governments. Besides, there are various technical challenges that we believe will become important with advancements in generative AI. For example, \synthid operates on text, images, audio or video. Thus, work on detection and handling fractional watermarks becomes important. Similarly, we likely need larger payloads as use cases of generative AI grow and we want to enable public detectability using cryptographic signatures~\citep{fairoze2025difficulty}. For open models, we need to work on improving security, particularly considering white-box threat models and we generally need to figure out proper versioning of watermarks, assuming that multiple watermarking solutions will be available soon.

\section*{Acknowledgments}

We would like to thank Demis Hassabis for starting, sponsoring and advising throughout this project. We thank Clement Wolf, Ian Goodfellow, Chris Bregler and Oriol Vinyals for their advice. We also thank many others who contributed across Google DeepMind and Google, including our partners at Google Research, Google Cloud, Google Global Business \& Operations and YouTube.

\bibliography{main}

\begin{thebibliography}{113}
\providecommand{\natexlab}[1]{#1}
\providecommand{\url}[1]{\texttt{#1}}
\expandafter\ifx\csname urlstyle\endcsname\relax
  \providecommand{\doi}[1]{doi: #1}\else
  \providecommand{\doi}{doi: \begingroup \urlstyle{rm}\Url}\fi

\bibitem[Adi et~al.(2018)Adi, Baum, Ciss{\'{e}}, Pinkas, and Keshet]{AdiUSENIX2018}
Y.~Adi, C.~Baum, M.~Ciss{\'{e}}, B.~Pinkas, and J.~Keshet.
\newblock Turning your weakness into a strength: Watermarking deep neural networks by backdooring.
\newblock In \emph{USENIX Security Symposium}, pages 1615--1631, 2018.

\bibitem[Agarwal et~al.(2019)Agarwal, Farid, Gu, He, Nagano, and Li]{AgarwalCVPR2019}
S.~Agarwal, H.~Farid, Y.~Gu, M.~He, K.~Nagano, and H.~Li.
\newblock Protecting world leaders against deep fakes.
\newblock In \emph{Proc. of the IEEE Conference on Computer Vision and Pattern Recognition (CVPR)}, 2019.

\bibitem[Ahmed et~al.(1974)Ahmed, Natarajan, and Rao]{AhmedTC1974}
N.~Ahmed, T.~R. Natarajan, and K.~R. Rao.
\newblock Discrete cosine transform.
\newblock \emph{{IEEE} Transactions on Computers}, 23\penalty0 (1):\penalty0 90--93, 1974.

\bibitem[Alaifari et~al.(2019)Alaifari, Alberti, and Gauksson]{AlaifariARXIV2018}
R.~Alaifari, G.~S. Alberti, and T.~Gauksson.
\newblock {ADef}: {An} iterative algorithm to construct adversarial deformations.
\newblock In \emph{Proc. of the International Conference on Learning Representations (ICLR)}, 2019.

\bibitem[An et~al.(2024)An, Ding, Rabbani, Agrawal, Xu, Deng, Zhu, Mohamed, Wen, Goldstein, and Huang]{AnARXIV2024}
B.~An, M.~Ding, T.~Rabbani, A.~Agrawal, Y.~Xu, C.~Deng, S.~Zhu, A.~Mohamed, Y.~Wen, T.~Goldstein, and F.~Huang.
\newblock {WAVES:} {Benchmarking} the robustness of image watermarks.
\newblock In \emph{Proc. of the International Conference on Machine Learning (ICML)}, 2024.

\bibitem[Anderson(1996)]{Anderson1996}
R.~J. Anderson, editor.
\newblock \emph{Information Hiding, First International Workshop, Cambridge, UK, May 30 - June 1, 1996, Proceedings}, volume 1174, 1996.

\bibitem[Balan et~al.(2023)Balan, Agarwal, Jenni, Parsons, Gilbert, and Collomosse]{BalanCVPRWORK2023}
K.~Balan, S.~Agarwal, S.~Jenni, A.~Parsons, A.~Gilbert, and J.~P. Collomosse.
\newblock {EKILA:} {Synthetic} media provenance and attribution for generative art.
\newblock In \emph{Proc. of the IEEE Conference on Computer Vision and Pattern Recognition (CVPR) Workshops}, 2023.

\bibitem[Barni et~al.(1998)Barni, Bartolini, Cappellini, and Piva]{BarniSP1998}
M.~Barni, F.~Bartolini, V.~Cappellini, and A.~Piva.
\newblock A {DCT}-domain system for robust image watermarking.
\newblock \emph{Signal Processing}, 66\penalty0 (3):\penalty0 357--372, 1998.

\bibitem[Barni et~al.(2001)Barni, Bartolini, and Piva]{BarniTIP2001}
M.~Barni, F.~Bartolini, and A.~Piva.
\newblock Improved wavelet-based watermarking through pixel-wise masking.
\newblock \emph{IEEE Trans. on Image Processing (TIP)}, 10\penalty0 (5):\penalty0 783--791, 2001.

\bibitem[Bates et~al.(2021)Bates, Candes, Lei, Romano, and Sesia]{BatesICMLWORK2021}
S.~Bates, E.~Candes, L.~Lei, Y.~Romano, and M.~Sesia.
\newblock Calibrated out-of-distribution detection with conformal p-values.
\newblock In \emph{Proc. of the International Conference on Machine Learning (ICML) Workshops}, 2021.

\bibitem[Bates et~al.(2023)Bates, Cand{\`{e}}s, Lei, Romano, and Sesia]{BatesARXIV2022}
S.~Bates, J.~Cand{\`{e}}s, L.~Lei, Y.~Romano, and M.~Sesia.
\newblock Testing for outliers with conformal p-values.
\newblock \emph{{Ann. of Statistics}}, 51\penalty0 (1):\penalty0 149--178, 2023.

\bibitem[Bharati et~al.(2021)Bharati, Moreira, Flynn, de~Rezende~Rocha, Bowyer, and Scheirer]{BharatiTIFS2021}
A.~Bharati, D.~Moreira, P.~J. Flynn, A.~de~Rezende~Rocha, K.~W. Bowyer, and W.~J. Scheirer.
\newblock Transformation-aware embeddings for image provenance.
\newblock \emph{Proc. of the IEEE Transactions on Information Forensics and Security}, 16, 2021.

\bibitem[Biggio et~al.(2013)Biggio, Corona, Maiorca, Nelson, Srndic, Laskov, Giacinto, and Roli]{BiggioECMLPKDD2013}
B.~Biggio, I.~Corona, D.~Maiorca, B.~Nelson, N.~Srndic, P.~Laskov, G.~Giacinto, and F.~Roli.
\newblock Evasion attacks against machine learning at test time.
\newblock In \emph{Proc. of the European Conference on Machine Learning and Knowledge Discovery in Databases (ECML PKDD)}, 2013.

\bibitem[Braudaway(1997)]{BraudawayICIP1997}
G.~W. Braudaway.
\newblock Protecting publicly-available images with an invisible image watermark.
\newblock In \emph{Proc. of the IEEE International Conference on Image Processing (ICIP)}, 1997.

\bibitem[Brown et~al.(2017)Brown, Man{\'{e}}, Roy, Abadi, and Gilmer]{BrownARXIV2017}
T.~B. Brown, D.~Man{\'{e}}, A.~Roy, M.~Abadi, and J.~Gilmer.
\newblock Adversarial patch.
\newblock \emph{arXiv.org}, abs/1712.09665, 2017.

\bibitem[Bryniarski et~al.(2022)Bryniarski, Hingun, Pachuca, Wang, and Carlini]{BryniarskiARXIV2021}
O.~Bryniarski, N.~Hingun, P.~Pachuca, V.~Wang, and N.~Carlini.
\newblock Evading adversarial example detection defenses with orthogonal projected gradient descent.
\newblock In \emph{Proc. of the International Conference on Learning Representations (ICLR)}, 2022.

\bibitem[Bui et~al.(2023)Bui, Agarwal, Yu, and Collomosse]{BuiCVPR2023}
T.~Bui, S.~Agarwal, N.~Yu, and J.~P. Collomosse.
\newblock {RoSteALS}: {Robust} steganography using autoencoder latent space.
\newblock In \emph{Proc. of the IEEE Conference on Computer Vision and Pattern Recognition (CVPR)}, 2023.

\bibitem[Bui et~al.(2025)Bui, Agarwal, and Collomosse]{BuiARXIV2023}
T.~Bui, S.~Agarwal, and J.~P. Collomosse.
\newblock {TrustMark}: {Universal} watermarking for arbitrary resolution images.
\newblock In \emph{Proc. of the IEEE International Conference on Computer Vision (ICCV)}, 2025.

\bibitem[Carlini and Farid(2020)]{CarliniARXIV2020}
N.~Carlini and H.~Farid.
\newblock Evading deepfake-image detectors with white- and black-box attacks.
\newblock In \emph{Proc. of the IEEE Conference on Computer Vision and Pattern Recognition (CVPR)}, 2020.

\bibitem[Carlini and Wagner(2017{\natexlab{a}})]{CarliniARXIV2017}
N.~Carlini and D.~Wagner.
\newblock Adversarial examples are not easily detected: Bypassing ten detection methods.
\newblock In \emph{Proc. of the ACM Workshop on Artificial Intelligence and Security}, 2017{\natexlab{a}}.

\bibitem[Carlini and Wagner(2017{\natexlab{b}})]{CarliniSP2017}
N.~Carlini and D.~Wagner.
\newblock Towards evaluating the robustness of neural networks.
\newblock In \emph{Proc. of the IEEE Symposium on Security and Privacy}, 2017{\natexlab{b}}.

\bibitem[Chen and Wornell(1998)]{ChenMMSP1998}
B.~Chen and G.~W. Wornell.
\newblock Digital watermarking and information embedding using dither modulation.
\newblock In \emph{{IEEE} Workshop on Multimedia Signal Processing ({MMSP})}, 1998.

\bibitem[Chen and Wornell(2001)]{ChenTIT2001}
B.~Chen and G.~W. Wornell.
\newblock Quantization index modulation: {A} class of provably good methods for digital watermarking and information embedding.
\newblock \emph{{IEEE} Transactions on Information Theory}, 47\penalty0 (4):\penalty0 1423--1443, 2001.

\bibitem[Christ et~al.(2024)Christ, Gunn, and Zamir]{ChristARXIV2023}
M.~Christ, S.~Gunn, and O.~Zamir.
\newblock Undetectable watermarks for language models.
\newblock In \emph{Proc. of the Conference on Learning Theory (COLT)}, 2024.

\bibitem[{Coalition for Content Provenance and Authenticity}(2023)]{c2pa2023}
{Coalition for Content Provenance and Authenticity}.
\newblock {C2PA} technical specifications version 1.3.
\newblock https://c2pa.org/specifications/specifications/1.3/specs/C2PA\_Specification.html, 2023.
\newblock Online; accessed Mar 15 2024.

\bibitem[Collomosse and Parsons(2024)]{collomosse2024authenticity}
J.~Collomosse and A.~Parsons.
\newblock To authenticity, and beyond! {B}uilding safe and fair generative {AI} upon the three pillars of provenance.
\newblock \emph{IEEE Computer Graphics and Applications}, 44\penalty0 (3):\penalty0 82--90, 2024.

\bibitem[Cox(2007)]{Cox2007}
I.~Cox.
\newblock Digital watermarking and steganography.
\newblock \emph{The Morgan Kaufmann Series in Multimedia Information and Systems}, 2:\penalty0 893--914, 2007.

\bibitem[Croce and Hein(2020)]{CroceICML2020}
F.~Croce and M.~Hein.
\newblock Reliable evaluation of adversarial robustness with an ensemble of diverse parameter-free attacks.
\newblock In \emph{Proc. of the International Conference on Machine Learning (ICML)}, 2020.

\bibitem[Dathathri et~al.(2024)Dathathri, See, Ghaisas, Huang, McAdam, Welbl, Bachani, Kaskasoli, Stanforth, Matejovicova, et~al.]{dathathri2024scalable}
S.~Dathathri, A.~See, S.~Ghaisas, P.-S. Huang, R.~McAdam, J.~Welbl, V.~Bachani, A.~Kaskasoli, R.~Stanforth, T.~Matejovicova, et~al.
\newblock Scalable watermarking for identifying large language model outputs.
\newblock \emph{Nature}, 634\penalty0 (8035):\penalty0 818--823, 2024.

\bibitem[Dekel et~al.(2017)Dekel, Rubinstein, Liu, and Freeman]{DekelCVPR2017}
T.~Dekel, M.~Rubinstein, C.~Liu, and W.~T. Freeman.
\newblock On the effectiveness of visible watermarks.
\newblock In \emph{Proc. of the IEEE Conference on Computer Vision and Pattern Recognition (CVPR)}, 2017.

\bibitem[Dumont et~al.(2018)Dumont, Maggio, and Montalvo]{DumontARXIV2018}
B.~Dumont, S.~Maggio, and P.~Montalvo.
\newblock Robustness of rotation-equivariant networks to adversarial perturbations.
\newblock \emph{arXiv.org}, abs/1802.06627, 2018.

\bibitem[Engstrom et~al.(2017)Engstrom, Tsipras, Schmidt, and Madry]{EngstromARXIV2017}
L.~Engstrom, D.~Tsipras, L.~Schmidt, and A.~Madry.
\newblock A rotation and a translation suffice: Fooling {CNNs} with simple transformations.
\newblock \emph{arXiv.org}, abs/1712.02779, 2017.

\bibitem[Fairoze et~al.(2025)Fairoze, Ortiz-Jimenez, Vecerik, Jha, and Gowal]{fairoze2025difficulty}
J.~Fairoze, G.~Ortiz-Jimenez, M.~Vecerik, S.~Jha, and S.~Gowal.
\newblock On the difficulty of constructing a robust and publicly-detectable watermark.
\newblock In \emph{Conference on Artificial Intelligence and Statistics (AISTATS)}, 2025.

\bibitem[Fazli and Khodaverdi(2009)]{FazliICMV2009}
S.~Fazli and G.~Khodaverdi.
\newblock Trade-off between imperceptibility and robustness of {LSB} watermarking using {SSIM} quality metrics.
\newblock In \emph{Proc. of the International Conference on Machine Vision}, 2009.

\bibitem[Fernandez et~al.(2023)Fernandez, Couairon, J{\'{e}}gou, Douze, and Furon]{FernandezICCV2023}
P.~Fernandez, G.~Couairon, H.~J{\'{e}}gou, M.~Douze, and T.~Furon.
\newblock The stable signature: Rooting watermarks in latent diffusion models.
\newblock In \emph{Proc. of the IEEE International Conference on Computer Vision (ICCV)}, 2023.

\bibitem[Fernandez et~al.(2024)Fernandez, Elsahar, Yalniz, and Mourachko]{fernandez2024video}
P.~Fernandez, H.~Elsahar, I.~Z. Yalniz, and A.~Mourachko.
\newblock {V}ideo {S}eal: Open and efficient video watermarking.
\newblock \emph{arXiv preprint arXiv:2412.09492}, 2024.

\bibitem[Gowal et~al.(2021)Gowal, Huang, van~den Oord, Mann, and Kohli]{GowalICLR2021}
S.~Gowal, P.-S. Huang, A.~van~den Oord, T.~Mann, and P.~Kohli.
\newblock Self-supervised adversarial robustness for the low-label, high-data regime.
\newblock In \emph{Proc. of the International Conference on Learning Representations (ICLR)}, 2021.

\bibitem[Gunn et~al.(2025)Gunn, Zhao, and Song]{gunn2024undetectable}
S.~Gunn, X.~Zhao, and D.~Song.
\newblock An undetectable watermark for generative image models.
\newblock In \emph{Proc. of the International Conference on Learning Representations (ICLR)}, 2025.

\bibitem[Hayes(2020)]{HayesARXIV2020}
J.~Hayes.
\newblock Provable trade-offs between private {\&} robust machine learning.
\newblock \emph{arXiv.org}, abs/2006.04622, 2020.

\bibitem[Hayes and Danezis(2017)]{HayesNIPS2017}
J.~Hayes and G.~Danezis.
\newblock Generating steganographic images via adversarial training.
\newblock In \emph{Advances in Neural Information Processing Systems (NeurIPS)}, 2017.

\bibitem[Hayes et~al.(2020)Hayes, Dvijotham, Chen, Dieleman, Kohli, and Casagrande]{HayesARXIV2020b}
J.~Hayes, K.~Dvijotham, Y.~Chen, S.~Dieleman, P.~Kohli, and N.~Casagrande.
\newblock Towards transformation-resilient provenance detection of digital media.
\newblock \emph{arXiv.org}, abs/2011.07355, 2020.

\bibitem[Hendrycks and Dietterich(2019)]{HendrycksARXIV2018}
D.~Hendrycks and T.~G. Dietterich.
\newblock Benchmarking neural network robustness to common corruptions and perturbations.
\newblock In \emph{Proc. of the International Conference on Learning Representations (ICLR)}, 2019.

\bibitem[Hendrycks et~al.(2019)Hendrycks, Mazeika, and Dietterich]{HendrycksARXIV2019b}
D.~Hendrycks, M.~Mazeika, and T.~G. Dietterich.
\newblock Deep anomaly detection with outlier exposure.
\newblock In \emph{Proc. of the International Conference on Learning Representations (ICLR)}, 2019.

\bibitem[Heusel et~al.(2017)Heusel, Ramsauer, Unterthiner, Nessler, and Hochreiter]{HeuselNIPS2017}
M.~Heusel, H.~Ramsauer, T.~Unterthiner, B.~Nessler, and S.~Hochreiter.
\newblock {GANs} trained by a two time-scale update rule converge to a local {N}ash equilibrium.
\newblock In \emph{Advances in Neural Information Processing Systems (NeurIPS)}, 2017.

\bibitem[Ho et~al.(2020)Ho, Jain, and Abbeel]{HoNIPS2020}
J.~Ho, A.~Jain, and P.~Abbeel.
\newblock Denoising diffusion probabilistic models.
\newblock In \emph{Advances in Neural Information Processing Systems (NeurIPS)}, 2020.

\bibitem[Hochberg(1988)]{HochbergB1988}
Y.~Hochberg.
\newblock A sharper {B}onferroni procedure for multiple tests of significance.
\newblock \emph{Biometrika}, 75\penalty0 (4):\penalty0 800--802, 1988.

\bibitem[Holm(1979)]{Holm1979}
S.~Holm.
\newblock A simple sequentially rejective multiple test procedure.
\newblock \emph{Scandinavian Journal of Statistics}, pages 65--70, 1979.

\bibitem[Hua et~al.(2016)Hua, Huang, Shi, Goh, and Thing]{HuaSP2016}
G.~Hua, J.~Huang, Y.~Q. Shi, J.~Goh, and V.~L.~L. Thing.
\newblock Twenty years of digital audio watermarking - a comprehensive review.
\newblock \emph{Signal Processing}, 128:\penalty0 222--242, 2016.

\bibitem[Jayasumana et~al.(2024)Jayasumana, Ramalingam, Veit, Glasner, Chakrabarti, and Kumar]{Jayasumana2023RethinkingFT}
S.~Jayasumana, S.~Ramalingam, A.~Veit, D.~Glasner, A.~Chakrabarti, and S.~Kumar.
\newblock Rethinking {FID}: Towards a better evaluation metric for image generation.
\newblock In \emph{Proc. of the IEEE Conference on Computer Vision and Pattern Recognition (CVPR)}, 2024.

\bibitem[Jovanovic et~al.(2024)Jovanovic, Staab, and Vechev]{JovanovicARXIV2024}
N.~Jovanovic, R.~Staab, and M.~T. Vechev.
\newblock Watermark stealing in large language models.
\newblock In \emph{Proc. of the International Conference on Machine Learning (ICML)}, 2024.

\bibitem[Kadian et~al.(2021)Kadian, Arora, and Arora]{KadianWPC2021}
P.~Kadian, S.~M. Arora, and N.~Arora.
\newblock Robust digital watermarking techniques for copyright protection of digital data: {A} survey.
\newblock \emph{Wireless Personal Communications}, 118\penalty0 (4), 2021.

\bibitem[Kashyap and Sinha(2012)]{Kashyap2012}
N.~Kashyap and G.~Sinha.
\newblock Image watermarking using 3-level discrete wavelet transform {(DWT)}.
\newblock \emph{International Journal of Modern Education and Computer Science}, 4\penalty0 (3):\penalty0 50, 2012.

\bibitem[Kirchenbauer et~al.(2023)Kirchenbauer, Geiping, Wen, Katz, Miers, and Goldstein]{KirchenbauerICML2023}
J.~Kirchenbauer, J.~Geiping, Y.~Wen, J.~Katz, I.~Miers, and T.~Goldstein.
\newblock A watermark for large language models.
\newblock In \emph{Proc. of the International Conference on Machine Learning (ICML)}, 2023.

\bibitem[Kurakin et~al.(2017)Kurakin, Goodfellow, and Bengio]{KurakinICLR2017b}
A.~Kurakin, I.~J. Goodfellow, and S.~Bengio.
\newblock Adversarial examples in the physical world.
\newblock In \emph{Proc. of the International Conference on Learning Representations (ICLR)}, 2017.

\bibitem[Lukas and Kerschbaum(2023)]{LukasUSENIX2023}
N.~Lukas and F.~Kerschbaum.
\newblock {PTW:} {Pivotal} tuning watermarking for pre-trained image generators.
\newblock In \emph{USENIX Security Symposium}, 2023.

\bibitem[Lukas et~al.(2024)Lukas, Diaa, Fenaux, and Kerschbaum]{LukasARXIV2023}
N.~Lukas, A.~Diaa, L.~Fenaux, and F.~Kerschbaum.
\newblock Leveraging optimization for adaptive attacks on image watermarks.
\newblock In \emph{Proc. of the International Conference on Learning Representations (ICLR)}, 2024.

\bibitem[Luo et~al.(2020)Luo, Zhan, Chang, Yang, and Milanfar]{LuoCVPR2020}
X.~Luo, R.~Zhan, H.~Chang, F.~Yang, and P.~Milanfar.
\newblock Distortion agnostic deep watermarking.
\newblock In \emph{Proc. of the IEEE Conference on Computer Vision and Pattern Recognition (CVPR)}, 2020.

\bibitem[Madry et~al.(2018)Madry, Makelov, Schmidt, Tsipras, and Vladu]{MadryICLR2018}
A.~Madry, A.~Makelov, L.~Schmidt, D.~Tsipras, and A.~Vladu.
\newblock Towards deep learning models resistant to adversarial attacks.
\newblock \emph{Proc. of the International Conference on Learning Representations (ICLR)}, 2018.

\bibitem[Maini et~al.(2020)Maini, Wong, and Kolter]{MainiARXIV2019}
P.~Maini, E.~Wong, and J.~Z. Kolter.
\newblock Adversarial robustness against the union of multiple perturbation models.
\newblock In \emph{Proc. of the International Conference on Machine Learning (ICML)}, 2020.

\bibitem[Mun et~al.(2017)Mun, Nam, Jang, Kim, and Lee]{MunARXIV2017}
S.~Mun, S.~Nam, H.~Jang, D.~Kim, and H.~Lee.
\newblock A robust blind watermarking using convolutional neural network.
\newblock \emph{arXiv.org}, abs/1704.03248, 2017.

\bibitem[Nguyen et~al.(2021)Nguyen, Bui, Swaminathan, and Collomosse]{NguyenICCV2021}
E.~Nguyen, T.~Bui, V.~V. Swaminathan, and J.~P. Collomosse.
\newblock {OSCAR-Net}: {Object}-centric scene graph attention for image attribution.
\newblock In \emph{Proc. of the IEEE International Conference on Computer Vision (ICCV)}, 2021.

\bibitem[Oh et~al.(2019)Oh, Schiele, and Fritz]{OhARXIV2017b}
S.~J. Oh, B.~Schiele, and M.~Fritz.
\newblock \emph{Towards Reverse-Engineering Black-Box Neural Networks}, pages 121--144.
\newblock 2019.
\newblock In book: Explainable AI: Interpreting, Explaining and Visualizing Deep Learning.

\bibitem[Orekondy et~al.(2019)Orekondy, Schiele, and Fritz]{OrekondyCVPR2019}
T.~Orekondy, B.~Schiele, and M.~Fritz.
\newblock Knockoff nets: Stealing functionality of black-box models.
\newblock In \emph{Proc. of the IEEE Conference on Computer Vision and Pattern Recognition (CVPR)}, 2019.

\bibitem[Orekondy et~al.(2020)Orekondy, Schiele, and Fritz]{OrekondyICLR2020}
T.~Orekondy, B.~Schiele, and M.~Fritz.
\newblock Prediction poisoning: Towards defenses against {DNN} model stealing attacks.
\newblock In \emph{Proc. of the International Conference on Learning Representations (ICLR)}, 2020.

\bibitem[Pan et~al.(2023)Pan, Zeng, Lin, Yu, Hsieh, and Jia]{PanNIPSWORK2023}
M.~Pan, Y.~Zeng, X.~Lin, N.~Yu, C.-J. Hsieh, and R.~Jia.
\newblock {AnchMark}: {Anchor}-contrastive watermarking vs {GenAI}-based image modifications.
\newblock In \emph{Advances in Neural Information Processing Systems (NeurIPS) Workshops}, 2023.

\bibitem[Pan et~al.(2024)Pan, Zeng, Lin, Yu, Hsieh, Henderson, and Jia]{PanARXIV2024}
M.~Pan, Y.~Zeng, X.~Lin, N.~Yu, C.-J. Hsieh, P.~Henderson, and R.~Jia.
\newblock {JIGMARK}: {A} black-box approach for enhancing image watermarks against diffusion model edits.
\newblock \emph{arXiv.org}, abs/406.03720, 2024.

\bibitem[Papernot et~al.(2017)Papernot, McDaniel, Goodfellow, Jha, Celik, and Swami]{PapernotASIACCS2017}
N.~Papernot, P.~McDaniel, I.~Goodfellow, S.~Jha, Z.~B. Celik, and A.~Swami.
\newblock Practical black-box attacks against machine learning.
\newblock In \emph{Proc. of the ACM on Asia Conference on Computer and Communications Security (AsiaCCS)}. ACM, 2017.

\bibitem[Papernot et~al.(2018)Papernot, McDaniel, Sinha, and Wellman]{PapernotEUROSP2018}
N.~Papernot, P.~D. McDaniel, A.~Sinha, and M.~P. Wellman.
\newblock {SoK}: {Security} and privacy in machine learning.
\newblock In \emph{Proc. of the {IEEE} European Symposium on Security and Privacy (EuroS{\&}P)}, 2018.

\bibitem[Parah et~al.(2016)Parah, Sheikh, Loan, and Bhat]{ParahDSP2016}
S.~A. Parah, J.~A. Sheikh, N.~A. Loan, and G.~M. Bhat.
\newblock Robust and blind watermarking technique in {DCT} domain using inter-block coefficient differencing.
\newblock \emph{Digital Signal Processing}, 53:\penalty0 11--24, 2016.

\bibitem[Petitcolas(2000)]{PetitcolasSPM2000}
F.~A.~P. Petitcolas.
\newblock Watermarking schemes evaluation.
\newblock \emph{Signal Processing Magazine}, 17\penalty0 (5), 2000.

\bibitem[Piet et~al.(2025)Piet, Sitawarin, Fang, Mu, and Wagner]{PietARXIV2023}
J.~Piet, C.~Sitawarin, V.~Fang, N.~Mu, and D.~A. Wagner.
\newblock {MarkMyWords}: Analyzing and evaluating language model watermarks.
\newblock In \emph{{IEEE} Conference on Secure and Trustworthy Machine Learning (SaTML)}, 2025.

\bibitem[Radford et~al.(2021)Radford, Kim, Hallacy, Ramesh, Goh, Agarwal, Sastry, Askell, Mishkin, Clark, et~al.]{RadfordICML2021}
A.~Radford, J.~W. Kim, C.~Hallacy, A.~Ramesh, G.~Goh, S.~Agarwal, G.~Sastry, A.~Askell, P.~Mishkin, J.~Clark, et~al.
\newblock Learning transferable visual models from natural language supervision.
\newblock In \emph{Proc. of the International Conference on Machine Learning (ICML)}, 2021.

\bibitem[Roman et~al.(2024)Roman, Fernandez, D{\'{e}}fossez, Furon, Tran, and Elsahar]{RomanARXIV2024}
R.~S. Roman, P.~Fernandez, A.~D{\'{e}}fossez, T.~Furon, T.~Tran, and H.~Elsahar.
\newblock Proactive detection of voice cloning with localized watermarking.
\newblock In \emph{Proc. of the International Conference on Machine Learning (ICML)}, 2024.

\bibitem[Rombach et~al.(2022)Rombach, Blattmann, Lorenz, Esser, and Ommer]{RombachARXIV2022}
R.~Rombach, A.~Blattmann, D.~Lorenz, P.~Esser, and B.~Ommer.
\newblock High-resolution image synthesis with latent diffusion models.
\newblock In \emph{Proc. of the IEEE Conference on Computer Vision and Pattern Recognition (CVPR)}, 2022.

\bibitem[Saberi et~al.(2024{\natexlab{a}})Saberi, Sadasivan, Rezaei, Kumar, Chegini, Wang, and Feizi]{SaberiARXIV2023}
M.~Saberi, V.~S. Sadasivan, K.~Rezaei, A.~Kumar, A.~M. Chegini, W.~Wang, and S.~Feizi.
\newblock Robustness of {AI}-image detectors: Fundamental limits and practical attacks.
\newblock In \emph{Proc. of the International Conference on Learning Representations (ICLR)}, 2024{\natexlab{a}}.

\bibitem[Saberi et~al.(2024{\natexlab{b}})Saberi, Sadasivan, Zarei, Mahdavifar, and Feizi]{SaberiARXIV2024}
M.~Saberi, V.~S. Sadasivan, A.~Zarei, H.~Mahdavifar, and S.~Feizi.
\newblock {DREW}: Towards robust data provenance by leveraging error-controlled watermarking.
\newblock \emph{arXiv.org}, abs/2406.02836, 2024{\natexlab{b}}.

\bibitem[Sander et~al.(2025)Sander, Fernandez, Durmus, Furon, and Douze]{sander2024watermark}
T.~Sander, P.~Fernandez, A.~Durmus, T.~Furon, and M.~Douze.
\newblock Watermark anything with localized messages.
\newblock In \emph{Proc. of the International Conference on Learning Representations (ICLR)}, 2025.

\bibitem[Scao et~al.(2022)Scao, Fan, Akiki, Pavlick, Ilic, Hesslow, Castagn{\'{e}}, Luccioni, Yvon, Gall{\'{e}}, Tow, Rush, Biderman, Webson, Ammanamanchi, Wang, Sagot, Muennighoff, del Moral, Ruwase, Bawden, Bekman, McMillan{-}Major, Beltagy, Nguyen, Saulnier, Tan, Suarez, Sanh, Lauren{\c{c}}on, Jernite, Launay, Mitchell, Raffel, Gokaslan, Simhi, Soroa, Aji, Alfassy, Rogers, Nitzav, Xu, Mou, Emezue, Klamm, Leong, van Strien, Adelani, and et~al.]{LescaoARXIV2022}
T.~L. Scao, A.~Fan, C.~Akiki, E.~Pavlick, S.~Ilic, D.~Hesslow, R.~Castagn{\'{e}}, A.~S. Luccioni, F.~Yvon, M.~Gall{\'{e}}, J.~Tow, A.~M. Rush, S.~Biderman, A.~Webson, P.~S. Ammanamanchi, T.~Wang, B.~Sagot, N.~Muennighoff, A.~V. del Moral, O.~Ruwase, R.~Bawden, S.~Bekman, A.~McMillan{-}Major, I.~Beltagy, H.~Nguyen, L.~Saulnier, S.~Tan, P.~O. Suarez, V.~Sanh, H.~Lauren{\c{c}}on, Y.~Jernite, J.~Launay, M.~Mitchell, C.~Raffel, A.~Gokaslan, A.~Simhi, A.~Soroa, A.~F. Aji, A.~Alfassy, A.~Rogers, A.~K. Nitzav, C.~Xu, C.~Mou, C.~Emezue, C.~Klamm, C.~Leong, D.~van Strien, D.~I. Adelani, and et~al.
\newblock {BLOOM:} {A} 176b-parameter open-access multilingual language model.
\newblock \emph{arXiv.org}, abs/2211.05100, 2022.

\bibitem[Shensa(1992)]{ShensaTSP1992}
M.~J. Shensa.
\newblock The discrete wavelet transform: {Wedding} the \`{a} trous and {Mallat} algorithms.
\newblock \emph{IEEE Trans. on Signal Processing (TSP)}, 40\penalty0 (10):\penalty0 2464--2482, 1992.

\bibitem[Simes(1986)]{SimesB1986}
R.~J. Simes.
\newblock An improved {Bonferroni} procedure for multiple tests of significance.
\newblock \emph{Biometrika}, 73\penalty0 (3):\penalty0 751--754, 1986.

\bibitem[Singh and Singh(2017)]{SinghMTA2017}
D.~Singh and S.~K. Singh.
\newblock {DWT-SVD} and {DCT} based robust and blind watermarking scheme for copyright protection.
\newblock \emph{Multimedia Tools and Applications}, 76\penalty0 (11):\penalty0 13001--13024, 2017.

\bibitem[Sohl{-}Dickstein et~al.(2015)Sohl{-}Dickstein, Weiss, Maheswaranathan, and Ganguli]{SohlDicksteinICML2015}
J.~Sohl{-}Dickstein, E.~A. Weiss, N.~Maheswaranathan, and S.~Ganguli.
\newblock Deep unsupervised learning using nonequilibrium thermodynamics.
\newblock In \emph{Proc. of the International Conference on Machine Learning (ICML)}, 2015.

\bibitem[Stutz et~al.(2020)Stutz, Hein, and Schiele]{StutzICML2020}
D.~Stutz, M.~Hein, and B.~Schiele.
\newblock Confidence-calibrated adversarial training: Generalizing to unseen attacks.
\newblock In \emph{Proc. of the International Conference on Machine Learning (ICML)}, 2020.

\bibitem[Stutz et~al.(2023)Stutz, Roy, Matejovicova, Strachan, Cemgil, and Doucet]{StutzTMLR2023}
D.~Stutz, A.~G. Roy, T.~Matejovicova, P.~Strachan, A.~T. Cemgil, and A.~Doucet.
\newblock Conformal prediction under ambiguous ground truth.
\newblock \emph{{Transactions on Machine Learning Research}}, 2023.

\bibitem[Su and Chen(2018)]{SuSC2018}
Q.~Su and B.~Chen.
\newblock Robust color image watermarking technique in the spatial domain.
\newblock \emph{Soft Computing}, 22\penalty0 (1):\penalty0 91--106, 2018.

\bibitem[Szegedy et~al.(2014)Szegedy, Zaremba, Sutskever, Bruna, Erhan, Goodfellow, and Fergus]{SzegedyICLR2014}
C.~Szegedy, W.~Zaremba, I.~Sutskever, J.~Bruna, D.~Erhan, I.~J. Goodfellow, and R.~Fergus.
\newblock Intriguing properties of neural networks.
\newblock In \emph{Proc. of the International Conference on Learning Representations (ICLR)}, 2014.

\bibitem[Tancik et~al.(2020)Tancik, Mildenhall, and Ng]{TancikCVPR2020}
M.~Tancik, B.~Mildenhall, and R.~Ng.
\newblock {StegaStamp}: {Invisible} hyperlinks in physical photographs.
\newblock In \emph{Proc. of the IEEE Conference on Computer Vision and Pattern Recognition (CVPR)}, 2020.

\bibitem[Touvron et~al.(2023)Touvron, Lavril, Izacard, Martinet, Lachaux, Lacroix, Rozi{\`{e}}re, Goyal, Hambro, Azhar, Rodriguez, Joulin, Grave, and Lample]{TouvronARXIV2023}
H.~Touvron, T.~Lavril, G.~Izacard, X.~Martinet, M.~Lachaux, T.~Lacroix, B.~Rozi{\`{e}}re, N.~Goyal, E.~Hambro, F.~Azhar, A.~Rodriguez, A.~Joulin, E.~Grave, and G.~Lample.
\newblock {LLaMA}: Open and efficient foundation language models.
\newblock \emph{arXiv.org}, abs/2302.13971, 2023.

\bibitem[Tram{\`{e}}r and Boneh(2019)]{TramerARXIV2019}
F.~Tram{\`{e}}r and D.~Boneh.
\newblock Adversarial training and robustness for multiple perturbations.
\newblock In \emph{Advances in Neural Information Processing Systems (NeurIPS)}, 2019.

\bibitem[Tram{\`{e}}r et~al.(2016)Tram{\`{e}}r, Zhang, Juels, Reiter, and Ristenpart]{TramerUSENIX2016}
F.~Tram{\`{e}}r, F.~Zhang, A.~Juels, M.~K. Reiter, and T.~Ristenpart.
\newblock Stealing machine learning models via prediction {API}s.
\newblock In \emph{USENIX Security Symposium}, 2016.

\bibitem[Vaccari and Chadwick(2020)]{Vaccari2020}
C.~Vaccari and A.~Chadwick.
\newblock Deepfakes and disinformation: Exploring the impact of synthetic political video on deception, uncertainty, and trust in news.
\newblock \emph{Social Media + Society}, 6\penalty0 (1), 2020.

\bibitem[Van~Schyndel et~al.(1994)Van~Schyndel, Tirkel, and Osborne]{VanSchyndelICIP1994}
R.~G. Van~Schyndel, A.~Z. Tirkel, and C.~F. Osborne.
\newblock A digital watermark.
\newblock In \emph{Proc. of the IEEE International Conference on Image Processing (ICIP)}, 1994.

\bibitem[Wan et~al.(2022)Wan, Wang, Zhang, Li, Yu, and Sun]{WanNEUROCOMPUTING2022}
W.~Wan, J.~Wang, Y.~Zhang, J.~Li, H.~Yu, and J.~Sun.
\newblock A comprehensive survey on robust image watermarking.
\newblock \emph{Neurocomputing}, 488, 2022.

\bibitem[Wang et~al.(2001)Wang, Lin, and Lin]{WangPR2001}
R.~Wang, C.~Lin, and J.~Lin.
\newblock Image hiding by optimal {LSB} substitution and genetic algorithm.
\newblock \emph{Pattern Recognition}, 34\penalty0 (3):\penalty0 671--683, 2001.

\bibitem[Wang et~al.(2004)Wang, Bovik, Sheikh, and Simoncelli]{WangTIP2004}
Z.~Wang, A.~C. Bovik, H.~R. Sheikh, and E.~P. Simoncelli.
\newblock Image quality assessment: From error visibility to structural similarity.
\newblock \emph{IEEE Trans. on Image Processing (TIP)}, 13\penalty0 (4):\penalty0 600--612, 2004.

\bibitem[Weidinger et~al.(2021)Weidinger, Mellor, Rauh, Griffin, Uesato, Huang, Cheng, Glaese, Balle, Kasirzadeh, Kenton, Brown, Hawkins, Stepleton, Biles, Birhane, Haas, Rimell, Hendricks, Isaac, Legassick, Irving, and Gabriel]{WeidingerARXIV2021}
L.~Weidinger, J.~Mellor, M.~Rauh, C.~Griffin, J.~Uesato, P.~Huang, M.~Cheng, M.~Glaese, B.~Balle, A.~Kasirzadeh, Z.~Kenton, S.~Brown, W.~Hawkins, T.~Stepleton, C.~Biles, A.~Birhane, J.~Haas, L.~Rimell, L.~A. Hendricks, W.~Isaac, S.~Legassick, G.~Irving, and I.~Gabriel.
\newblock Ethical and social risks of harm from language models.
\newblock \emph{arXiv.org}, abs/2112.04359, 2021.

\bibitem[Wen and Ayd{\"{o}}re(2019)]{WenARXIV2019}
B.~Wen and S.~Ayd{\"{o}}re.
\newblock {ROMark}: {A} robust watermarking system using adversarial training.
\newblock \emph{arXiv.org}, abs/1910.01221, 2019.

\bibitem[Wen et~al.(2023)Wen, Kirchenbauer, Geiping, and Goldstein]{WenARXIV2023}
Y.~Wen, J.~Kirchenbauer, J.~Geiping, and T.~Goldstein.
\newblock Tree-ring watermarks: Fingerprints for diffusion images that are invisible and robust.
\newblock \emph{arXiv.org}, abs/2305.20030, 2023.

\bibitem[Xu et~al.(2025)Xu, Hu, Lei, Li, Lowe, Gorevski, Wang, Ching, Deng, et~al.]{xu2024invismark}
R.~Xu, M.~Hu, D.~Lei, Y.~Li, D.~Lowe, A.~Gorevski, M.~Wang, E.~Ching, A.~Deng, et~al.
\newblock {InvisMark:} invisible and robust watermarking for {AI}-generated image provenance.
\newblock In \emph{Proc. of the IEEE Winter Conference on Applications of Computer Vision (WACV)}, 2025.

\bibitem[Xuehua(2010)]{Xuehua2010}
J.~Xuehua.
\newblock Digital watermarking and its application in image copyright protection.
\newblock In \emph{Proc. of the International Conference on Intelligent Computation Technology and Automation}, volume~2, pages 114--117. IEEE, 2010.

\bibitem[Yang et~al.(2024)Yang, Zeng, Chen, Fang, Zhang, and Yu]{YangARXIV2024}
Z.~Yang, K.~Zeng, K.~Chen, H.~Fang, W.~Zhang, and N.~H. Yu.
\newblock Gaussian shading: Provable performance-lossless image watermarking for diffusion models.
\newblock In \emph{Proc. of the IEEE Conference on Computer Vision and Pattern Recognition (CVPR)}, 2024.

\bibitem[Yu et~al.(2021)Yu, Skripniuk, Abdelnabi, and Fritz]{YuICCV2021b}
N.~Yu, V.~Skripniuk, S.~Abdelnabi, and M.~Fritz.
\newblock Artificial fingerprinting for generative models: Rooting deepfake attribution in training data.
\newblock In \emph{Proc. of the IEEE International Conference on Computer Vision (ICCV)}, 2021.

\bibitem[Yu et~al.(2001)Yu, Tsai, and Lin]{YuSP2001}
P.~Yu, H.~Tsai, and J.~Lin.
\newblock Digital watermarking based on neural networks for color images.
\newblock \emph{Signal Processing}, 81\penalty0 (3):\penalty0 663--671, 2001.

\bibitem[Zellers et~al.(2019)Zellers, Holtzman, Rashkin, Bisk, Farhadi, Roesner, and Choi]{ZellersNIPS2019}
R.~Zellers, A.~Holtzman, H.~Rashkin, Y.~Bisk, A.~Farhadi, F.~Roesner, and Y.~Choi.
\newblock Defending against neural fake news.
\newblock In \emph{Advances in Neural Information Processing Systems (NeurIPS)}, 2019.

\bibitem[Zhang et~al.(2020{\natexlab{a}})Zhang, Chen, Xiao, Li, Boning, and Hsieh]{ZhangARXIV2019}
H.~Zhang, H.~Chen, C.~Xiao, B.~Li, D.~S. Boning, and C.~Hsieh.
\newblock Towards stable and efficient training of verifiably robust neural networks.
\newblock In \emph{Proc. of the International Conference on Learning Representations (ICLR)}, 2020{\natexlab{a}}.

\bibitem[Zhang et~al.(2023)Zhang, Edelman, Francati, Venturi, Ateniese, and Barak]{ZhangARXIV2023b}
H.~Zhang, B.~L. Edelman, D.~Francati, D.~Venturi, G.~Ateniese, and B.~Barak.
\newblock Watermarks in the sand: Impossibility of strong watermarking for generative models.
\newblock In \emph{Proc. of the International Conference on Learning Representations (ICLR) Workshops}, 2023.

\bibitem[Zhang et~al.(2019)Zhang, Xu, Cuesta{-}Infante, and Veeramachaneni]{ZhangARXIV2019e}
K.~A. Zhang, L.~Xu, A.~Cuesta{-}Infante, and K.~Veeramachaneni.
\newblock Robust invisible video watermarking with attention.
\newblock \emph{arXiv.org}, abs/1909.01285, 2019.

\bibitem[Zhang et~al.(2024)Zhang, Liu, Martin, Bearfield, Brun, and Guan]{ZhangARXIV2024}
L.~Zhang, X.~Liu, A.~V. Martin, C.~X. Bearfield, Y.~Brun, and H.~Guan.
\newblock Robust image watermarking using stable diffusion.
\newblock \emph{arXiv.org}, abs/2401.04247, 2024.

\bibitem[Zhang et~al.(2018)Zhang, Isola, Efros, Shechtman, and Wang]{ZhangARXIV2018}
R.~Zhang, P.~Isola, A.~A. Efros, E.~Shechtman, and O.~Wang.
\newblock The unreasonable effectiveness of deep features as a perceptual metric.
\newblock In \emph{Proc. of the IEEE Conference on Computer Vision and Pattern Recognition (CVPR)}, 2018.

\bibitem[Zhang et~al.(2020{\natexlab{b}})Zhang, Sun, Karaman, and Chang]{Zhang2020}
X.~Zhang, Z.~H. Sun, S.~Karaman, and S.~Chang.
\newblock Discovering image manipulation history by pairwise relation and forensics tools.
\newblock \emph{{IEEE} Journal of Selected Topics in Signal Processing}, 14\penalty0 (5):\penalty0 1012--1023, 2020{\natexlab{b}}.

\bibitem[Zhao et~al.(2024)Zhao, Zhang, Su, Vasan, Grishchenko, Kruegel, Vigna, Wang, and Li]{ZhaoARXIV2023a}
X.~Zhao, K.~Zhang, Z.~Su, S.~Vasan, I.~Grishchenko, C.~Kruegel, G.~Vigna, Y.-X. Wang, and L.~Li.
\newblock Invisible image watermarks are provably removable using generative {AI}.
\newblock In \emph{Advances in Neural Information Processing Systems (NeurIPS)}, 2024.

\bibitem[Zhao et~al.(2025)Zhao, Gunn, Christ, Fairoze, Fabrega, Carlini, Garg, Hong, Nasr, Tramer, et~al.]{ZhaoARXIV2024}
X.~Zhao, S.~Gunn, M.~Christ, J.~Fairoze, A.~Fabrega, N.~Carlini, S.~Garg, S.~Hong, M.~Nasr, F.~Tramer, et~al.
\newblock {SoK}: Watermarking for {AI}-generated content.
\newblock In \emph{Proc. of the IEEE Symposium on Security and Privacy}, 2025.

\bibitem[Zhu et~al.(2018)Zhu, Kaplan, Johnson, and Fei{-}Fei]{ZhuECCV2018}
J.~Zhu, R.~Kaplan, J.~Johnson, and L.~Fei{-}Fei.
\newblock {HiDDeN}: Hiding data with deep networks.
\newblock In \emph{Proc. of the European Conference on Computer Vision (ECCV)}, 2018.

\end{thebibliography}

\clearpage
\appendix
\section{Additional robustness evaluations}

\subsection{Native image resolution evaluation \label{app:native_image}}

While Section~\ref{sec:results} evaluated all models at a 512$\times$512 resolution, this section evaluates all baselines as well as  \synthidom using the native resolution of images in \imagenet.
The average resolution of \imagenet images is 469$\times$387, which means that methods that operate at larger resolutions may come at a slight disadvantage since images will generally be downscaled compared to the resolution at which they are watermarked (due to the intrinsic resizing stemming from variable image resolutions).

\noindent Table~\ref{tab:results_tpr_0.1_globalfpr_aggregated_source_image_size} summarizes the results.
It reports the true positive rate (TPR) at a fixed false positive rate of 0.1\% in average on \textit{worst}-case transformations.
We observe that over aggregated \textit{random} and aggregated \textit{worst} transformations, %
\external{\synthidom continues to outperform} all other methods by significant margins (i.e., %
\external{$+10.64$ and $+17.17$} percentage points respectively).
In the \textit{worst} setting,
\external{\synthidom is the only models surpassing 97\% TPR in worst-case aggregate (which is the most challenging setting).
}
\vspace{1cm}

\external{%
\begin{table}[h]
\resizebox{\textwidth}{!}{%
\begin{tabular}{lcccccccc}
\toprule
\rowcolor[HTML]{EFEFEF}  & \rotatebox[origin=c]{45}{~\synthidom~} & \rotatebox[origin=c]{45}{~\invismark~} & \rotatebox[origin=c]{45}{~\stegastamp~} & \rotatebox[origin=c]{45}{~\trustmarkp~} & \rotatebox[origin=c]{45}{~\trustmarkq~} & \rotatebox[origin=c]{45}{~\videosealzero~} & \rotatebox[origin=c]{45}{~\videosealone~} & \rotatebox[origin=c]{45}{~\wam~} \\
\midrule
Aggregated & \textbf{98.72\%} & [65.60\%, 65.88\%] & [72.24\%, 72.43\%] & [64.81\%, 65.31\%] & [77.24\%, 77.70\%] & [76.60\%, 77.30\%] & \underline{[87.93\%, 88.08\%]} & 86.79\% \\
Aggregated Worst & \textbf{97.22\%} & [56.72\%, 56.98\%] & [66.21\%, 66.48\%] & [52.78\%, 53.36\%] & [68.06\%, 68.21\%] & [58.69\%, 59.38\%] & [76.31\%, 76.49\%] & \underline{80.05\%} \\
\bottomrule
\end{tabular}
}
\caption{TPR at 0.1\% FPR aggregated across \textit{random} and \textit{worst-case} categories. For each model, the detection threshold is calibrated to reach 0.1\% FPR across all \textit{worst} transformations in average. When brackets are used, we display the range of TPR values that can accommodate the target FPR target(this number is not unique when there are ties). We highlight the best model in bold and the second best with an underline.\label{tab:results_tpr_0.1_globalfpr_aggregated_source_image_size}}
\end{table}
}

\clearpage
\subsection{Native model resolution evaluation \label{app:native_model}}

While the previous section evaluated all models using the native resolution of images in \imagenet, this section evaluates all baselines as well as  \synthidom at their preferred resolution.
Table~\ref{tab:results_tpr_0.1_globalfpr_aggregated_model_size} reports the true positive rate (TPR) at a fixed false positive rate of 0.1\% in average on \textit{worst}-case transformations.
\vspace{1cm}

\external{%
\begin{table}[h]
\resizebox{\textwidth}{!}{%
\begin{tabular}{lcccccccc}
\toprule
\rowcolor[HTML]{EFEFEF}  & \rotatebox[origin=c]{45}{\shortstack[c]{~\synthidom~\\(512$\times$512)}} & \rotatebox[origin=c]{45}{\shortstack[c]{~\invismark~\\(256$\times$256)}} & \rotatebox[origin=c]{45}{\shortstack[c]{~\stegastamp~\\(400$\times$400)}} & \rotatebox[origin=c]{45}{\shortstack[c]{~\trustmarkp~\\(256$\times$256)}} & \rotatebox[origin=c]{45}{\shortstack[c]{~\trustmarkq~\\(256$\times$256)}} & \rotatebox[origin=c]{45}{\shortstack[c]{~\videosealzero~\\(256$\times$256)}} & \rotatebox[origin=c]{45}{\shortstack[c]{~\videosealone~\\(256$\times$256)}} & \rotatebox[origin=c]{45}{\shortstack[c]{~\wam~\\(256$\times$256)}} \\
\midrule
Aggregated & \textbf{99.98\%} & [61.45\%, 61.77\%] & [72.42\%, 72.58\%] & [58.91\%, 59.54\%] & [77.69\%, 78.11\%] & [64.53\%, 65.45\%] & \underline{[86.56\%, 86.78\%]} & 82.32\% \\
Aggregated Worst & \textbf{99.72\%} & [51.11\%, 51.40\%] & [66.49\%, 66.78\%] & [47.89\%, 48.38\%] & [67.61\%, 67.78\%] & [52.08\%, 52.60\%] & [78.73\%, 78.95\%] & \underline{79.38\%} \\
\bottomrule
\end{tabular}
}
\caption{TPR at 0.1\% FPR aggregated across each transformation category when images are resized to the model input size. For each model, the detection threshold is calibrated to reach 0.1\% FPR across all \textit{worst} transformations in average. When brackets are used, we display the range of TPR values that can accommodate the target FPR target(this number is not unique when there are ties). We highlight the best model in bold and the second best with an underline.\label{tab:results_tpr_0.1_globalfpr_aggregated_model_size}}
\end{table}
}

\clearpage

\section{Human evaluation prompts}
\label{sec:prompts}

Below is the list of prompts used to generated images for our quality evaluations. We used \imagen~2 to generate 1000 images (i.e., 4 images per prompt).

{
\tiny
\begin{longtable}{p{.8\textwidth}p{.2\textwidth}}
\rowcolor[HTML]{EFEFEF} 
Realistic 3d render of a happy, furry and cute baby tiger smiling with big eyes looking straight at you, Pixar style, full body shot with a light blue background & 3D Rendering \\
\rowcolor[HTML]{EFEFEF} 
Detailed robot portrait, shading details & 3D Rendering \\
\rowcolor[HTML]{EFEFEF} 
A 3D rendering of a spaceship. & 3D Rendering \\
\rowcolor[HTML]{EFEFEF} 
3d rendering of a sleek, silver futuristic car with a glass roof and a glowing blue light on the front. & 3D Rendering \\
\rowcolor[HTML]{EFEFEF} 
Isometric 3D render of a small Japanese house with a thatched roof, surrounded by bamboo and cherry blossoms. & 3D Rendering \\
\rowcolor[HTML]{EFEFEF} 
3D render of 3 planets orbiting around themselves in space. The planets are made of different colors and textures. The background is black. & 3D Rendering \\
\rowcolor[HTML]{EFEFEF} 
3D pixelated render of a panda in the style of Doom, standing on a rocky platform, with a lava background. & 3D Rendering \\
\rowcolor[HTML]{EFEFEF} 
Ray-traced 3D rendering of a crystal ball on a table in a dark room. The ball is reflecting the light from a nearby candle. & 3D Rendering \\
\rowcolor[HTML]{EFEFEF} 
3D render of Gandalf fighting Balrog in a dark fantasy setting. The scene is rendered in Unity with realistic graphics. & 3D Rendering \\
\rowcolor[HTML]{EFEFEF} 
Exciting car race from a 3D computer game, rendered in Unity, & 3D Rendering \\
Abstract art, very regular shapes and lines. & Abstract Art \\
Cubist painting, digital art, colourful, patchwork & Abstract Art \\
An abstract expressionist painting using bold brush strokes, dripping paint, and vibrant colors, reflecting the inner emotions and turmoil & Abstract Art \\
A regular black and white triangle pattern on a white background. The pattern is made of equilateral triangles, and the triangles are arranged in a grid. & Abstract Art \\
Abstract painting with all the colors of the rainbow, made with acrylic paint on canvas. & Abstract Art \\
Abstract art in greyscale. & Abstract Art \\
Abstract art piece of several overlapping circles in bright colors & Abstract Art \\
An art piece consisting of a lot of eyes painting in different style, abstract & Abstract Art \\
Drawing of different traffic signs chaotically arranged, abstract art & Abstract Art \\
Abstract art painting made with oil, different shades of green in a triangle shapes & Abstract Art \\
\rowcolor[HTML]{EFEFEF} 
Painting of bear hiking in the mountains & Animal Painting \\
\rowcolor[HTML]{EFEFEF} 
Ukiyo-e style cat & Animal Painting \\
\rowcolor[HTML]{EFEFEF} 
Blacklight painting style cat, painting, blacklight painting & Animal Painting \\
\rowcolor[HTML]{EFEFEF} 
A water color painting of a flamingo, Kilimanjaro in the background & Animal Painting \\
\rowcolor[HTML]{EFEFEF} 
A panda in the style of Miro & Animal Painting \\
\rowcolor[HTML]{EFEFEF} 
Panda in the style of Picasso, Cubism & Animal Painting \\
\rowcolor[HTML]{EFEFEF} 
Abstract painting of a horse with broad brush strokes using only 2 colors & Animal Painting \\
\rowcolor[HTML]{EFEFEF} 
Oil painting of a close-up of a cat eye & Animal Painting \\
\rowcolor[HTML]{EFEFEF} 
A photorealistic painting of a family of African elephants with long tusks & Animal Painting \\
\rowcolor[HTML]{EFEFEF} 
An impressionist painting of a butterfly & Animal Painting \\
Black and white picture of a swan swimming & Animal Photo \\
A picture of a cow on an alpine meadow. & Animal Photo \\
A herd of zebras standing on the surface of mars. The sun is shining. High quality photograph. 4k. & Animal Photo \\
Photograph of a coral reef with lots of colorful fishes swimming around. The reef is lit by sunlight coming through the water. & Animal Photo \\
Close up photo of a dog with the pop rocket instagram color filter. The dog is a golden retriever, and it is sitting on a red blanket. The background is a blur of green grass. & Animal Photo \\
Polaroid photo of a horse, taken in a park on a sunny day. 70s photo. Polaroid photo with yellowed edges. & Animal Photo \\
Close-up of a cat eye & Animal Photo \\
A full-body photograph of a Sumatran tiger roaring, with blurred background & Animal Photo \\
A photograph of a Humpback whale and its calf playing underwater & Animal Photo \\
Close up photograph of a group of ants. & Animal Photo \\
\rowcolor[HTML]{EFEFEF} 
Anime drawing, portrait sunny weather, wearing sun hat, shadow hiding part of the face & Cartoon \\
\rowcolor[HTML]{EFEFEF} 
A still from an 80s cartoon show about a butterfly with superpowers. The butterfly is wearing a cape and has a glowing red heart on its chest. It is flying through the air, surrounded by stars. & Cartoon \\
\rowcolor[HTML]{EFEFEF} 
A male anime character with black hair runs through a deep forest. He is wearing a red ninja suit and has a sword in his hand. The sun is shining through the trees, creating a dappled pattern on the ground. anime, cartoon & Cartoon \\
\rowcolor[HTML]{EFEFEF} 
A black and white cartoon of a giraffe driving a red car. The giraffe is wearing a top hat and a bow tie, and he is smiling. The car is driving down a road, and there is a palm tree in the background. & Cartoon \\
\rowcolor[HTML]{EFEFEF} 
A cartoony, colorful icon of a mobile game app. The icon is of a smiling cat wearing a hat and sunglasses, and it is sitting on a pile of gold coins. The background is a bright blue sky with clouds. & Cartoon \\
\rowcolor[HTML]{EFEFEF} 
A close-up portrait of a cute character in the style of Studio Ghibli, with big expressive eyes and a bright smile. & Cartoon \\
\rowcolor[HTML]{EFEFEF} 
A dark and atmospheric scene from a seinen anime, with a lone character walking through a rain-soaked city, lit only by the neon lights of storefronts. & Cartoon \\
\rowcolor[HTML]{EFEFEF} 
A whimsical and imaginative scene from a children's cartoon, with a group of animal friends exploring a magical forest filled with talking trees and playful fairies. & Cartoon \\
\rowcolor[HTML]{EFEFEF} 
A humorous and satirical scene from an cartoon, with a family sitting around the dinner table, engaging in witty banter and absurd situations. & Cartoon \\
\rowcolor[HTML]{EFEFEF} 
A 1970s-style cartoon of a futuristic car flying through the sky. The car is bright yellow and has a big rocket engine on the back. The sky is a deep blue and there are clouds floating by. & Cartoon \\
Defocused specks of blurry swaying coral in a landscape colorful & Defocused Blurry \\
Close-up of out-of-focus grass & Defocused Blurry \\
street lights, beautiful bokeh & Defocused Blurry \\
a wasp in front of out-of-focus meadow & Defocused Blurry \\
A black and white photo of a microphone stand with the band members standing out of focus in the back. The band members are wearing black and white clothes, and the background is a dark blue. The photo is taken in a studio & Defocused Blurry \\
Underwater photo of a sunken ship. Out of focus. Very blurry. & Defocused Blurry \\
A supermarket selection of fruits. Out of focus and very blurry. & Defocused Blurry \\
A wall of text, out of focus, unreadable, blurry & Defocused Blurry \\
Sunset at a beach, out of focus, blurry & Defocused Blurry \\
A large group of people in a street, out of focus, blurry & Defocused Blurry \\
\rowcolor[HTML]{EFEFEF} 
A portrait of a mysterious person sitting in an ornate, gothic throne room, with subtle elements of fantasy and magical realism, painted with the dark and dramatic style of the Baroque period & Fantasy Painting \\
\rowcolor[HTML]{EFEFEF} 
A dynamic action scene of a knight battling a dragon on a cliffside, with roaring waves below, inspired by the Romanticism movement, reminiscent of the works of Eugène Delacroix & Fantasy Painting \\
\rowcolor[HTML]{EFEFEF} 
A haunting gothic horror scene of an abandoned mansion on a hill, with fog, full moon, and shadowy figures, in a digital painting style reminiscent of the Castlevania game series. & Fantasy Painting \\
\rowcolor[HTML]{EFEFEF} 
Dramatic fantasy painting of a Balrog facing a wizard. & Fantasy Painting \\
\rowcolor[HTML]{EFEFEF} 
Fantasy painting of a pirate ship resting inside a canyon, with a waterfall cascading down the rocks. The ship is lit by a full moon, and the water is reflecting the stars. & Fantasy Painting \\
\rowcolor[HTML]{EFEFEF} 
A group of orcs with axes in a forest, oil painting & Fantasy Painting \\
\rowcolor[HTML]{EFEFEF} 
Detailed painting of a wizard casting a spell, she has a magic wand and sparkles fly out of it & Fantasy Painting \\
\rowcolor[HTML]{EFEFEF} 
A beautiful painting of a unicorn grazing in a fairy tale forest & Fantasy Painting \\
\rowcolor[HTML]{EFEFEF} 
A fantasy dwarf with a long beard sitting at a camp fire, painting & Fantasy Painting \\
\rowcolor[HTML]{EFEFEF} 
Dramatic and rich painting of a medieval city with fireballs falling from the sky & Fantasy Painting \\
A bear dancing disco. Flat art style. & Flat art \\
Hedgehog in the grass and a tree in flat art style & Flat art \\
A large cruise ship with lots of flags on it. Flat art style. & Flat art \\
Flat art style postcard of a sailor. & Flat art \\
All planets of our solar system in flat art style. & Flat art \\
Picture of a flag of a fictional country. 5 diagonal blue and green stripes and a silver spoon in the top right corner. Made in flat art style. & Flat art \\
A steak with potatoes, carrots and green beans. Flat art style. & Flat art \\
Abstract flat art style image of a red radio tower with yellow lines emitting from it. White background. & Flat art \\
Lightning coming from a cloud and hitting a tree. Abstract. Minimalist. Flat art style. Cartoon in flat art style. & Flat art \\
Minimalist flat art style depiction of a woman sitting on a bench. The woman is black and wearing red clothes. Flat art style. Cartoon. Abstract. & Flat art \\
\rowcolor[HTML]{EFEFEF} 
Black and white photo of a train conductor in a blue uniform standing on a train platform in 1875 & Historic Photo \\
\rowcolor[HTML]{EFEFEF} 
A historic photograph of the Parthenon temple in Athens, Greece, taken before 1900 & Historic Photo \\
\rowcolor[HTML]{EFEFEF} 
A historic photo of a family of 5. Mother, father, 2 sons and a daughter. & Historic Photo \\
\rowcolor[HTML]{EFEFEF} 
A farmer in a field of sun flowers. Historic photo from 1912. & Historic Photo \\
\rowcolor[HTML]{EFEFEF} 
Historic photo from 1874 of a personal computer. & Historic Photo \\
\rowcolor[HTML]{EFEFEF} 
Cows in a barn. Black and white photo from 1930. & Historic Photo \\
\rowcolor[HTML]{EFEFEF} 
Historic photo of a burning house. Taken in 1921. Grainy, bad quality. & Historic Photo \\
\rowcolor[HTML]{EFEFEF} 
Black and white photo of a 1900s Model T Ford parked on a dirt road in front of a barn. The sky is cloudy and the sun is setting. & Historic Photo \\
\rowcolor[HTML]{EFEFEF} 
Vintage photo of the Chicago skyline from the 1950s, with washed out colors and a sepia tone. & Historic Photo \\
\rowcolor[HTML]{EFEFEF} 
Vintage photo of a steam boat on a large river, with a city in the background. & Historic Photo \\
A digital painting of a serene Japanese garden in autumn, with a tea house, koi pond, and maple trees with falling leaves, with an anime-inspired art style reminiscent of Studio Ghibli films. & Landscape Painting \\
A wide-angle realistic painting of mount Bromo volcano in Indonesia erupting, with visible lava and volcanic fumes coming out of the crater & Landscape Painting \\
A dreamy watercolour painting of many cherry blossom trees in front of Mount Fuji in Japan, mainly in pastel colours & Landscape Painting \\
A realistic painting of a stormy night in the mountains. It's raining and there are lightning visible in the background. & Landscape Painting \\
Oil painting of a savanna. The painting is in the Renaissance style, with a lot of detail and bright colors. & Landscape Painting \\
Impressionist painting of the ice cold plains of Antarctica. The sky is a deep blue, and the sun is setting. There is a sense of loneliness and isolation. & Landscape Painting \\
Expressionist painting of the Grand Canyon in shades of red, orange, and yellow. The painting is done on canvas and is 3 feet by 4 feet. & Landscape Painting \\
Baroque painting of a camel caravan crossing the dunes in the desert of Tunisia. The painting is done in a realistic style, with the dunes depicted in shades of yellow and brown. The sky is a deep blue, and the sun is setting, casting long shadows across the sand. & Landscape Painting \\
A painting of the city of London in the early medieval period, with a castle in the foreground and a river in the background. The painting is done in a realistic style, with the colors muted and the details precise. & Landscape Painting \\
A modern painting of a deserted beach in the Caribbean. The painting is done in a vibrant, fauvist style, and the colors are saturated and bright. The beach is deserted, with only a few palm trees and a few rocks. The sky is a deep blue, and the water is a clear turquoise. & Landscape Painting \\
\rowcolor[HTML]{EFEFEF} 
High quality photo of a sun set on the beach. & Landscape Photo \\
\rowcolor[HTML]{EFEFEF} 
A serene autumn landscape with a reflective lake surrounded by trees with colorful fall foliage, in the style of the Hudson River School, capturing the romantic essence and fine details. & Landscape Photo \\
\rowcolor[HTML]{EFEFEF} 
A black and white street photography of a busy New York City intersection with the Empire State Building in the background, capturing the energy and chaos of city life & Landscape Photo \\
\rowcolor[HTML]{EFEFEF} 
A realistic landscape shot of the Northern Lights dancing over a snowy mountain range in Iceland, with long exposure to capture the motion and vibrant colors. & Landscape Photo \\
\rowcolor[HTML]{EFEFEF} 
A realistic photo of the Milky Way with diagonal orientation, next to the Tre Cime di Lavaredo mountain peaks in Italy & Landscape Photo \\
\rowcolor[HTML]{EFEFEF} 
A photo of a Baroque chapel on a green valley catching rays of the rising sun, with defined mountain peaks in the background & Landscape Photo \\
\rowcolor[HTML]{EFEFEF} 
A photograph of the Golden Gate bridge on a foggy sunset, with its pillars and the water underneath being illuminated by red light & Landscape Photo \\
\rowcolor[HTML]{EFEFEF} 
A dramatic black and white photo of a stone bridge leading to a lighthouse on a rocky cliff, with ocean waves crashing onto the cliff & Landscape Photo \\
\rowcolor[HTML]{EFEFEF} 
A photograph capturing the movement of black horses running on the side of an empty road which leads towards a large snowy mountain range & Landscape Photo \\
\rowcolor[HTML]{EFEFEF} 
A beautiful photo of a medieval castle in the mountains at sunset. The castle is made of gray stone and has turrets and towers. The mountains are covered in snow and the sky is a deep blue. & Landscape Photo \\
Figurative line drawing & Line Drawing \\
Blueprint for a rocket ship heading to Mars, blueprint, pencil drawing, drawing & Line Drawing \\
City skyline picasso pencil & Line Drawing \\
Sketch of the ying yang & Line Drawing \\
Architectural drawing of bank floor plan, line drawing & Line Drawing \\
A line drawing of a woman's face, with geometric patterns used to define facial features, line drawing & Line Drawing \\
Line drawing of two people arguing, pencil. & Line Drawing \\
A black and white line drawing of a skyscraper driving with wheels on a road. & Line Drawing \\
Colorful line drawing of a rooster, line drawing & Line Drawing \\
Pencil drawing of a complex machine. Line drawing, pencil drawing & Line Drawing \\
\rowcolor[HTML]{EFEFEF} 
A mascot logo of a robot, simple, vector & Logo \\
\rowcolor[HTML]{EFEFEF} 
A minimalist logo of a research company, consisting of a magnifying glass and a DNA strand. The logo is black and white, and the background is a light blue. Just the logo. & Logo \\
\rowcolor[HTML]{EFEFEF} 
A logo of a fashion company. The logo is a stylized letter "Y" in bright colors. The logo is on a white background. & Logo \\
\rowcolor[HTML]{EFEFEF} 
A complex logo made of geometric shapes and lines in a dark blue color. The logo is on a white background. & Logo \\
\rowcolor[HTML]{EFEFEF} 
A red and white logo of a games company. The logo is a wolf with sunglasses on black background. & Logo \\
\rowcolor[HTML]{EFEFEF} 
A playful logo of a AI startup making shoes. The logo is a cartoon of a shoe with a big smile on its face. The shoe is wearing a hard hat and a tool belt.  Logo on a white background. & Logo \\
\rowcolor[HTML]{EFEFEF} 
A simple, modern logo for a company called Logo. The logo is a blue circle with a white letter L in the center. The logo is on a white background. & Logo \\
\rowcolor[HTML]{EFEFEF} 
Logo for an early 2000s dot-com bubble company that sells books online. The logo is a stylized book with a rocket ship on top. & Logo \\
\rowcolor[HTML]{EFEFEF} 
A modern, minimalist logo for a new supermarket chain. The logo is a green leaf with a wavy black line circling around it. On a pink background. & Logo \\
\rowcolor[HTML]{EFEFEF} 
Logo for a new online bank. The logo has several currency symbols in triangle pattern with the word bank above it. The background is black. & Logo \\
A cubist painting of a still life with geometric shapes, fragmented forms, and multiple perspectives, inspired by the works of Pablo Picasso and Georges Braque, using a muted color palette & Object Painting \\
Minimalist painting of a flower on black background. Very minimalistic. Extremely minimalistic. Minimalism. & Object Painting \\
A painting of a knight in the style of fauvism, with bold colors and simplified shapes. & Object Painting \\
A Horse Drawn Carriage. Simplistic woodcut from 1568. & Object Painting \\
Still life painting of a calculator. Oil on canvas & Object Painting \\
Surrealist painting of a toaster melting on a white background. The toaster is made of blue and yellow plastic, and the melting is done in a realistic style. & Object Painting \\
A baroque painting of a chest full with treasure. & Object Painting \\
A painting of a pineapple in the cubist style. & Object Painting \\
An impressionist painting of a shoe. Impressionist style. & Object Painting \\
Oil painting of an very intrinsically decorated vase & Object Painting \\
\rowcolor[HTML]{EFEFEF} 
Chess board, high contrast, top view & Object Photo \\
\rowcolor[HTML]{EFEFEF} 
A close-up macro shot of morning dew on a spider’s web, with the focus on the intricate patterns and water droplets, giving it an ethereal quality, high resolution. & Object Photo \\
\rowcolor[HTML]{EFEFEF} 
Large fireworks in the night sky. & Object Photo \\
\rowcolor[HTML]{EFEFEF} 
A close up photograph of snowflakes & Object Photo \\
\rowcolor[HTML]{EFEFEF} 
Food photo of a steak with potatoes, carrots and green beans. & Object Photo \\
\rowcolor[HTML]{EFEFEF} 
A photo of a house built of match sticks. & Object Photo \\
\rowcolor[HTML]{EFEFEF} 
A table covered in various tools, high quality photograph & Object Photo \\
\rowcolor[HTML]{EFEFEF} 
Photo of an astronauts helmet, you can see the surface of the moon in the reflection & Object Photo \\
\rowcolor[HTML]{EFEFEF} 
A photograph of a broken glass lying across a marble floor. & Object Photo \\
\rowcolor[HTML]{EFEFEF} 
Very close up photo of a purple light diode & Object Photo \\
Pencil drawing of a cloud & Pencil Drawing \\
1950s cat illustration & Pencil Drawing \\
Architectural pencil drawing of a residential home. & Pencil Drawing \\
Pencil drawing of steam train. & Pencil Drawing \\
A pencil drawing of a street in Berlin. There is a döner kebap shop next to a bar. & Pencil Drawing \\
A pencil drawing of a cityscape, with shadows on the glass and concrete of tall buildings & Pencil Drawing \\
An extremely detailed pencil portrait of a black woman's face & Pencil Drawing \\
Pencil drawing of a rainbow & Pencil Drawing \\
Extremely realistic pencil drawing of an eye & Pencil Drawing \\
An inkjet printer drawn with pencil, pencil drawing & Pencil Drawing \\
\rowcolor[HTML]{EFEFEF} 
A man with a blue hat jumping on a two legged turtle. Pixel Art. & Pixel art \\
\rowcolor[HTML]{EFEFEF} 
A female doctor jumping over an abyss. Pixel art game screenshot. & Pixel art \\
\rowcolor[HTML]{EFEFEF} 
A pixelated screenshot of a 8-bit game, showing a spaceship shooting at asteroids. Pixel Art. & Pixel art \\
\rowcolor[HTML]{EFEFEF} 
Pixel art of a steampunk machine with gears, cogs, and valves. Pixel art. Pixelated. Super Nintendo graphics. & Pixel art \\
\rowcolor[HTML]{EFEFEF} 
A pirate ship on the sea, pixel art, 16-bit & Pixel art \\
\rowcolor[HTML]{EFEFEF} 
16-bit pixel art of a supermarket & Pixel art \\
\rowcolor[HTML]{EFEFEF} 
A knight on a horse riding over a draw bridge, pixel art & Pixel art \\
\rowcolor[HTML]{EFEFEF} 
A group of crocodiles in a river, pixel art & Pixel art \\
\rowcolor[HTML]{EFEFEF} 
Pixel art drawing of cars on a highway, top view, pixel art & Pixel art \\
\rowcolor[HTML]{EFEFEF} 
A desert scene with cacti and tumble weeds, 32-bit pixel art & Pixel art \\
A portrait of a man in glasses, psychedelic painting, primary colors, water color & Portrait Painting \\
Renaissance painting of a king wearing a crown, reneaissance painting & Portrait Painting \\
Abstract painting showing the portrait of women in her 20s on the left with bright colors and in her 80s on the right with darker colors & Portrait Painting \\
Painting with thick lines and bold colors of a doctor's face & Portrait Painting \\
Portrait of a female african-american professor using pointillism & Portrait Painting \\
Official painted portrait of the first Indian-American President of the United States & Portrait Painting \\
Painted portrait of a happy child, oil painting & Portrait Painting \\
Portrait painting of a female baker in traditional white clothing & Portrait Painting \\
Baroque portrait of a blacksmith in his workshop, baroque painting, painting & Portrait Painting \\
Very detailed painting of a female Chinese bank manager & Portrait Painting \\
\rowcolor[HTML]{EFEFEF} 
Close-up portrait of a man, Sepia & Portrait Photo \\
\rowcolor[HTML]{EFEFEF} 
Colorful vivid portrait of a woman, bokeh, kodachrome & Portrait Photo \\
\rowcolor[HTML]{EFEFEF} 
A powerful portrait of an elderly person, capturing the wrinkles, texture of the skin, and expressive eyes that tell a story of a lifetime, in high resolution and natural lighting. & Portrait Photo \\
\rowcolor[HTML]{EFEFEF} 
A black and white portrait of a middle-aged Vietnamese lady smiling, and wearing a traditional outfit and bamboo hat & Portrait Photo \\
\rowcolor[HTML]{EFEFEF} 
A portrait of a kid, balloons floating in the  background. & Portrait Photo \\
\rowcolor[HTML]{EFEFEF} 
Bokeh portrait shot of a group of 3 adult brothers, in a garden & Portrait Photo \\
\rowcolor[HTML]{EFEFEF} 
Portrait photo of a man in his 50s with sun glasses on, he has green colored hair and smiles & Portrait Photo \\
\rowcolor[HTML]{EFEFEF} 
A airplane pilot in her full uniform, bokeh photo & Portrait Photo \\
\rowcolor[HTML]{EFEFEF} 
Wedding photo of a couple & Portrait Photo \\
\rowcolor[HTML]{EFEFEF} 
College graduation photo of a native american man & Portrait Photo \\
Movie Poster of a 1980s comedy movie about racoons living on a college campus & Poster Art \\
A poster of a Maneki-Neko (Japanese lucky cat) and a bowl of ramen noodles on a grey background, with the cat holding a few noodles in its raised paw & Poster Art \\
A coffee guide poster with colourful icons of different kinds of coffees and brewing methods & Poster Art \\
A poster showing the dramatic cliffs and colourful houses in the Amalfi Cost, Italy, with text rendered on top “ITALY, AMALFI COAST” & Poster Art \\
Abstract poster of a pattern of thick black lines looking like a maze & Poster Art \\
Poster of an animated movie about a funny bear with title & Poster Art \\
A poster for an action movie featuring a a panda in sunglasses, holding a bamboo stick. The background is dark and stormy, and the panda is standing in front of a burning building. & Poster Art \\
An advertisement poster for a new detergent & Poster Art \\
A scientific poster explaining neural networks to children & Poster Art \\
Advertisement poster of a new smartphone from the 1960s & Poster Art \\
\rowcolor[HTML]{EFEFEF} 
A futuristic cyberpunk cityscape at night with towering neon-lit skyscrapers, flying cars, and a diverse crowd of humans and androids, in a highly detailed digital painting reminiscent of Blade Runner. & SciFi Photo \\
\rowcolor[HTML]{EFEFEF} 
A portrait of a post-apocalyptic warrior with battle scars, armor, and weapons, standing in a desolate wasteland, in a detailed digital painting with gritty textures, inspired by the Mad Max franchise. & SciFi Photo \\
\rowcolor[HTML]{EFEFEF} 
Empty spaceship interior. Dramatic light. & SciFi Photo \\
\rowcolor[HTML]{EFEFEF} 
A futuristic hover craft with cyan and orange tones & SciFi Photo \\
\rowcolor[HTML]{EFEFEF} 
A spacesuit-clad astronaut walking through a forest, surrounded by trees and ferns. The astronaut is carrying a backpack. & SciFi Photo \\
\rowcolor[HTML]{EFEFEF} 
Photo of a green alien in Paris dressed like a tourist & SciFi Photo \\
\rowcolor[HTML]{EFEFEF} 
Photo of a woman in a futuristic spacesuit floating in space & SciFi Photo \\
\rowcolor[HTML]{EFEFEF} 
Dramatic photo of a UFO hovering over a fast food restaurant & SciFi Photo \\
\rowcolor[HTML]{EFEFEF} 
Photo of a man in a futuristic jumpsuit riding a gigantic grasshopper & SciFi Photo \\
\rowcolor[HTML]{EFEFEF} 
Close up photo of a futuristic laser gun & SciFi Photo \\
Black matte cube in the middle of room of a poorly lit room & Shapes \\
A dark but slightly transparent sphere, fre-floating in front of clear sky & Shapes \\
A solid prism on a table in front of a forest & Shapes \\
An abstract sculpture out of dark metal in the middle of a traffic circle & Shapes \\
A closeup photo of a honeycomb with hexagonal prismatic cells of varying size & Shapes \\
A page of a notebook with a pencil drawing of a triangle. The notebook is open to a blank page, and the triangle is drawn in the center of the page. The pencil is resting on the page next to the triangle. & Shapes \\
Two circles overlapping (Venn diagram) & Shapes \\
An image filled with concentric circles & Shapes \\
Multiple irregular polygons in black and white & Shapes \\
A smooth white pyramid on a white floor & Shapes \\
\rowcolor[HTML]{EFEFEF} 
A scientist writing an equation on a blackboard & Stock Photo \\
\rowcolor[HTML]{EFEFEF} 
A trader looking at stock price graphs on the screen & Stock Photo \\
\rowcolor[HTML]{EFEFEF} 
A group of business professionals with different nationalities having a meeting & Stock Photo \\
\rowcolor[HTML]{EFEFEF} 
A female hacker sitting in front of a computer, she is wearing a dark hoody and the room is badly lit & Stock Photo \\
\rowcolor[HTML]{EFEFEF} 
Stock photo of a man eating a salad & Stock Photo \\
\rowcolor[HTML]{EFEFEF} 
Two men holding hands smiling at the camera & Stock Photo \\
\rowcolor[HTML]{EFEFEF} 
An Asian man in a cafe ordering, looking pensive, stock photo & Stock Photo \\
\rowcolor[HTML]{EFEFEF} 
Two women high fiving. One is African American and one is Indian American & Stock Photo \\
\rowcolor[HTML]{EFEFEF} 
Stock photo of a person having a headache & Stock Photo \\
\rowcolor[HTML]{EFEFEF} 
A woman having an idea, stock photo & Stock Photo \\
The word provenance in a funny font & Text \\
Restaurant menu written on a yellow page, written menu, full of text & Text \\
A large blackboard with lecture notes from a computer science lecture & Text \\
A small whiteboard with Maxwell's equations & Text \\
3D rendering of letters forming "HELLO", with the front surface of each 3D letter having a different inclination & Text \\
Logo of well-known web search company. White background and colorful letters. & Text \\
A white letter with the words "I love you" written on them in red & Text \\
All the letters of the alphabet & Text \\
A soup with letters in it forming the word "Soup" & Text \\
A page in a very old book & Text \\
\rowcolor[HTML]{EFEFEF} 
A vertical wallpaper that is painted to look like a bird's wing, with the wing in different shades of gray and streaks of light in the sky. & Wallpaper \\
\rowcolor[HTML]{EFEFEF} 
Golden tree branches densely packed against a dark background & Wallpaper \\
\rowcolor[HTML]{EFEFEF} 
Clouds in a blue sky, wallpaper photo & Wallpaper \\
\rowcolor[HTML]{EFEFEF} 
Matte colors in triangle shapes, wallpaper & Wallpaper \\
\rowcolor[HTML]{EFEFEF} 
Regular pattern of stick figures holding hands & Wallpaper \\
\rowcolor[HTML]{EFEFEF} 
Wallpaper of stars on a black background & Wallpaper \\
\rowcolor[HTML]{EFEFEF} 
Stones of different colors and shapes, wallpaper & Wallpaper \\
\rowcolor[HTML]{EFEFEF} 
An empty beach from above, perfect photography, clear blue water & Wallpaper \\
\rowcolor[HTML]{EFEFEF} 
Wooden texture & Wallpaper \\
\rowcolor[HTML]{EFEFEF} 
Image of a seastar, made as a tile mosaic, blue and white, phone wallpaper & Wallpaper
\end{longtable}
}

\clearpage

\end{document}